\documentclass[onecolumn,11pt]{IEEEtran}

\IEEEoverridecommandlockouts

\usepackage[final]{graphicx}
\usepackage{balance}
\usepackage[T1]{fontenc}
\usepackage{ifthen}
\usepackage[cmex10]{amsmath} 

\usepackage[lined,boxed,commentsnumbered,linesnumbered,ruled]{algorithm2e}
\usepackage[url,hyperrefblack,notheorems,IEEEtran]{research17}

\hypersetup{final} 
\usepackage{lipsum} 
\usepackage{comment}
\usepackage{booktabs} 
\usepackage{mathtools}
\usepackage{amsmath,amssymb,amsfonts,amsthm}
\usepackage{dsfont}
\usepackage{thmtools} 
\renewcommand{\thmcontinues}[1]{continued} 
\usepackage{tikz} 
\usetikzlibrary{shapes, patterns,decorations.text, decorations.pathreplacing}
\newcommand{\tikzmark}[1]{\tikz[overlay,remember picture] \node (#1){};}
\newtheorem{lemma}{\mylemmaname}
\newtheorem{theorem}{\mytheoremname}
\newtheorem{definition}{\mydefinitionname}
\newtheorem{proposition}{\mypropositionname}
\newtheorem{corollary}{\mycorollaryname}
\newtheorem{example}{\myexamplename} 

\newcommand*{\QEDA}{\null\nobreak\hfill\ensuremath{\bigtriangledown}} 

\newtheorem{remark}{\myremarkname}

\usepackage[capitalize]{cleveref}
\crefname{equation}{\unskip}{\unskip}
\crefname{claim}{Claim}{Claims} 
\usepackage{array}
\usepackage{multirow}
\newcolumntype{C}[1]{>{\centering\arraybackslash}p{#1}}
\renewcommand{\deg}[1]{\operatorname{deg}\left(#1\right)} 
\renewcommand{\vect}[1]{\vectg{#1}} 
\renewcommand{\vmat}[1]{\bm{\mat{#1}}} 
\newcommand{\code}[1]{\mathscr{#1}} 
\newcommand{\collect}[1]{\mathscr{#1}} 
\newcommand{\Hwt}[1]{w_\mathsf{H}\left(#1\right)} 
\newcommand*{\Resize}[2][4]{\resizebox{#1}{!}{\ensuremath{#2}}} 
\makeatletter
\renewcommand*\env@matrix[1][*\c@MaxMatrixCols c]{%
  \hskip -\arraycolsep
  \let\@ifnextchar\new@ifnextchar
  \array{#1}}
\makeatother

\newcommand{\HP}[1]{\HH\left(#1\right)} 
\newcommand{\eHP}[1]{\HH(#1)} 
\newcommand{\bigHP}[1]{\HH\bigl(#1\bigr)}

\newcommand{\HPcond}[2]{\HH\left(#1 \kern0.1em\middle|\kern0.1em #2\right)}
\newcommand{\eHPcond}[2]{\HH(#1 \kern0.1em|\kern0.1em #2)} 
\newcommand{\bigHPcond}[2]{\HH\bigl(#1 \kern-0.1em \bigm| \kern-0.1em#2\bigr)}
\newcommand{\BigHPcond}[2]{\HH\Bigl(#1 \kern-0.1em \Bigm| \kern-0.1em#2\Bigr)}
\newcommand{\MI}[2]{\II\left(#1 \kern0.1em{;}\kern0.1em #2\right)} 
\newcommand{\eMI}[2]{\II(#1 \kern0.1em{;}\kern0.1em #2)} 
\newcommand{\bigMI}[2]{\II\bigl(#1 \kern0.1em{;}\kern0.1em #2\bigr)}
\newcommand{\BigMI}[2]{\II\Bigl(#1 \kern0.1em{;}\kern0.1em #2\Bigr)}
\newcommand{\MIcond}[3]{\II\left(#1 \kern0.1em{;}\kern0.1em #2 \kern0.1em\middle|\kern0.1em #3\right)}
\newcommand{\eMIcond}[3]{\II(#1 \kern0.1em{;}\kern0.1em #2 \kern0.1em|\kern0.1em #3)} 
\newcommand{\bigMIcond}[3]{\II\bigl(#1 \kern0.1em{;}\kern0.1em #2 \kern-0.1em \bigm| \kern-0.1em#3\bigr)}
\newcommand{\BigMIcond}[3]{\II\Bigl(#1 \kern0.1em{;}\kern0.1em #2 \kern-0.1em \Bigm| \kern-0.1em#3\Bigr)}
\renewcommand{\r}{\color{red}} 
\renewcommand{\b}{\color{blue}} 
\newcommand{\lin}{\color{teal}} 
\DeclareSymbolFont{matha}{OML}{txmi}{m}{it}
\DeclareMathSymbol{\varv}{\mathord}{matha}{118}
\usepackage{upgreek}
\usepackage{caption}
\usepackage{subcaption}
\usepackage[normalem]{ulem}

\begin{document}
%
\sloppy \title{Private Function Computation for Noncolluding Coded Databases \thanks{This work is supported by US NSF grant CNS-1526547 and the Research Council of Norway (grant 240985/F20).}}

\author{\thanks{Parts of the material in this paper were presented at the 56th Annual Allerton Conference on Communications, Control, and Computing, Monticello, IL, USA, October 2018 \cite{ObeadLinRosnesKliewer18_1}, and at the IEEE International Symposium on Information Theory (ISIT), Paris, France, July 2019 \cite{ObeadLinRosnesKliewer19_1}.}
		
  \IEEEauthorblockN{Sarah A.~Obead\IEEEauthorrefmark{2}, Hsuan-Yin Lin\IEEEauthorrefmark{3}, Eirik Rosnes\IEEEauthorrefmark{3}, and J{\"o}rg Kliewer\IEEEauthorrefmark{2}\\
    \IEEEauthorblockA{\IEEEauthorrefmark{2}Helen and John C.~Hartmann Department of Electrical and Computer Engineering\\ New Jersey Institute of Technology, Newark, New Jersey 07102, USA}\\
    \IEEEauthorblockA{\IEEEauthorrefmark{3}Simula UiB, N--5006 Bergen, Norway}}}

\maketitle

\begin{abstract}
  Private computation in a distributed storage system (DSS) is a generalization of the private information retrieval (PIR) problem. In such setting a user wishes to compute a function of $f$ messages stored in $n$ noncolluding coded databases, i.e., databases storing data encoded with an $[n,k]$ linear storage code, while revealing no information about the desired function to the databases. We consider the problem of private polynomial computation (PPC). In PPC, a user wishes to compute a multivariate polynomial of degree at most $g$ over $f$ variables (or messages) stored in multiple databases. First, we consider the private computation of polynomials of degree $g=1$, i.e., private linear computation (PLC) for coded databases. In PLC, a user wishes to compute a linear combination over the $f$ messages while keeping the coefficients of the desired linear combination hidden from the database. For a linearly encoded DSS, we present a capacity-achieving PLC scheme and show that the PLC capacity, which is the ratio of the desired amount of information and the total amount of downloaded information, matches the maximum distance separable coded capacity of PIR for a large class of linear storage codes. 
  Then, we consider private computation of higher degree polynomials, i.e., $g>1$. For this setup, we construct two novel PPC schemes. In the first scheme, we consider Reed-Solomon coded databases with Lagrange encoding, which leverages ideas from recently proposed star-product PIR and Lagrange coded computation. The second scheme considers the special case of coded databases with systematic Lagrange encoding. Both schemes yield improved rates, while asymptotically, as $f\rightarrow \infty$, the systematic scheme gives a significantly better computation retrieval rate compared to all known schemes up to some storage code rate that depends on the maximum degree of the candidate polynomials.
\end{abstract}


\section{Introduction}
\label{sec:introduction}

The problem of private information retrieval (PIR) from public databases, introduced by Chor \emph{et al.} \cite{ChorGoldreichKushilevitzSudan95_1}, has been the focus of attention for several decades in the computer science community (see, e.g., \cite{ChorKushilevitzGoldreichSudan98_1,Gasarch04_1,Yekhanin10_1}). The goal of PIR is to allow a user to privately access an arbitrary message stored in a set of databases, i.e., without revealing any information of the identity of the requested message to each database. If the users do not have any side information on the data stored in the databases, the best strategy is to store the messages in at least two databases while ensuring PIR. Hence, the design of PIR protocols has focused on the case when multiple databases store the messages. This connects to the active and renowned research area of distributed storage systems (DSSs), where the data is encoded by an $[n,k]$ linear code and then distributed and stored across $n$ storage nodes \cite{DimakisRamchandranWuSuh11_1}, usually referred to as \emph{coded DSSs}. Using coding techniques, coded DSSs possess many practical features and benefits such as high reliability, efficient repairability, robustness, and security \cite{RawatKoyluogluSilbersteinVishwanath14_1}. Recently, the aspect of minimizing the communication cost, e.g., the required rate or bandwidth of privately querying the databases with the desired requests and downloading the corresponding information from the databases has attracted a great deal of attention in the information theory and coding communities. Thus, the renewed interest in PIR primarily focused on the study and design of efficient PIR protocols for coded DSSs  (see, e.g., \cite{ShahRashmiRamchandran14_1,ChanHoYamamoto15_1,SunJafar17_1,SunJafar18_2,BanawanUlukus18_1,TajeddineGnilkeElRouayheb18_1,FreijGnilkeHollantiKarpuk17_1,KumarLinRosnesGraellAmat19_1}).

\begin{figure}[t!]
  \centering
  \includegraphics[scale=0.16]{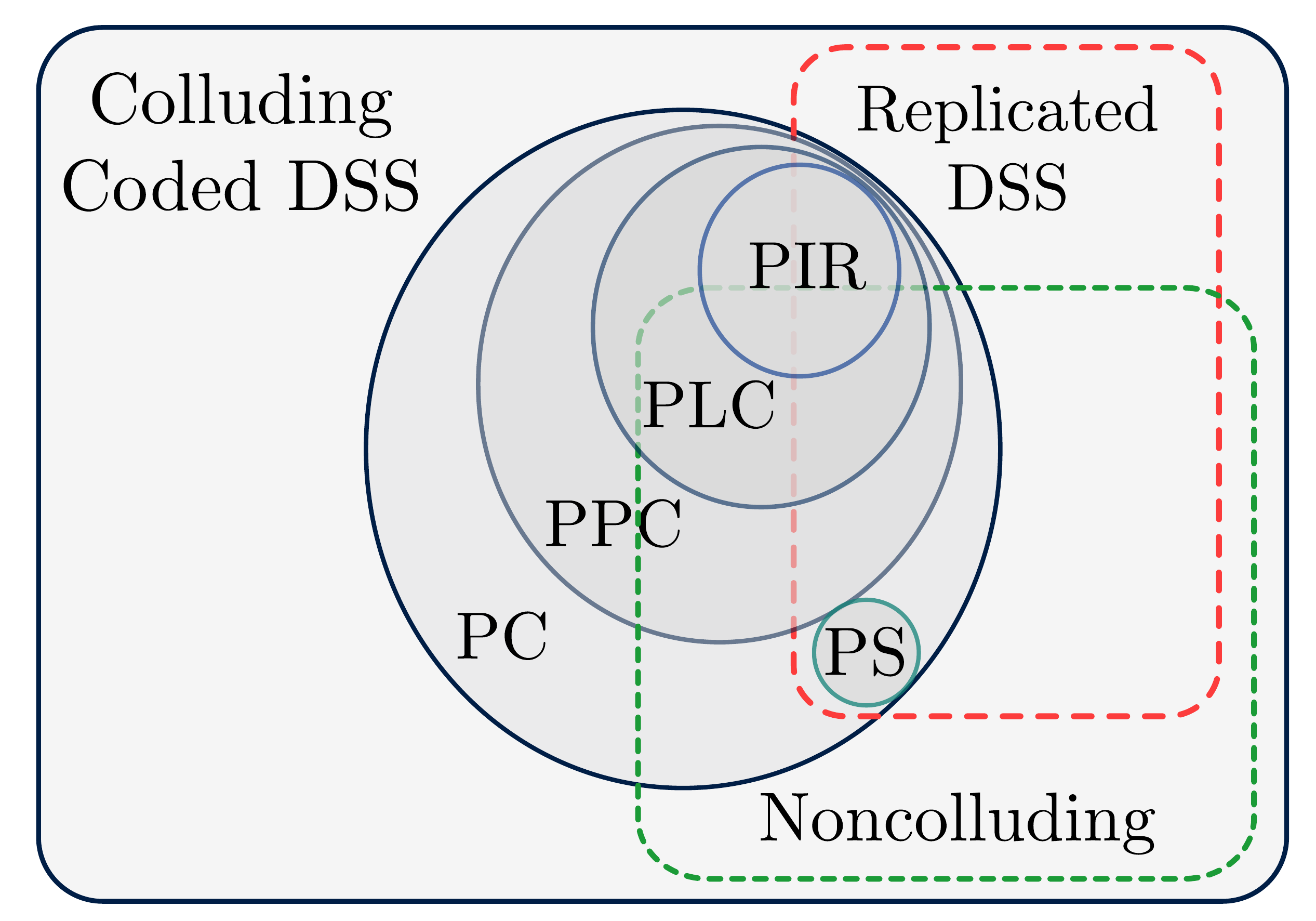}
  \caption{Simple overview of PIR problem extensions and variations.}
  \label{fig:Venn}
  \vspace{-2ex}
\end{figure}

A recently proposed generalization of the PIR problem \cite{SunJafar19_2,MirmohseniMaddahAli18_1,ChenWangJafar18_1,ObeadKliewer18_1,Karpuk18_1,RavivKarpuk19_2} addresses the \emph{private computation (PC)} for functions of the stored messages, also denoted as private function retrieval. In PC a user has access to a given number of databases and intends to compute a function of messages stored in these databases. This function is kept private from the databases, as they may be under the control of an adversary. In \cite{SunJafar19_2,MirmohseniMaddahAli18_1}, the scenario of private linear functions computation is considered for noncolluding replicated databases. In these works, the capacity and achievable rates for the communication overhead needed to privately compute a given \emph{linear} function, called private linear computation (PLC), were derived as a function of the number of messages and the number of databases, respectively. Interestingly, the obtained PLC capacity is equal to the PIR capacity of \cite{SunJafar17_1}. The extension to the coded case is addressed in \cite{ObeadKliewer18_1,Karpuk18_1,RavivKarpuk19_2}. In particular, in \cite{ObeadKliewer18_1} we proposed a PLC scheme based on maximum distance separable (MDS) coded storage. The presented scheme is able to achieve the MDS-coded PIR capacity established in \cite{BanawanUlukus18_1}, referred to as the MDS-PIR capacity in the sequel. In \cite{Karpuk18_1}, private polynomial computation (PPC) over $t$ colluding and systematically coded databases is considered by generalizing the star-product PIR scheme of \cite{FreijGnilkeHollantiKarpuk17_1}. In that work, the functions to be computed are polynomials of degree at most $g$, and a PC rate equal to the best asymptotic PIR rate of MDS-coded storage (when the number of messages tends to infinity) is achieved for $g=t=1$ (the case of linear function retrieval and noncolluding databases). An alternative PPC approach was recently proposed in \cite{RavivKarpuk19_2} for polynomials with higher degree, i.e.,~$g>1$, by employing Reed-Solomon (RS) coded databases with Lagrange encoding. For low code rates, the scheme improves on the PC rate of \cite{Karpuk18_1}. Finally, a separate but relevant form of PC, the private search (PS) problem \cite{ChenWangJafar18_1} considers mapping records replicated over $n$ noncolluding databases to binary search patterns. Each pattern represents the search result of one value out of a set of candidate alphabets. The asymptotic capacity, i.e.,~information retrieval rate for PS with large alphabet size, of privately retrieving one search pattern is found to match the asymptotic capacity of PIR for the special case of \emph{balanced} PS. In a balanced PS scenario, the nonlinearly dependent search patterns are assumed to contain equal amount of information. An overview of how these extensions align together can be seen in \cref{fig:Venn}.

In another line of research, for the case of noncolluding databases, two PIR protocols for a DSS where data is stored using a non-MDS linear code, are proposed in \cite{KumarLinRosnesGraellAmat19_1}, and their protocols are shown to achieve both the asymptotic and the nonasymptotic MDS-PIR capacity for a large class of linear codes. The first family of non-MDS codes for which the PIR capacity is known is found in \cite{LinKumarRosnesGraellAmat18_2,LinKumarRosnesGraellAmat18_3sub}. Further, PIR on linearly-coded databases for the case of colluding databases is also addressed in \cite{FreijGnilkeHollantiKarpuk17_1,TajeddineGnilkeElRouayheb18_1,FreijGnilkeHollantiHTrautmannKarpukKubjas19_1,KumarLinRosnesGraellAmat19_1}. For the PC case with noncolluding databases, however, capacity results for arbitrary linearly-coded DSSs have not been addressed so far in the open literature to the best of our knowledge.

In this work, we intend to fill this void by proposing three PC schemes and deriving an outer bound on the PC rate over all possible PC protocols. Our contributions are outlined as follow. 
\begin{itemize}
\item For the capacity of PPC, we adapt the converse proof of \cite[Thm.~4]{LinKumarRosnesGraellAmat18_3sub} to the coded PPC problem and derive an outer bound on the PPC rate (see Theorem~\ref{thm:converse_bound}). From this outer bound, as a special case of PPC when $g=1$, we prove a converse bound for the coded PLC capacity (see Theorem~\ref{thm:PLCconverse_MDS-PIRcapacity-achieving-codes}). The significance of our PLC converse is that, in contrast to \cite{SunJafar19_2}, it is valid for any number of messages $f$ and any number of candidate linear combinations $\mu$. In addition, our converse result depends on the rank of the coefficient matrix obtained from all $\mu$ linear combinations.
 
\item A capacity-achieving PLC scheme for a large class of linearly-coded DSSs with noncolluding databases is proposed. 
Essentially, the proposed PLC scheme jointly extends the optimal PIR scheme from DSSs coded with the MDS-PIR capacity-achieving codes of \cite{KumarLinRosnesGraellAmat19_1} and the PLC scheme from MDS-coded DSSs of \cite{ObeadKliewer18_1}, strictly generalizing the replication-based PC schemes of \cite{SunJafar19_2,MirmohseniMaddahAli18_1}. As for the optimality of the achievable PLC rate, we prove that the achievable rate matches the PLC converse bound of Theorem \ref{thm:PLCconverse_MDS-PIRcapacity-achieving-codes} and settle the coded PLC capacity (see Theorem~\ref{thm:PLCcapacity_MDS-PIRcapacity-achieving-codes}).

\item For higher degree PPC, i.e.,~$g>1$, we present two new approaches for PPC from RS-coded DSSs by generalizing the presented capacity-achieving PLC scheme and leveraging ideas from star-product PIR \cite{FreijGnilkeHollantiKarpuk17_1} and Lagrange coded computation \cite{YuLiRavivKalanSoltanolkotabiAvestimehr19_1}. Although the problem of PPC from Lagrange encoded DSSs was recently studied in \cite{RavivKarpuk19_2}, the authors were mainly concerned with constructing explicit PPC schemes with focus on preserving privacy against colluding DSSs. We, on the other hand, aim our attention at providing PPC solutions that minimize the download cost and we focus on establishing the capacity of the PPC setup. Towards that aim, we propose two PPC schemes from  RS-coded noncolluding databases with Lagrange encoding (one for systematic encoding) that improve on the rate of the PPC schemes presented in \cite{Karpuk18_1,RavivKarpuk19_2} (see Theorems \ref{thm:PMCrate_LagrangeCoded-DSS} and \ref{thm:RS_PPC_rate}).  The systematic scheme  is an improved version of the systematic scheme presented in \cite{ObeadLinRosnesKliewer19_1}. To demonstrate the performance of our proposed PPC schemes, a number of examples and numerical results are presented. We show that, compared to the schemes in \cite{Karpuk18_1,RavivKarpuk19_2}, both proposed PPC schemes yield a larger PC rate, i.e., lower download cost,  when the number of messages is small. As the number of messages tends to infinity, the achievable rate of our RS-coded (nonsystematic) PPC scheme approaches the rate of \cite{RavivKarpuk19_2} (see \cref{cor:PMCrate_LagrangeCoded-DSS}), while our systematic scheme outperforms all known schemes up to some storage code rate that depends on the maximum degree of the candidate polynomials (see \cref{rem:1} and \cref{cor:RS_PPC_rate}).
\end{itemize}

The remainder of the paper is organized as follows. \cref{sec:preliminaries} outlines the notation and basic definitions, then the problem of PPC from coded DSSs and the system model are presented.  
We derive the converse bound for an arbitrary number of messages and polynomial functions in \cref{sec:converse-proof_coded-PC}. A generic query generation scheme for PC for linearly-coded storage  with an MDS-PIR capacity-achieving code is presented in~\cref{sec:generic-query-generation_PC_coded-DSSs}. This scheme acts as a building block for the three schemes constructed in the following sections. In \cref{sec:private-linear-computation_coded-DSSs} we present the capacity-achieving PLC scheme. In Sections~\ref{sec:achievable-scheme_PPC} and \ref{sec:achievable-scheme_PPC_systematic}, we propose two PPC schemes for RS-coded storage and higher degree polynomials with examples. Then, in \cref{sec:numerical-results}, numerical results for the proposed PPC schemes and the converse bound from \cref{sec:converse-proof_coded-PC} are presented,  establishing the achievability of larger retrieval rates compared with PPC schemes from the literature. %
Some conclusions are drawn in Section~\ref{sec:conclusion}.

\section{Preliminaries}
\label{sec:preliminaries}

\subsection{Notation}
\label{sec:notation}

We denote by $\Naturals$ the set of all positive integers and let $\Naturals_0 \eqdef\{ 0 \} \cup \Naturals$, $ [a]\eqdef\{1,2,\ldots,a\}$, and $[a:b]\eqdef\{a,a+1,\ldots,b\}$ for $a,b\in \Naturals$, $a \leq b$. Random and deterministic quantities are carefully distinguished as follows. A random variable is denoted by a capital Roman letter, e.g., $X$, while its realization is denoted by the corresponding small Roman letter, e.g., $x$. Vectors are boldfaced, e.g., $\vect{X}$ denotes a random vector and $\vect{x}$ denotes a deterministic vector, respectively. The notation $\vect{X} \sim \vect{Y}$ is used to indicate that $\vect{X}$ and $\vect{Y}$ are identically distributed. Random matrices are represented by bold sans serif letters, e.g., $\vmat{X}$, where $\mat{X}$ represents its realization. In addition, sets are denoted by calligraphic uppercase letters, e.g., $\set{X}$, and $\comp{\set{X}}$ denotes the complement of a set $\set{X}$ in a universe set.  We denote a submatrix of $\mat{X}$ that is restricted in columns by the set $\set{I}$ by $\mat{X}|_{\set{I}}$. For a given index set $\set{S}$, we also write $\vmat{X}^\set{S}$ and $Y_\set{S}$ to represent $\bigl\{\vmat{X}^{(v)}\colon v\in\set{S}\bigr\}$ and $\bigl\{Y_j\colon j\in\set{S}\bigr\}$, respectively. 
Furthermore, some constants and functions are also depicted by Greek letters or a special font, e.g., $\const{X}$. 
The function $\HP{X}$ represents the entropy of $X$, and $\MI{X}{Y}$ the mutual information between $X$ and $Y$. The binomial coefficient of $a$ over $b$, $a,b \in \Naturals_0$, is denoted by $a \choose b$ where  ${a \choose b} =0$ if $a < b$. The notation $\lfloor \cdot \rfloor$ denotes the floor function and ${\mathds{1}}(\cdot)$ represents the indicator function, i.e., ${\mathds{1}}(\text{statement})$ equals to $1$ if the statement holds, and $0$ otherwise.

We use the customary code parameters $[n,k]$ to denote a code $\code{C}$ over the finite field $\Field_q$ of blocklength $n$ and dimension $k$. A generator matrix of $\code{C}$ is denoted by $\mat{G}^{\code{C}}$. A set of coordinates of $\code{C}$, $\set{I}\subseteq[n]$, of size $k$ is said to be an \emph{information set} if and only if $\mat{G}^\code{C}|_\set{I}$ is invertible. $\trans{(\cdot)}$ denotes the transpose operator, while $\rank{\mat{V}}$ denotes the rank of a matrix $\mat{V}$. The function $\chi(\vect{x})$ denotes the support of a vector $\vect{x}$, and the linear span of a set of vectors $\{\vect{x}_1,\ldots,\vect{x}_a\}$, $a\in \Naturals$, is denoted by $\textnormal{span}\{\vect{x}_1,\ldots,\vect{x}_a\}$. 
Finally, $\Field_q[z]$ denotes the set of all univariate polynomials over $\Field_q$ in the variable $z$, and we denote by $\deg{\phi(z)}$ the degree of a polynomial $\phi(z) \in \Field_q[z]$.

We now proceed with a general description for the problem statement of private function computation from linearly-coded DSSs.

\subsection{Problem Statement and System Model}
\label{sec:problem-SystemModel}

The PC problem for coded DSSs is described as follows. We consider a DSS that stores in total $f$ independent messages $\vmat{W}^{(1)},\ldots,\vmat{W}^{(f)}$, where each message $\vmat{W}^{(m)}=\bigl(W_{i,j}^{(m)}\bigr)$, $m\in [f]$, is a random $\beta\times k$ matrix over  $\Field_q$ with some $\beta,k \in\Naturals$. Let $\const{L}\eqdef \beta k$. Then, each message $\vmat{W}^{(m)}$, $m\in[f]$, can also be seen as a random vector variable $\vmat{W}^{(m)}=(W^{(m)}_1,\ldots,W^{(m)}_\const{L})$ of $\const{L}$ symbols that are chosen independently and uniformly at random from $\Field_q$. Thus, $\eHP{\vmat{W}^{(m)}}=\beta k,\,\forall\,m\in[f]$ (in $q$-ary units). Each message is encoded using an $[n,k]$ code as follows. Let $\vect{W}^{(m)}_i=\bigl(W^{(m)}_{i,1},\ldots,W^{(m)}_{i,k}\bigr)$, $i\in [\beta]$, be a message vector corresponding to the $i$-th row of $\vmat{W}^{(m)}$. Each $\vect{W}^{(m)}_i$ is encoded by an $[n,k]$ code $\code{C}$ over $\Field_q$ into a length-$n$ codeword $\vect{C}^{(m)}_i=\bigl(C^{(m)}_{i,1},\ldots,C^{(m)}_{i,n}\bigr)$. The $\beta f$ generated codewords $\vect{C}_i^{(m)}$ are then arranged in the array $\vmat{C}=\trans{\bigl(\trans{(\vmat{C}^{(1)})}|\ldots|\trans{(\vmat{C}^{(f)})}\bigr)}$ of dimensions $\beta f \times n$, where $\vmat{C}^{(m)}=\trans{\bigl(\trans{(\vect{C}^{(m)}_1)}|\ldots|\trans{(\vect{C}^{(m)}_{\beta})}\bigr)}$. The code symbols $C_{1,j}^{(m)},\ldots,C_{\beta,j}^{(m)}$, $m\in[f]$, for all $f$ messages are stored on the $j$-th database, $j\in[n]$.

\begin{figure}[t!]
  \centering
  \includegraphics[scale=0.18]{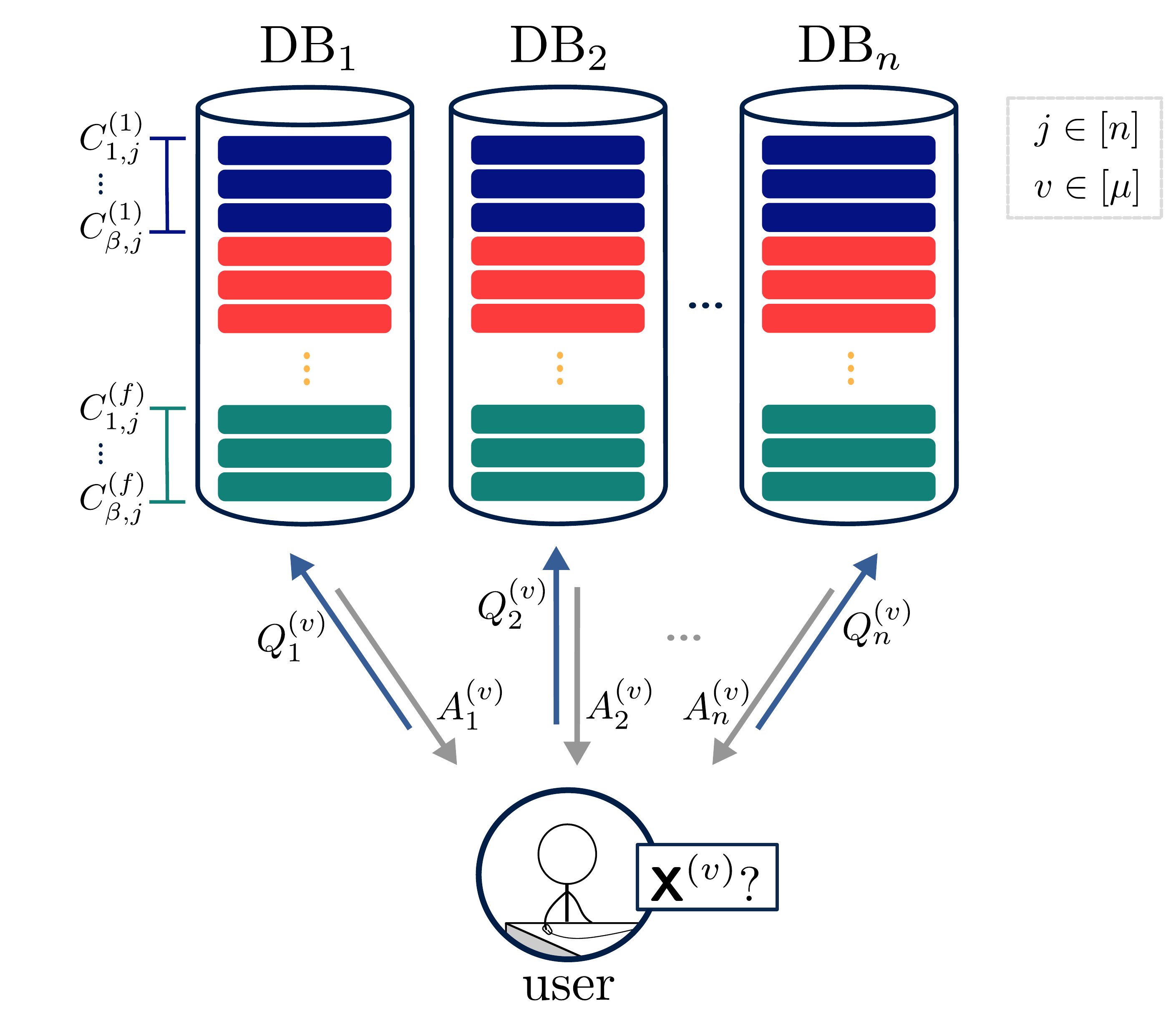}
  \caption{System model.}
  \label{fig:SysModel}
  \vspace{-2ex}
\end{figure}

We consider the case of $n$ noncolluding databases. In private function computation, a user wishes to privately compute exactly one function image $X_i^{(v)} \triangleq \phi^{(v)}(\vect{W}_{i})$, where $\vect{W}_{i} = (W_{i}^{(1)},\ldots,W_{i}^{(f)})$, $\forall\, i\in [ \const{L}]$,
out of $\mu$ arbitrary \emph{candidate} functions $\phi^{(1)},\ldots,\phi^{(\mu)}\colon \Field_q^f \to\Field_q$ from the coded DSS. Let $\vmat{X}^{(v)}=\bigl(X^{(v)}_1,\ldots,X^{(v)}_\const{L}\bigr)$, where $X^{(v)}_1,\ldots,X^{(v)}_{\const{L}}$ are independent and identically distributed according to a prototype random variable $X^{(v)}$ with probability mass function $P_{X^{(v)}}$. Thus, $\eHP{\vmat{X}^{(v)}}=\const{L}\eHP{X^{(v)}},\,\forall\,v\in[\mu]$, $\eHP{\vmat{X}^{(1)},\ldots,\vmat{X}^{(\mu)}}=\const{L}\HP{X^{(1)},\ldots,X^{(\mu)}}$, and we let $\HH_\textnormal{min}\eqdef \min_{v\in[\mu]}{\HP{X^{(v)}}}$ and $\HH_\textnormal{max} \eqdef\max_{v\in[\mu]}{\HP{X^{(v)}}} $.

The user privately selects an index $v\in[\mu]$ and wishes to compute the $v$-th function while keeping the requested function index $v$ private from each database.  In order to retrieve the desired function evaluation $\vmat{X}^{(v)}$, $v\in[\mu]$,  from the coded DSS, the user sends a query $Q^{(v)}_j$ to the $j$-th database for all $j\in[n]$ as illustrated in \cref{fig:SysModel}. Since the queries are generated by the user without any prior knowledge of the realizations of the candidate functions, the queries are independent of the candidate functions evaluations. In other words, we have
\begin{IEEEeqnarray*}{c}
  \MI{\vmat{X}^{(1)},\ldots,\vmat{X}^{(\mu)}}{Q^{(v)}_1,\ldots,Q^{(v)}_n}=0,\,\forall\,v\in[\mu].
\end{IEEEeqnarray*}
In response to the received query, database $j$ sends the answer $A^{(v)}_j$ back to the user. $A^{(v)}_j$ is a deterministic function of $Q^{(v)}_j$ and the data stored in the database. Thus, $\forall\,v\in[\mu]$, 
\begin{IEEEeqnarray*}{rCl}
  \BigHPcond{A^{(v)}_j}{Q^{(v)}_j,\vect{C}_j}=0,\,\forall\,j\in[n],
\end{IEEEeqnarray*}
 where $\vect{C}_j\eqdef\trans{\bigl(C^{(1)}_{1,j},\ldots,C^{(1)}_{\beta,j},C^{(2)}_{1,j},\ldots,C^{(f)}_{\beta,j}\bigr)}$ denotes the $f$ coded chunks that are stored in the $j$-th database.
 
To preserve user's privacy, the query-answer function must be identically distributed for each possible desired function index $v\in[\mu]$ from the perspective of each database $j\in[n]$. In other words, the scheme's queries and answer strings must be independent from the desired function index. Moreover, the user must be able to reliably decode the desired polynomial function evaluation $\vmat{X}^{(v)}$. Accordingly, we define a PC protocol for $[n,k]$ coded DSSs as follows.

Consider a DSS with $n$ noncolluding databases storing $f$ messages using an $[n,k]$ code. The user wishes to retrieve the $v$-th function evaluation $\vmat{X}^{(v)}$,  $v\in[\mu]$, from the available information $Q^{(v)}_j$ and $A^{(v)}_j$, $j\in[n]$. For a PC protocol, the following conditions must be satisfied $\forall\,v, v'\in[\mu]$, $v \neq v'$, and $\forall\,j\in[n]$,
\begin{IEEEeqnarray}{rCl}    
  \text{[Privacy]} \qquad\qquad\quad%
  && (Q^{(v)}_j,A^{(v)}_j,\vmat{X}^{[\mu]})  \sim (Q^{(v')}_j,A^{(v')}_j,\vmat{X}^{[\mu]}),
  \IEEEyesnumber\IEEEyessubnumber\IEEEeqnarraynumspace \label{eq:privacy}
  \\
  \text{[Recovery]} \qquad\qquad\quad%
  && \bigHPcond{\vmat{X}^{(v)}}{A^{(v)}_1,\ldots,A^{(v)}_n,Q^{(v)}_1,\ldots,Q^{(v)}_n}=0.
  \IEEEyessubnumber \label{eq:recovery}
\end{IEEEeqnarray}

From an information-theoretic perspective, the efficiency of a PC protocol is measured by the \emph{PC rate}, which is defined as follows.

\begin{definition}[PC rate and capacity for linearly-coded DSSs]
  \label{def:def_info-PCrate}
  The exact information-theoretic rate of a PC scheme, denoted by $\const{R}$, is defined as the ratio of the minimum desired function size $\const{L}\HH_\textnormal{min}$ over the total required download cost, i.e.,
  \begin{IEEEeqnarray*}{c}
    \const{R}\eqdef\frac{\const{L}\HH_\textnormal{min}}{\const{D}},
  \end{IEEEeqnarray*}
  where $\const{D}$ is the total required download cost. The PC \emph{capacity} $\const{C}_\textnormal{PC}$ is the maximum of all achievable PC rates over all possible PC protocols for a given $[n,k]$ storage code. %
\end{definition}

\subsection{Background}

A monomial ${\vect{z}}^{\vect{i}}$ in $m$ variables $z_1,\ldots,z_m$ with degree $g$ is written as ${\vect{z}}^{\vect{i}} = z_1^{i_1} z_2^{i_2}\cdots z_m^{i_m}$, where $\vect{i}\eqdef(i_1,\ldots,i_m)\in \Naturals_0^{m}$ is the exponent vector with $\textnormal{wt}(\vect{i})\eqdef\sum_{j=1}^{m}i_j = g$. The set $\{\vect{z}^{\vect{i}}: \vect{i} \in \Naturals_0^{m},\, 1\leq \textnormal{wt}(\vect{i})\leq g\}$ of all monomials in $m$ variables of degree at most $g$ has size
\begin{IEEEeqnarray*}{rCl}
  \const{M}_{g}(m)\eqdef\sum\limits_{h=1}^{g}\binom{h+m-1}{h}=\binom{g+m}{g}-1.
\end{IEEEeqnarray*}

Moreover, a polynomial $\phi(\vect{z})$ of degree at most $g$ is  represented as
$\phi(\vect{z})=\sum_{\vect{i}:\textsf{wt}(\vect{i}) \leq g} a_{\vect{i}}{\vect{z}}^{\vect{i}}$, $a_{\vect{i}} \in \Field_q$. The total number of polynomials in $m$ variables of degree at most $g$ generated with all possible distinct (up to scalar multiplication) $\const{M}_g(m)$-dimensional coefficients vectors defined over $\Field_q$ is equal to $\upmu_{g}(m) \eqdef \frac{q^{\const{M}_g(m)}-1}{q-1}$.

\begin{definition}[Star-product]
  Let $\code{C}$ and $\code{D}$ be two linear codes of length $n$ over $\Field_q$. The star-product (Hadamard product) of $\vect{v}=(v_1,\ldots,v_n)  \in\code{C}$ and $\vect{u} = (u_1,\ldots,u_n) \in\code{D}$ is defined as $\vect{v}\star\vect{u}=(v_1u_1,\ldots,v_nu_n)\in \Field_q^n$. Further, the star-product of $\code{C}$ and $\code{D}$, denoted by $\code{C}\star\code{D}$, is defined by $\textnormal{span}\{\vect{v}\star\vect{u}:\vect{v}\in\code{C}, \vect{u}\in\code {D}\}$ and the $g$-fold star-product of  $\code{C}$ with itself is given by $\code{C}^{\star g}=\textnormal{span}\{\vect{v}_1\star\cdots\star\vect{v}_g:\vect{v}_i\in\code{C}, i\in[g]\}.$
\end{definition}

\begin{definition}[RS code]
  Let $\vect{\alpha}=(\alpha_1,\ldots,\alpha_n)$ be a vector of $n$ distinct elements of $\Field_q$. For $n\in\Naturals$, $k \in [n]$, and $q \geq n$, the $[n,k]$ RS code (over $\Field_q$) is defined as
  \begin{IEEEeqnarray}{rCl}
    \set{RS}_{k}(\vect{\alpha})
    & \eqdef & \{(\phi(\alpha_1),\ldots,\phi(\alpha_n))\colon \phi\in\Field_q[z],\, \deg{\phi}<k \}.
    \IEEEeqnarraynumspace\label{eq:RScodeDef}
  \end{IEEEeqnarray}
\end{definition}

It is well-known that RS codes are MDS codes that behave well under the star-product. We state the following proposition that was introduced in \cite{FreijGnilkeHollantiKarpuk17_1}.

\begin{proposition}\label{prop:RSstarProduct}
  Let ${\cal RS}_{k}(\vect{\alpha})$ be a length-$n$ RS code. Then, for $g\in\Naturals$, the $g$-fold star-product of ${\cal RS}_{k}(\vect{\alpha})$ with itself is the RS code given by ${\cal RS}_k^{\star g}(\vect{\alpha})={\cal RS}_{\min{\{g(k-1)+1,n\}}}(\vect{\alpha})$.
\end{proposition}

Let $\vect{\gamma}=(\gamma_1,\ldots,\gamma_k)$ be a vector of $k$ distinct elements of $\Field_q$. For a message vector $\vect{W}=(W_1,\ldots,W_k)$, let $\ell(z)\in \Field_q[z]$ be a polynomial of degree at most $k-1$ such that $\ell(\gamma_i)=W_{i}$ for all $i\in[k]$. Using the Lagrange interpolation formula we present this polynomial as $\ell(z)=\sum_{i\in[k]} W_i \iota_i(z)$, where $\iota_i(z)$ is the Lagrange basis polynomial 
\vspace{-0.4ex}
\begin{IEEEeqnarray*}{rCl}
  \iota_{i}(z)=\prod_{t\in[k]\setminus \{i\}}\frac{z-\gamma_t}{\gamma_i-\gamma_t}.
\end{IEEEeqnarray*}

It has been shown in \cite{RavivKarpuk19_2} that Lagrange encoding is equivalent to the choice of a specific basis for an RS code. Thus, for encoding we choose the set of Lagrange basis polynomials as the code generating polynomials of \eqref{eq:RScodeDef} \cite{YuLiRavivKalanSoltanolkotabiAvestimehr19_1}. Thus, a generator matrix of ${\cal RS}_{k}(\vect{\alpha})$ is $\mat{G}_{\mathcal{RS}_{k}}(\vect{\alpha,\gamma}) = (\iota_{i}(\alpha_j))$, $i\in[k]$, $j\in[n]$. Note that if we choose $\gamma_i=\alpha_i$ for $i\in[k]$, then the generator matrix $\mat{G}_{\mathcal{RS}_{k}}(\vect{\alpha,\gamma})$ becomes systematic.

\subsection{MDS-PIR Capacity-Achieving Codes}
\label{sec:MDS-PIRcapacity-achieving-codes_PIR}

In \cite{KumarLinRosnesGraellAmat19_1}, a PIR protocol for any linearly-coded DSS that uses an $[n,k]$ code to store $f$ messages, named Protocol~1, is proposed. The PIR rate of Protocol~1 can be derived by finding a \emph{PIR achievable rate matrix} of the underlying storage code $\code{C}$, which is defined as follows.
\begin{definition}[{\cite[Def.~10]{KumarLinRosnesGraellAmat19_1}}]
  \label{def:PIRachievable-rate-matrix}
  Let $\code{C}$ be an arbitrary $[n,k]$ code. A $\nu\times n$ binary matrix $\mat{\Lambda}_{\kappa,\nu}^{\textnormal{PIR}}(\code{C})$ is said to be a PIR achievable rate matrix for $\code{C}$ if the following conditions are satisfied. %
  \begin{enumerate}
  \item \label{item:1} The Hamming weight of each column of $\mat{\Lambda}^{\textnormal{PIR}}_{\kappa,\nu}$ is $\kappa$, and
  \item \label{item:2} for each matrix row $\vect{\lambda}_i$, $i\in[\nu]$, $\chi(\vect{\lambda}_i)$ always contains
    an information set.
  \end{enumerate}
  In other words, each coordinate $j$ of $\code{C}$, $j\in [n]$, appears exactly $\kappa$ times in $\{\chi(\vect{\lambda}_i)\}_{i\in [\nu]}$, and every set $\chi(\vect{\lambda}_i)$ contains an information set.
\end{definition}
\begin{example}
		\label{ex:MDSCapCodeex_n4k2}
		Consider a $[4,2]$ code $\code{C}$ with generator matrix
		\begin{IEEEeqnarray*}{c}
		\mat{G}^{\code{C}}=
		\begin{pmatrix}
		1 & 0 &1 & 1
		\\ 
		0 & 1 &1 & 1  
		\end{pmatrix}.
			\end{IEEEeqnarray*}  
		One can verify that  
		\begin{IEEEeqnarray*}{c}
		\mat{\Lambda}^{\textnormal{PIR}}_{1,2}=
		\begin{pmatrix}
		1 & 0 &1 & 0
		\\
		0 & 1 &0 & 1
		\end{pmatrix}
		\end{IEEEeqnarray*}  
		is a valid PIR achievable rate matrix for $\code{C}$ with $(\kappa,\nu)=(1,2)$. This is true given that, column-wise, the Hamming weight of each column in $\mat{\Lambda}^{\textnormal{PIR}}_{1,2}$ is $\kappa=1$. %
		On the other hand, row-wise, %
		$\chi(\vect{\lambda}_1)=\{1,3\}$ and $\chi(\vect{\lambda}_2)=\{2,4\}$ are two information sets of $\code{C}$. \QEDA%
\end{example} 

In \cite{KumarLinRosnesGraellAmat19_1}, it is shown that the MDS-PIR capacity \cite{BanawanUlukus18_1} can be achieved using Protocol~1 for a special class of $[n,k]$ codes. In particular, to achieve the MDS-PIR capacity using Protocol~1, the $[n,k]$ storage code should possess a specific underlying structure as given by the following theorem.

\begin{theorem}[{\cite[Cor.~1]{KumarLinRosnesGraellAmat19_1}}]
  \label{thm:MDS-PIRcapacity-achieving-matrix}
  Consider a DSS that uses an $[n,k]$ code $\code{C}$ to store $f$ messages. If a PIR achievable rate matrix $\mat{\Lambda}^{\textnormal{PIR}}_{\kappa,\nu}(\code{C})$ with $\frac{\kappa}{\nu}=\frac{k}{n}$ exists, then the MDS-PIR capacity
  \begin{IEEEeqnarray*}{rCl}
    \const{C}_{\textnormal{MDS-PIR}}\eqdef\Bigl(1-\frac{k}{n}\Bigr)\inv{\left[1-\Bigl(\frac{k}{n}\Bigr)^f\right]}
    \label{eq:PIRcapacity}  
  \end{IEEEeqnarray*}
  is achievable.
\end{theorem}

This gives rise to the following definition.
\begin{definition}[{\cite[Def.~13]{KumarLinRosnesGraellAmat19_1}}]
  \label{def:MDS-PIRcapacity-achieving-codes}
  Given an $[n,k]$ code $\code{C}$, if a PIR achievable rate matrix $\mat{\Lambda}_{\kappa,\nu}^{\textnormal{PIR}}(\code{C})$ with $\frac{\kappa}{\nu}=\frac{k}{n}$ exists, then the code $\code{C}$ is referred to as an \emph{MDS-PIR capacity-achieving} code, and the matrix $\mat{\Lambda}_{\kappa,\nu}^{\textnormal{PIR}}(\code{C})$ is called an \emph{MDS-PIR capacity-achieving} matrix.
\end{definition}

Accordingly, one can easily see that the $[4,2]$ code $\code{C}$ given in \cref{ex:MDSCapCodeex_n4k2} is an MDS-PIR capacity-achieving code. 
Note that the class of MDS-PIR capacity-achieving codes includes MDS codes, cyclic codes, Reed-Muller codes, and certain classes of distance-optimal local reconstruction codes \cite{KumarLinRosnesGraellAmat19_1}.


\section{Converse Bound}
\label{sec:converse-proof_coded-PC}

In \cite{LinKumarRosnesGraellAmat18_2,LinKumarRosnesGraellAmat18_3sub}, the PIR capacity for a coded DSS using an MDS-PIR capacity-achieving code is shown to be equal to the MDS-PIR capacity. In this section, we derive an outer bound on the PPC rate (\cref{thm:converse_bound} below) by adapting the converse proof of \cite[Thm.~4]{LinKumarRosnesGraellAmat18_3sub} to the scenario of the linearly-coded PPC problem, where the storage code is MDS-PIR capacity-achieving. The converse is valid for any number of messages $f$ and candidate functions $\mu$. Then, we state the converse bound for PLC, as a special case, in  \cref{thm:PLCconverse_MDS-PIRcapacity-achieving-codes} and show that it matches the MDS-PIR capacity (i.e., the PIR capacity for a DSS where data is encoded and stored using an MDS code).

We first define an effective rank for the PC problem as follows.
\begin{definition}
  \label{def:rank-function_PC}  
  Let $\vmat{X}^{[\mu]}=\{\vmat{X}^{(1)},\ldots,\vmat{X}^{(\mu)}\}$ denote the set of candidate functions evaluations where $\vmat{X}^{(\ell)} = \bigl(X^{(\ell)}_1,\ldots,X^{(\ell)}_\const{L}\bigr)$, $\ell \in [\mu]$.  %
  The \emph{effective rank} $r\bigl(\vmat{X}^{[\mu]}\bigr)$ is defined as 
  \begin{IEEEeqnarray}{c}
    r\bigl(\vmat{X}^{[\mu]}\bigr)\eqdef \min \bigl\{s\colon\bigHP{{X}^{(\ell_1)}_l,\ldots,{X}^{(\ell_s)}_l}=\bigHP{X^{[\mu]}_l},\, \{\ell_1,\ldots,\ell_s\}\subseteq [\mu],\, s\in[\mu],\, l \in [\const{L}]\bigr\},
    \label{eq:rank-function_PC}
  \end{IEEEeqnarray}
  and we define the set $\set{L}\triangleq\{\ell_1,\ldots,\ell_r\}\subseteq [\mu]$ to be a minimum set that satisfies \eqref{eq:rank-function_PC}.
\end{definition}

Note that, when the candidate functions are of degree at most $g=1$, it can be seen that there is a deterministic linear mapping $\mat{V}$ of size $\mu\times f$ between $\bigl(X^{(1)}_l,\ldots,X^{(\mu)}_l\bigr)$ and $\bigl(W^{(1)}_l,\ldots,W^{(f)}_l)$, i.e., it reduces to the PLC problem where
\begin{IEEEeqnarray}{rCl}
  \begin{pmatrix}
    X^{(1)}_l
    \\
    \vdots
    \\
    X^{(\mu)}_l
  \end{pmatrix}=\mat{V}_{\mu\times f}
  \begin{pmatrix}
    W^{(1)}_l
    \\
    \vdots
    \\
    W^{(f)}_l
  \end{pmatrix},\label{eq:linear-mappingV}
\end{IEEEeqnarray}
and $r\bigl(\vmat{X}^{[\mu]}\bigr)=\rank{\mat{V}}\leq\min\{\mu,f\}$.
	
Accordingly, an upper bound on the capacity of PPC for a coded DSS where data is encoded and stored using an MDS-PIR capacity-achieving code introduced in Definition~\ref{def:MDS-PIRcapacity-achieving-codes}, is stated in the following. %

\subsection{General Converse}
\begin{theorem} 
  \label{thm:converse_bound}
  Consider a DSS with $n$ noncolluding databases that uses an $[n,k]$ MDS-PIR capacity-achieving code $\code{C}$ to store $f$ messages. Then, the maximum achievable PPC rate over all possible PPC protocols, i.e., the PPC capacity $\const{C}_\textnormal{PPC}$, is upper bounded by
  \begin{IEEEeqnarray*}{c}
    \const{C}_\textnormal{PPC}\leq\frac{\HH_\textnormal{min}}{\HH^{(\textnormal{B})}_\textnormal{min} +\sum_{v=1}^{r-1}\bigl(\frac{k}{n}\bigr)^{v} \bigHPcond{{X}^{(\ell_{v+1})}}{{X}^{(\ell_1)},\ldots,{X}^{(\ell_{v})}}},
    \label{eq:R_upperbound}
  \end{IEEEeqnarray*}
  for any effective rank $r\bigl(\vmat{X}^{[\mu]}\bigr)=r$, where  $\HH^{(\textnormal{B})}_\textnormal{min}\eqdef \min_{\ell\in \set{L}}{\bigHP{X^{(\ell)}}}$.
\end{theorem}
Here, we remark that Theorem~\ref{thm:converse_bound} generalizes \cite[Thm.~1]{ChenWangJafar18_1}, which is a converse bound on the capacity of dependent PIR (DPIR) for noncolluding replicated databases. 

Before we proceed with the converse proof, we provide some general results that are useful for the proof.

\begin{enumerate}
\item From the condition of privacy,
  \begin{IEEEeqnarray}{rCl}
    \bigHPcond{A_j^{(v)}}{\vmat{X}^{(v)},\set{Q}}=\bigHPcond{A_j^{(v')}}{\vmat{X}^{(v)},\set{Q}},
    \label{eq:indistinct-answers}
  \end{IEEEeqnarray}
  where $v\neq v'$, $v,v'\in [\mu]$, and $\set{Q}\eqdef\bigl\{Q^{(v)}_j\colon v\in[\mu], j\in[n]\bigr\}$ denotes the set of all possible queries  made by the user. Although this seems to be intuitively true, a proof of this property is still required and can be found in \cite{BanawanUlukus18_1}.
	
\item Consider a PPC protocol for a coded DSS that uses an $[n,k]$ code $\code{C}$ to store $f$ messages. For any subset of function evaluations $\vmat{X}^{\set{V}}$, $\set{V}\subseteq[\mu]$, and for any information set $\set{I}$ of $\code{C}$, we have
  \begin{IEEEeqnarray}{rCl}
    \bigHPcond{A^{(v)}_\set{I}}{\vmat{X}^{\set{V}},\set{Q}}& = &
    \sum_{j\in\set{I}}\bigHPcond{A^{(v)}_j}{\vmat{X}^{\set{V}},\set{Q}}.
    \label{eq:independent_kAnswers}
  \end{IEEEeqnarray}
  The proof uses the linear independence of the columns of a generator matrix of $\code{C}$ corresponding to an information set, and can be seen as a simple extension of \cite[Lem.~1]{BanawanUlukus18_1}. This argument applies to the case of PPC due to the fact that $A^{(v)}_\set{I}$ is still a deterministic function of independent random variables $\{\vect{C}_j\colon j \in \set{I} \}$ and $\set{Q}$. %
\end{enumerate}

Next, we state Shearer's Lemma, which represents a very useful entropy method for combinatorial problems.
\begin{lemma}[Shearer's Lemma \cite{Radhakrishnan03_1}]
  \label{lem:Shearer-lemma}
  Let $\collect{S}$ be a collection of subsets of $[n]$, with each $j\in[n]$ included in at least $\kappa$ members
  of $\collect{S}$. For random variables $Z_1,\ldots,Z_n$, we have
  \begin{IEEEeqnarray*}{rCl}
    \sum_{\set{S}\in\collect{S}}\eHP{Z_\set{S}}\geq \kappa\eHP{Z_1,\ldots,Z_n}.
  \end{IEEEeqnarray*}
\end{lemma}

Now, we are ready for the converse proof. By \cite[Lem.~2]{KumarLinRosnesGraellAmat19_1}, since the code $\code{C}$ is MDS-PIR capacity-achieving, there exist $\nu$ information sets $\set{I}_1,\ldots,\set{I}_\nu$ such that each coordinate $j\in[n]$ is included in exactly $\kappa$ members of $\collect{I}=\{\set{I}_1,\ldots,\set{I}_\nu\}$ with $\frac{\kappa}{\nu}=\frac{k}{n}$.

Applying the chain rule of entropy we have
\begin{IEEEeqnarray*}{rCl}
  \bigHPcond{A^{(v)}_{[n]}}{\vmat{X}^\set{V},\set{Q}}\geq\bigHPcond{A^{(v)}_{\set{I}_i}}{\vmat{X}^\set{V},\set{Q}},
  \quad\forall\,i\in [\nu].
\end{IEEEeqnarray*}

Let $v\in\set{V}$ and $v'\in\cset{\set{V}}\eqdef[\mu]\setminus\set{V}$. Following similar steps as in the proof given in \cite{BanawanUlukus18_1,XuZhang18_1}, we get
\begin{IEEEeqnarray}{rCl}
  \nu\bigHPcond{A^{(v)}_{[n]}}{\vmat{X}^\set{V},\set{Q}} 
  &\geq &\sum_{i=1}^\nu\bigHPcond{A^{(v)}_{\set{I}_i}}{\vmat{X}^\set{V},\set{Q}}
  \nonumber\\
  & = &\sum_{i=1}^\nu\left(\sum_{j\in\set{I}_i}\bigHPcond{A^{(v)}_j}{\vmat{X}^\set{V},\set{Q}}\right)
  \label{eq:use_kAnswers1}\\[1mm]
  & = &\sum_{i=1}^\nu\left(\sum_{j\in\set{I}_i}\bigHPcond{A^{(v')}_j}{\vmat{X}^\set{V},\set{Q}}\right)
  \label{eq:use_indistinct-answers}\\
  & = &\sum_{i=1}^\nu\bigHPcond{A^{(v')}_{\set{I}_i}}{\vmat{X}^\set{V},\set{Q}}
  \label{eq:use_kAnswers2}\\[1mm]  	
  & \geq &\kappa\bigHPcond{A^{(v')}_{[n]}}{\vmat{X}^\set{V},\set{Q}}
  \label{eq:use_Shearer-lemma}\\[1mm]
  & = & \kappa\Bigl[\bigHPcond{A^{(v')}_{[n]},\vmat{X}^{(v')}}{\vmat{X}^\set{V},\set{Q}}
  -\bigHPcond{\vmat{X}^{(v')}}{A^{(v')}_{[n]},\vmat{X}^\set{V},\set{Q}}\Bigr]
  \nonumber\\
  & = &\kappa\Bigl[\bigHPcond{\vmat{X}^{(v')}}{\vmat{X}^\set{V},\set{Q}}
  +\>\bigHPcond{A^{(v')}_{[n]}}{\vmat{X}^\set{V},\vmat{X}^{(v')},\set{Q}}-0\Bigr]
  \label{eq:use_recovery1}\\
  & = & \kappa\Bigl[\bigHPcond{\vmat{X}^{(v')}}{\vmat{X}^\set{V}}
  \!+\!\bigHPcond{A^{(v')}_{[n]}}{\vmat{X}^\set{V},\vmat{X}^{(v')},\set{Q}}\Bigr],\label{eq:use_independence}
\end{IEEEeqnarray}
where \eqref{eq:use_kAnswers1} and \eqref{eq:use_kAnswers2} follow from \eqref{eq:independent_kAnswers}; \eqref{eq:use_indistinct-answers} is because of \eqref{eq:indistinct-answers}; \eqref{eq:use_Shearer-lemma} is due to the Shearer's Lemma; \eqref{eq:use_recovery1} is from the fact that the $v'$-th function evaluation $\vmat{X}^{(v')}$ is determined by the answers $A^{(v')}_{[n]}$ and all possible queries $\set{Q}$; and finally, \eqref{eq:use_independence} follows from the independence between all possible queries and the messages. Therefore, we can conclude that
\begin{IEEEeqnarray}{rCl}
  \bigHPcond{A^{(v)}_{[n]}}{\vmat{X}^\set{V},\set{Q}}
  & \geq & \frac{\kappa}{\nu}\bigHPcond{\vmat{X}^{(v')}}{\vmat{X}^\set{V}}+\frac{\kappa}{\nu}
  \bigHPcond{A^{(v')}_{[n]}}{\vmat{X}^\set{V},\vmat{X}^{(v')},\set{Q}}
  \nonumber\\
  & = & \frac{k}{n}
  \bigHPcond{\vmat{X}^{(v')}}{\vmat{X}^\set{V}}
  +\frac{k}{n}\bigHPcond{A^{(v')}_{[n]}}{\vmat{X}^\set{V},\vmat{X}^{(v')},\set{Q}},
  \IEEEeqnarraynumspace\label{eq:use_MDS-PIRcapaciy-achieving-codes}
\end{IEEEeqnarray}
where we have used Definition~\ref{def:MDS-PIRcapacity-achieving-codes} to obtain \eqref{eq:use_MDS-PIRcapaciy-achieving-codes}.

Since there are in total $\mu$ function evaluations, by Definition~\ref{def:rank-function_PC} we can recursively use \eqref{eq:use_MDS-PIRcapaciy-achieving-codes} $r-1$ times with $\set{L}=\{\ell_1,\ldots,\ell_r\}\subseteq [\mu]$ to obtain
\begin{IEEEeqnarray}{rCl}    
  \bigHPcond{A_{[n]}^{(\ell_1)}}{\vmat{X}^{(\ell_1)},\set{Q}} 
  & \geq &\sum_{v=1}^{r-1}\Bigl(\frac{k}{n}\Bigr)^v
  \bigHPcond{\vmat{X}^{(\ell_{v+1})}}{\vmat{X}^{\{\ell_1,\ldots,\ell_{v}\}}}
  +\Bigl(\frac{k}{n}\Bigr)^{r-1}\bigHPcond{A_{[n]}^{(\ell_r)}}{\vmat{X}^{\{\ell_1,\ldots,\ell_r\}},\set{Q}}
  \nonumber\\
  & \geq &\sum_{v=1}^{r-1}\Bigl(\frac{k}{n}\Bigr)^v
  \bigHPcond{\vmat{X}^{(\ell_{v+1})}}{\vmat{X}^{\{\ell_1,\ldots,\ell_{v}\}}}
  \label{eq:use_nonegative-entropy}
\end{IEEEeqnarray}
where \eqref{eq:use_nonegative-entropy} follows from the nonnegativity of entropy. Here, we also remark that the recursive steps follow the same principle of the general converse for DPIR from \cite[Thm.~1]{ChenWangJafar18_1}. In \cite{ChenWangJafar18_1}, the authors claim that the general converse for the DPIR problem strongly depends on the chosen permutation of the indices of the candidate functions. Here, we also recognize a similar observation and assume that the order of indices $\{\ell_1,\dots,\ell_r\}$ is the permutation that maximizes the summation term of \eqref{eq:use_nonegative-entropy} and consider that $\vmat{X}^{(\ell_1)}$ is the polynomial function evaluation with the minimum entropy, i.e., $\bigHP{\vmat{X}^{(\ell_1)}}=\const{L}\HH^{(\textnormal{B})}_\textnormal{min}$. Now, 
\begin{IEEEeqnarray}{rCl}
  \const{L} \eHP{X^{(\ell_1)}} & = &\bigHP{\vmat{X}^{(\ell_1)}}
  \nonumber\\
  & = &\bigHPcond{\vmat{X}^{(\ell_1)}}{\set{Q}}-\underbrace{\bigHPcond{\vmat{X}^{(\ell_1)}}{A^{(\ell_1)}_{[n]},\set{Q}}}_{=0}
  \label{eq:use_recovery2}\\
  & = &\bigMIcond{\vmat{X}^{(\ell_1)}}{A^{(\ell_1)}_{[n]}}{\set{Q}}
  \nonumber\\[1mm]
  & = &\BigHPcond{A^{(\ell_1)}_{[n]}}{\set{Q}}-\BigHPcond{A^{(\ell_1)}_{[n]}}{\vmat{X}^{(\ell_1)},\set{Q}}
  \nonumber\\
  & \leq &\BigHPcond{A^{(\ell_1)}_{[n]}}{\set{Q}}-\sum_{v=1}^{r-1}\Bigl(\frac{k}{n}\Bigr)^v\bigHPcond{\vmat{X}^{(\ell_{v+1})}}{\vmat{X}^{(\ell_1)},\ldots,\vmat{X}^{(\ell_{v})}}, 
  \label{eq:use_recursive-step}
\end{IEEEeqnarray}
where \eqref{eq:use_recovery2} holds since any message is independent of the queries $\set{Q}$, and knowing the answers $A^{(\ell_1)}_{[n]}$ and the queries $\set{Q}$, one can determine $\vmat{X}^{(\ell_1)}$, and \eqref{eq:use_recursive-step} follows directly from \eqref{eq:use_nonegative-entropy}.

Finally, the converse proof is completed by showing that
\begin{IEEEeqnarray}{rCl}
  \const{R}& = &\frac{\const{L}\HH_\textnormal{min}}{\sum_{j=1}^n\bigHP{A^{(\ell_1)}_j}}
  \nonumber\\
  & \leq &\frac{\const{L}\HH_\textnormal{min}}{\bigHP{A^{(\ell_1)}_{[n]}}}
  \label{eq:use_chain-rule}\\
  & \leq &\frac{\const{L}\HH_\textnormal{min}}{\bigHPcond{A^{(\ell_1)}_{[n]}}{\set{Q}}}
  \label{eq:use_conditioning-entropy}\\
  & \leq &\frac{\HH_\textnormal{min}}{\HH^{(\textnormal{B})} _\textnormal{min}+\sum_{v=1}^{r-1}\bigl(\frac{k}{n}\bigr)^{v} \bigHPcond{{X}^{(\ell_{v+1})}}{{X}^{(\ell_1)},\ldots,{X}^{(\ell_{v})}}},\label{eq:converse_codedPC}
\end{IEEEeqnarray}%
where \eqref{eq:use_chain-rule} holds because of the chain rule of entropy, \eqref{eq:use_conditioning-entropy} is due to the fact that conditioning reduces entropy, and we apply \eqref{eq:use_recursive-step} to obtain \eqref{eq:converse_codedPC}.

\subsection{Special Case: PLC Converse}
\label{sec:PLCcapacity_MDS-PIRcapacity-achieving-codes}

Restricting the candidate polynomial set to degree $g=1$ polynomials gives rise to an interesting property following the linear dependencies between the function evaluations. In this subsection, we show how this property will reduce the general coded PPC converse bound to the coded PLC converse stated in the following theorem. %

\begin{theorem}
	\label{thm:PLCconverse_MDS-PIRcapacity-achieving-codes}
	Consider a DSS with $n$ noncolluding databases that uses an $[n,k]$ MDS-PIR capacity-achieving code $\code{C}$ to store $f$ messages. Then, the maximum achievable PLC rate over all possible PLC protocols, i.e., the PLC capacity $\const{C}_\textnormal{PLC}$, is upper bounded by
	\begin{IEEEeqnarray*}{c}
\const{C}_\textnormal{PLC}\leq\frac{1}{1+\sum_{v=1}^{r-1}\bigl(\frac{k}{n}\bigr)^{v}}=\Bigl(1-\frac{k}{n}\Bigr)\inv{\left[1-\Bigl(\frac{k}{n}\Bigr)^r\right]},  
	\end{IEEEeqnarray*}
	where $r$ is the rank of the linear mapping from \eqref{eq:linear-mappingV}.
\end{theorem}

To this end, we need the following lemma, whose proof is presented in Appendix~\ref{sec:proof_uniform-dist}. 
\begin{lemma}
  \label{lem:uniform_dist}
  Consider the linear mapping $\mat{V}=(v_{i,j})$ defined in \eqref{eq:linear-mappingV} with $\rank{\mat{V}}=r$ where $v_{i_1,j_1},\ldots,v_{i_r,j_r}$ are the entries corresponding to the pivot elements of $\mat{V}$. It follows that $\bigl(\vmat{X}^{(i_1)},\ldots,\vmat{X}^{(i_h)}\bigr)$ and $\bigl(\vmat{W}^{(j_1)},\ldots,\vmat{W}^{(j_h)}\bigr)$ are identically distributed, for some $h\in [r]$. In other words, $\bigHP{\vmat{X}^{(i_1)},\ldots,\vmat{X}^{(i_h)}}=\const{L}\bigHP{{X}^{(i_1)},\ldots,{X}^{(i_h)}}=h\const{L}$, $h\in [r]$.
\end{lemma}
\begin{IEEEproof}[Proof of \cref{thm:PLCconverse_MDS-PIRcapacity-achieving-codes}]
Now, from \eqref{eq:converse_codedPC}, we have
\begin{IEEEeqnarray}{rCl}
  \const{R}& \leq & \frac{\HH_\textnormal{min}}{\HH^{(\textnormal{B})} _\textnormal{min}+\sum_{v=1}^{r-1}\bigl(\frac{k}{n}\bigr)^{v} \bigHPcond{{X}^{(\ell_{v+1})}}{{X}^{(\ell_1)},\ldots,{X}^{(\ell_{v})}}}
  \nonumber\\
  & = &\frac{\HH_\textnormal{min}}{\HH^{(\textnormal{B})}_\textnormal{min}+\sum_{v=1}^{r-1}\bigl(\frac{k}{n}\bigr)^v},\label{eq:use_rank-lemma}
\end{IEEEeqnarray}%
where \eqref{eq:use_rank-lemma} holds since it follows from Lemma~\ref{lem:uniform_dist} that
 $\bigHPcond{{X}^{(\ell_{v+1})}}{{X}^{\{\ell_1,\ldots,\ell_{v}\}}}=\bigHP{{X}^{(\ell_{v+1})}}=1$.
For the PLC case, $\HH_\textnormal{min}=\HH^{(\textnormal{B})}_\textnormal{min}=1$, and the claim follows.
\end{IEEEproof}

It can be easily seen that the converse bound of Theorem~\ref{thm:PLCconverse_MDS-PIRcapacity-achieving-codes} matches the MDS-PIR capacity $\const{C}_{\textnormal{MDS-PIR}}$ for $f=r$ files given in  Theorem~\ref{thm:MDS-PIRcapacity-achieving-matrix}.


\section {Generic Query Generation for PC  From Coded DSSs} 
\label{sec:generic-query-generation_PC_coded-DSSs}

In this section, we construct a generic query generation algorithm for a PIR-like scheme, where its \emph{dependent virtual} messages represent the evaluations of the $\mu$ candidate polynomial functions. 
A PIR-like scheme achieves a private retrieval of the desired \emph{virtual} message by following three important design principle:
	\begin{itemize}
		\item Enforcing symmetry across databases. Each database is queried for an equal number of symbols and the query structure does not depend on the individual database, i.e., the scheme structure is fixed for all databases. 
		\item Enforcing symmetry across virtual messages.
		\item Exploiting side information represented by undesired information downloaded to maintain message symmetry. 
	\end{itemize}

The constructed generic algorithm is a generalized version of our query generation algorithm for PLC from coded DSSs, that first appeared in \cite{ObeadLinRosnesKliewer18_1}, and will act as the main building block for the PC schemes presented in this work.

\subsection{Generic PC Achievable Rate Matrix}
\label{sec:pc-achievable-rate}

Similar to Definition~\ref{def:PIRachievable-rate-matrix}, we now extend the notion of a PIR achievable rate matrix for the coded PIR problem to a coded generic PC problem.

\begin{definition}
  \label{def:generic-PCachievable-rate}  
  A $\nu\times n$ binary matrix ${\mat{\Lambda}}_{\kappa,\nu}^{\textnormal{PC}}$ is called a \emph{generic PC achievable rate matrix} 
  if it is a $\kappa$-column regular matrix, i.e., its column sums are equal to $\kappa$.
\end{definition}

Clearly, a PIR achievable rate matrix $\mat{\Lambda}^{\textnormal{PIR}}_{\kappa,\nu}$ is a generic PC achievable rate matrix. In general, the condition for  each row for a generic PC achievable rate matrix is not given, since it is not needed for generating the queries from our proposed algorithm. The required condition for each row of a particular PC achievable rate matrix will be specified in the subsequent sections, depending on the specific PC scheme  considered.

In~\cite[Def.~11]{KumarLinRosnesGraellAmat19_1}, two PIR interference matrices are defined from a PIR achievable rate matrix. Similar to the notion of PIR interference matrices, given a generic PC achievable rate matrix $\mat{\Lambda}^{\textnormal{PC}}_{\kappa,\nu}$, we can also formally define the \emph{PC interference matrices} $\mat{A}_{\kappa\times n}$ and $\mat{B}_{(\nu-\kappa)\times n}$, which are given by the following definition.

\begin{definition}
  \label{def:PCinterference-matrices}
  For a given $\nu\times n$ generic PC achievable rate matrix $\mat{\Lambda}^{\textnormal{PC}}_{\kappa,\nu}(\code{C})=(\lambda_{u,j})$, we define the PC interference matrices $\mat{A}_{\kappa\times n}=(a_{i,j})$ and $\mat{B}_{(\nu-\kappa)\times n}=(b_{i,j})$ for the code $\code{C}$ as
  \begin{IEEEeqnarray*}{rCl}
    a_{i,j}& \eqdef &u \text{ if } \lambda_{u,j}=1,\,\forall j \in [n], i \in[\kappa], u \in  [\nu], \\ 
    b_{i,j}& \eqdef &u \text{ if } \lambda_{u,j}=0,\,\forall j \in [n], i \in[\nu-\kappa], u \in  [\nu].
  \end{IEEEeqnarray*}
\end{definition}

Note that in \cref{def:PCinterference-matrices}, for each $j \in [n]$, distinct values of $u \in [\nu]$ should be assigned for all $i$. Thus, the assignment is not unique in the sense that the order of the entries of each column of $\mat{A}$ and $\mat{B}$ can be permuted.
	\begin{example} 
		\label{ex:MDSCapCodeex_n3k2}
		Consider the generic PC achievable rate matrix 
\begin{IEEEeqnarray*}{c}
	\mat{\Lambda}^{\textnormal{PC}}_{2,3}=
	\begin{pmatrix}
		0 &1 & 1
		\\
		1 &0 & 1
		\\
		1 &1 & 0
	\end{pmatrix}
\end{IEEEeqnarray*}  
with $(\kappa,\nu)=(2,3)$. One can readily see that 
the corresponding interference matrices are given by 
\begin{IEEEeqnarray*}{c}
	\mat{A}_{2\times 3}=
	\begin{pmatrix}
		2 &1 & 1
		\\
		3&3 & 2
	\end{pmatrix},
\quad \qquad
	\mat{B}_{1\times 3}=
	\begin{pmatrix}
		1 &2 & 3
	\end{pmatrix}. 
\end{IEEEeqnarray*} \QEDA
\end{example}

For $j\in [n]$, let $\set{A}_j\eqdef\{a_{i,j}\colon i\in [\kappa]\}$ and $\set{B}_j\eqdef\{b_{i,j}\colon i\in [\nu-\kappa]\}$. Note that the $j$-th column of $\mat{A}_{\kappa\times n}$ contains the row indices of $\mat{\Lambda}_{\kappa,\nu}$ whose entries in the $j$-th column are equal to $1$, while $\mat{B}_{(\nu-\kappa)\times n}$ contains the remaining row indices of $\mat{\Lambda}_{\kappa,\nu}$. Hence, it can be observed that $\set{B}_j=[\nu]\setminus\set{A}_j$, $\forall\,j\in [n]$. 

Next, for the sake of illustrating our generic query generation algorithm, we make use of the following definition.
\begin{definition}
  \label{def:uCoordinateSet_A}
  By $\set{S}(u|\mat{A}_{\kappa \times n})$ we denote the set of column coordinates of matrix
  $\mat{A}_{\kappa{\times}n}=(a_{i,j})$ in which at least one of its entries is equal to $u$, i.e.,
  \begin{IEEEeqnarray*}{rCl}
    \set{S}(u|\mat{A}_{\kappa\times n})\eqdef\{j\in [n]\colon\exists\,a_{i,j}=u,i\in [\kappa]\}.
  \end{IEEEeqnarray*}
\end{definition}
As a result, we require the size of the message to be $\const{L}= \nu^{\mu}\cdot k$ (i.e.,~${\beta=\nu^\mu}$). %

\subsection{Generic Query Generation}
\label{sec:query-generation}

In this subsection, we construct the generic queries that will be used in a coded PC scheme for $\mu$ \emph{dependent virtual} messages, which represent the evaluations of the $\mu$ candidate functions. Before running the main algorithm to generate the query sets, the following index preparation for the coded symbols stored in each database is performed.

{\bf \textit{1) Index Preparation}:} The goal is to make the symbols queried from each database to appear to be chosen randomly and independently from the desired function index. Note that the function is computed separately for the $t$-th row of all messages, $t \in [\beta]$. Therefore, similar to the PLC scheme in \cite{SunJafar19_2} and the MDS-coded PLC scheme in \cite{ObeadKliewer18_1}, we apply a permutation that is fixed across all coded symbols for the $t$-th row to maintain the dependency across the associated message elements. Let $\pi(\cdot)$ be a random permutation function over $[\beta]$, and let 
\begin{IEEEeqnarray}{c}
  U^{(v')}_{t,j}\triangleq \phi^{(v')}(\vect{C}_{\pi(t),j}),\,t\in[\beta], \, j\in[n], \, v'\in[\mu], \label{eq:IndexPrep}
\end{IEEEeqnarray}
denote the $t$-th permuted symbol associated with the $v'$-th virtual message ${\vmat{X}^{(v')}}$ stored in the $j$-th database, 
 where $\vect{C}_{t,j}\eqdef\trans{\bigl(C_{t,j}^{(1)},\ldots,C_{t,j}^{(f)}\bigr)}$. %
The permutation $\pi(\cdot)$ is randomly selected privately and uniformly by the user.

{\bf \textit{2) Preliminaries}:} The query generation procedure is subdivided into $\mu$ rounds, where in each round $\tau$ we generate the queries based on the concept of \emph{$\tau$-sums} as defined in the following.
\begin{definition}[$\tau$-sum]
  \label{def:tau-sum}
  For $\tau\in[\mu]$, a sum $U^{(v_1)}_{i_1,j} + U^{(v_2)}_{i_2,j} + \cdots + U^{(v_\tau)}_{i_\tau,j}$, $j\in[n]$, of $\tau$ distinct symbols is called a $\tau$-sum for any $(i_1,\ldots,i_{\tau})\in [\beta]^\tau$, and $\{v_1,\ldots,v_{\tau}\}\subseteq [\mu]$ determines the \emph{type} of the $\tau$-sum.
\end{definition}
Since we have $\binom{\mu}{\tau}$ different selections of $\tau$ distinct elements out of $\mu$ elements, a $\tau$-sum can have $\binom{\mu}{\tau}$ different \emph{types}.
For a requested function evaluation indexed by $v\in[\mu]$, a query set $Q^{(v)}_j$, $j\in[n]$, is composed of $\mu$ disjoint subsets of queries, each subset of queries is generated by the operations of each round $\tau\in[\mu]$. In a round we generate the queries for all possible $\binom{\mu}{\tau}$ types of $\tau$-sums. 
For each round $\tau\in [\mu]$ the  corresponding query subset is further subdivided into two subsets $Q^{(v)}_j(\set{D};\tau)$ and $Q^{(v)}_j(\set{U};\tau)$. The first subset $Q^{(v)}_j(\set{D};\tau)$ consists of $\tau$-sums with a single symbol from the \emph{desired} function evaluation and $\tau-1$ symbols from the evaluations of \emph{undesired} functions, while the second subset $Q^{(v)}_j(\set{U};\tau)$ contains $\tau$-sums with symbols only from the evaluations of undesired functions. Here, $\set{D}$ is an indicator for ``desired function evaluations'', while  $\set{U}$ an indicator for ``undesired functions evaluations''.
  Note that we require $\kappa^{\mu-(\tau-1)}(\nu-\kappa)^{\tau-1}$ distinct instances of each $\tau$-sum type for every query set $Q_j^{(v)}$. To this end, the algorithm will generate $\kappa n$ auxiliary query sets $Q^{(v)}_j(a_{i,j},\set{D};\tau)$, $i\in[\kappa]$, where each query consists of a distinct symbol from the desired function evaluation and $\tau-1$ symbols from undesired functions evaluations, and $(\nu-\kappa) n$ auxiliary query sets $Q^{(v)}_j(b_{i,j},\set{U};\tau)$, $i\in[\nu-\kappa]$, to represent the query sets of symbols from the undesired functions evaluations for each database $j\in [n]$. We utilize these sets to generate the query sets of each round according to the PC interference matrices $\mat{A}_{\kappa\times n}$ and $\mat{B}_{(\nu-\kappa)\times n}$. 
  
  	To illustrate the key concepts of the generic query generation algorithm, we use the following Example~\ref{ex:RunningMDSCapCodeex_n3k2} as a running example for this section. 
  	\begin{example} 
  		\label{ex:RunningMDSCapCodeex_n3k2}
  		Consider two messages $\vmat{W}^{(1)}$ and $\vmat{W}^{(2)}$ that are stored in a DSS using a length-$3$ code according to the system model in \cref{sec:problem-SystemModel}. In the following, we use the  generic PC achievable rate matrix from Example~\ref{ex:MDSCapCodeex_n3k2} with $(\kappa,\nu)=(2,3)$. 
  		Suppose that the user wishes to obtain the function evaluation $\vmat{X}^{(v)}$ from a set of $\mu=3$ candidate functions evaluations, i.e., $v\in[3]$. 
  We simplify notation by letting $x_{t,j}=U^{(1)}_{t,j}$, $y_{t,j}=U^{(2)}_{t,j}$, and $z_{t,j}=U^{(3)}_{t,j}$,  for all ${t\in[\beta]}$, $j\in[n]$, where $\beta=\nu^\mu=27.$ Let the desired function evaluation index be $v=1$. \QEDA
\end{example}

The query sets for all databases are generated by Algorithm~\ref{alg:generation_QuerySet} through the following procedures.\footnote{Note that a query $Q_j^{(v)}$ sent to the $j$-th database usually indicates the row indices of the symbols that the user requests, while the answer $A_j^{(v)}$ to the query $Q_j^{(v)}$ refers to the particular symbols requested through the query. In Algorithm~\ref{alg:generation_QuerySet}, with some abuse of notation for the sake of simplicity, the generated queries are sets containing their answers.}

{\bf \textit{3) Initialization (Round ${\tau=1}$)}:} In the initialization step, the algorithm generates the auxiliary queries for the first round. This round is described in lines {5} to {11} of Algorithm~\ref{alg:generation_QuerySet}, where we have $\tau=1$ for the $\tau$-sum. At this point, Algorithm~\ref{alg:generation_QuerySet} invokes the subroutine \texttt{Initial-Round} given in Algorithm~\ref{alg:initial-round} to generate $Q^{(v)}_j(a_{i,j},\set{D};1)$, $i\in[\kappa]$, such that each of these query sets contains $\alpha_1=\kappa^{\mu-1}$ distinct symbols. Furthermore, to maintain function symmetry, the algorithm asks each database for the same number of distinct symbols of all other functions evaluations in $Q^{(v)}_j(a_{i,j},\set{U};1)$, $i\in[\kappa]$, resulting in a total number of $\binom{\mu-1}{1}\kappa^{\mu-1}$ symbols. As a result, the queried symbols in the auxiliary query sets for each database are symmetric with respect to all function evaluation vectors indexed by $v'\in [\mu]$. In the following steps, we will associate the symbols of undesired functions evaluations in $\kappa$ groups, each placed in the undesired query sets $Q^{(v)}_j(a_{i,j},\set{U};1)$, $i\in[\kappa]$. Since this procedure produces $\kappa$ undesired query sets for each database, database symmetry is maintained. 
\begin{example}[continues=ex:RunningMDSCapCodeex_n3k2]
The initialization step is described in the following.
Algorithm~\ref{alg:generation_QuerySet} starts with ${\tau=1}$ to generate auxiliary query sets $Q^{(v)}_j(a_{i,j},\set{D};1)$,  $Q^{(v)}_j(a_{i,j},\set{U};1)$,  $i\in[\kappa]$, for each database $j\in[n]$. 
Starting at line {6} of Algorithm~\ref{alg:generation_QuerySet}, since $\nu=3$, we have the row indicator 
$u\in[3]$. 
This indicator is first used to identify the code coordinates pertaining 
to different entries $u=a_{i,j},$ 
as specified by the interference matrix $\mat{A}_{2\times 3}$
. For example, when $u=1$, following \cref{def:uCoordinateSet_A}, we have  $\set{S}(1|\mat{A}_{\kappa \times n})=\{2,3\}$. 
In line {7} of Algorithm~1, for $j\in\{2,3\}$, algorithm \texttt{Initial-Round} is invoked to generate the desired and undesired query subsets $Q^{(1)}_j(1,\set{D};1)$ and $Q^{(1)}_j(1,\set{U};1)$. The set $Q^{(1)}_j(1,\set{D};1)$ queries $\alpha_1=\kappa^{\mu-1}=4$ distinct instances of the desired function evaluation $x_{t,j}$  and the set $Q^{(1)}_j(1,\set{U};1)$ $\alpha_1=4$ distinct instances of the remaining functions evaluations $y_{t,j}$ and $z_{t,j}$. 
To this end, the row 
indicator $u$ is passed to the subroutine \texttt{Initial-Round}, i.e.,~Algorithm~\ref{alg:initial-round}, where it is used to determine the indices of the queried symbols.  For example, the first auxiliary query set for $u=1$ generated by Algorithm~\ref{alg:initial-round} is given by $Q^{(1)}_j(1,\set{D};1)=\{U^{(1)}_{({\r 1}-1)\cdot4+1,j}, U^{(1)}_{({\r 1}-1)\cdot4+2,j},U^{(1)}_{({\r 1}-1)\cdot4+3,j},U^{(1)}_{({\r 1}-1)\cdot4+4,j}\} =\{x_{1,j},x_{2,j}, x_{3,j}, x_{4,j}\}$, $j\in \{2,3\}$. A similar process is followed for $Q^{(1)}_j(1,\set{U};1)$. The same process is then repeated for the remaining 
$u=2$ and $u=3$. 
By the end of this step, we have queried $\nu\alpha_1=12$ distinct instances of the desired function evaluation $x_{t,j}$ and by message symmetry, $\nu\alpha_1=12$ distinct instances of the remaining functions evaluations $y_{t,j}$ and $z_{t,j}$. In total, the first round of queries comprises $n\kappa\alpha_{1}\mu=72$ symbols, which can be written in the form $n\binom{\mu}{1}\kappa^{\mu-1+1}(\nu-\kappa)^{1-1}$.
The resulting auxiliary query sets for the first round of queries are shown in Table~\ref{tab:axQ-tableA1}. \QEDA
\end{example}

\renewcommand{\baselinestretch}{0.95}
\begin{algorithm}[htbp]
  \caption{\texttt{Q-Gen}}
  \label{alg:generation_QuerySet}
  \SetInd{0.35em}{0.35em}
  \small
  \SetKwFunction{InitialRound}{Initial-Round}
  \SetKwFunction{Partition}{Partition}
  \SetKwFunction{DesiredQ}{Desired-Q}
  \SetKwFunction{ExploitSI}{Exploit-SI}
  \SetKwFunction{SetAddition}{SetAddition}
  \SetKwFunction{MSym}{M-Sym}  
  \SetKwInOut{Input}{Input}
  \SetKwInOut{Output}{Output}
  \SetKwComment{Comment}{$\triangleright$\ }{}{}
  \DontPrintSemicolon
  
  \Input{$v$, $\mu$, $\kappa$, $\nu$, $n$, $\mat{A}_{\kappa\times n}$, and $\mat{B}_{(\nu-\kappa)\times n}$}
  \Output{${Q}^{(v)}_1,\ldots,{Q}^{(v)}_n$}
  \For{$\tau\in [\mu]$}{    
  $Q^{(v)}_j(\mathcal{D};\tau) \leftarrow\emptyset$, $Q^{(v)}_j(\mathcal{U};\tau) \leftarrow\emptyset$,  $j\in [n]$\;
    $\alpha_{\tau}\leftarrow\kappa^{\mu-1}+\sum_{h=1}^{\tau-1}\binom{\mu-1}{h}\kappa^{\mu-(h+1)}(\nu-\kappa)^{h}$\;
    \Comment{\b Generate query sets for the initial round}
    \If{$\tau=1$}{
      \For{$u\in [\nu]$}{
        \For{$j\in \set{S}(u|\mat{A}_{\kappa\times n})$}{
          $Q^{(v)}_j(u,\set{D};\tau),Q^{(v)}_j(u,\set{U};\tau)\leftarrow\InitialRound(u,\alpha_\tau,j,v,\tau)$\;     
        }      
      }
    }
    \Comment{\b Generate query sets for the following rounds $\tau>1$}
    \Else{
      \For{$u\in [\nu]$}{
        \Comment{\b Generate desired symbols for the following rounds $\tau>1$}
        \For{$j\in \set{S}(u|\mat{A}_{\kappa\times n})$}{
          $Q^{(v)}_j(u,\set{D};\tau)\leftarrow\DesiredQ(u,\alpha_\tau,j,v,\tau)$\;}
        \Comment{\b Generate side information for the following rounds $\tau>1$}
        \For{$j\in\set{S}(u|\mat{B}_{(\nu-\kappa)\times n})$}{
          $Q^{(v)}_j(u,\set{U};\tau-1)\leftarrow
          \ExploitSI(u,Q^{(v)}_1(u,\mathcal{U},\tau-1),\ldots,Q^{(v)}_n(u,\mathcal{U},\tau-1),j,v,\tau)$\;}
      }
      \Comment{\b Generate the final desired query sets for the following rounds $\tau>1$}
        \For{$j\in [n]$}{
          $\tilde{Q}^{(v)}_j(\set{U};\tau-1)\leftarrow\bigcup\limits_{i\in [\nu-\kappa] }Q^{(v)}_j(b_{i,j},\set{U};\tau-1)$\;
          $\tilde{Q}^{(v)}_j(1,\set{U};\tau-1),\ldots, \tilde{Q}^{(v)}_j(\kappa,\set{U};\tau-1) \leftarrow
          \Partition\bigl(\tilde{Q}^{(v)}_j(\set{U};\tau-1)\bigr)$\;
          \For{$i\in [\kappa]$}{
            $Q^{(v)}_j(a_{i,j},\set{D};\tau)\leftarrow
            \SetAddition\bigl(Q_j^{(v)}(a_{i,j},\set{D};\tau),\tilde{Q}^{(v)}_j(i,\set{U};\tau-1)\bigr)$\;          
          }      
        }
      \Comment{\b Generate the query sets of undesired symbols by forcing message symmetry for the following rounds $\tau>1$}  
      \For{$u\in [\nu] $}{
        \For{$j\in\set{S}(u|\mat{A}_{\kappa\times n})$}{
          $Q^{(v)}_j(u,\set{U};\tau)\leftarrow\MSym\bigl(Q_j^{(v)}(u,\set{D};\tau),j,v,\tau\bigr)$\;}
      }
    }
      \For{$u\in [\nu]$}{
        \For{$j\in \set{S}(u|\mat{A}_{\kappa\times n})$}{
          $Q^{(v)}_j(\set{D};\tau)\leftarrow Q^{(v)}_j(\set{D};\tau) \cup Q^{(v)}_j(u,\set{D};\tau)$\;
          $Q^{(v)}_j(\set{U};\tau)\leftarrow Q^{(v)}_j(\set{U};\tau) \cup Q^{(v)}_j(u,\set{U};\tau)$\;
        }
      }
  }  
  \For{$j\in [n]$}{
    $Q^{(v)}_j\leftarrow\bigcup\limits_{\tau\in [\mu]}\Bigl(Q^{(v)}_j(\set{D};\tau)\cup 
    Q^{(v)}_j(\set{U};\tau)\Bigr)$\;
  }
\end{algorithm}
\renewcommand{\baselinestretch}{1}

\begin{algorithm}[htbp]
  \caption{\texttt{Initial-Round}}
  \label{alg:initial-round}
  \SetInd{0.35em}{0.35em}
  \small
  \SetKwFunction{new}{new}
  \SetKwInOut{Input}{Input}
  \SetKwInOut{Output}{Output}
  \SetKwComment{Comment}{$\triangleright$\ }{}
  \DontPrintSemicolon	 
  
  \Input{$u$, $\alpha_\tau$, $j$, $v$, and $\tau$}
  \Output{$\varphi^{(v)}(u,\set{D};\tau), \varphi^{(v)}(u,\set{U};\tau)$}
  
  $\varphi^{(v)}(u,\set{D};\tau)\leftarrow\emptyset$, $\varphi^{(v)}(u,\set{U};\tau)\leftarrow\emptyset$\;
  \For{$l\in [\alpha_{\tau}]$}{
    $\varphi^{(v)}(u,\set{D};\tau)\leftarrow\varphi^{(v)}(u,\set{D};\tau)\cup\bigl\{U^{(v)}_{({u}-1)
      \cdot\alpha_\tau+l,j}\bigr\}$\;
    $\varphi^{(v)}(u,\set{U};\tau)\leftarrow\varphi^{(v)}(u,\set{U};\tau)\ \cup$
    $
    \left(\bigcup\limits_{v'=1}^{\mu}\bigl\{U^{(v')}_{({u}-1)\cdot\alpha_\tau+l,j}\bigr\}
      \setminus\bigl\{U^{(v)}_{({u}-1)\cdot\alpha_\tau+l,j}\bigr\}\right)$\;
  }
\end{algorithm}

\begin{table}[htbp!]  
	\centering
	\caption{Auxiliary query sets for the first round. Highlighted in red is the row 
            indicator $u\in[\nu]$ used in determining the indices of the queried symbols. The cyan arrows indicate message symmetry. }
	 \label{tab:axQ-tableA1}
	\begin{IEEEeqnarraybox}[
		\IEEEeqnarraystrutmode
		\IEEEeqnarraystrutsizeadd{3pt}{2pt}]{v/c/v/c/v/c/v/c/v}
		\IEEEeqnarrayrulerow\\
		& j && 1  && 2 && 3\\
		\hline\hline
		& Q^{(1)}_j({\r 1},\set{D};1)
		&& \phantom{x_{1,1},  x_{2,1} ,  x_{3,1} ,  x_{4,1} } && x_{1,2},  x_{2,2} ,  x_{3,2} ,  x_{4,2} \tikzmark{x112e} && \tikzmark{x113b} x_{1,3},  x_{2,3} ,  x_{3,3} ,  x_{4,3} \tikzmark{x113e}&
		\\*\hline
			& \multirow{2}{*}{$Q^{(1)}_j({\r 1},\set{U};1)$}
		&& \phantom{x_{1,1},  x_{2,1} ,  x_{3,1} ,  x_{4,1} } && y_{1,2},  y_{2,2} ,  y_{3,2} ,  y_{4,2}  \tikzmark{y112e} && \tikzmark{y113b} y_{1,3},  y_{2,3} ,  y_{3,3} ,  y_{4,3} \tikzmark{y113e}&
		\\
		& && \phantom{x_{1,1},  x_{2,1} ,  x_{3,1} ,  x_{4,1} } && z_{1,2},  z_{2,2} ,  z_{3,2} ,  z_{4,2}  \tikzmark{z112e}  && z_{1,3},  z_{2,3} ,  z_{3,3} ,  z_{4,3} \tikzmark{z113}&	
		\\*\hline
		& Q^{(1)}_j({\r 2},\set{D};1)
		&& x_{5,1},  x_{6,1} ,  x_{7,1} ,  x_{8,1} \tikzmark{x121e}  && \phantom{x_{9,2},  x_{10,2} ,  x_{11,2} ,  x_{12,2}}  && \tikzmark{x123b} x_{5,3},  x_{6,3} ,  x_{7,3} ,  x_{8,3} \tikzmark{x123e}&
		\\*\hline
		& \multirow{2}{*}{$Q^{(1)}_j({\r 2},\set{U};1)$}
		&&    y_{5,1},  y_{6,1} ,  y_{7,1} ,  y_{8,1} \tikzmark{y121e}  && \phantom{ y_{1,2},  y_{2,2} ,  y_{3,2} ,  y_{4,2} } && \tikzmark{y123b} y_{5,3},  y_{6,3} ,  y_{7,3} ,  y_{8,3} \tikzmark{y123e}&
		\\
		& && z_{5,1},  z_{6,1} ,  z_{7,1} ,  z_{8,1} \tikzmark{z121e} &&\phantom{ z_{1,2},  z_{2,2} ,  z_{3,2} ,  z_{4,2}}  && z_{5,3},  z_{6,3} ,  z_{7,3} ,  z_{8,3} \tikzmark{z123e} &	
		\\*\hline
		& Q^{(1)}_j({\r 3},\set{D};1)
		&& x_{9,1},  x_{10,1} ,  x_{11,1} ,  x_{12,1} \, \tikzmark{x131e} && \tikzmark{x132b}\, x_{9,2},  x_{10,2} ,  x_{11,2} ,  x_{12,2}  \tikzmark{x132e}&& \phantom{x_{1,3},  x_{2,3} ,  x_{3,3} ,  x_{4,3}}&
		\\*\hline
		& \multirow{2}{*}{$Q^{(1)}_j({\r 3},\set{U};1)$}
		&&     y_{9,1},  y_{10,1} ,  y_{11,1} ,  y_{12,1}   \tikzmark{y131e} && \tikzmark{y132b} \, y_{9,2},  y_{10,2} ,  y_{11,2} ,  y_{12,2} \, \tikzmark{y132e}  && \phantom{y_{9,3},  y_{10,3} ,  y_{11,3} ,  y_{12,3} }&
		\\
		& && z_{9,1},  z_{10,1} ,  z_{11,1} ,  z_{12,1} \tikzmark{z131e}  && \tikzmark{z132b} z_{9,2},  z_{10,2} ,  z_{11,2} ,  z_{12,2} \, \tikzmark{z132e}  && \phantom{z_{1,3},  z_{2,3} ,  z_{3,3} ,  z_{4,3}}&	
		\\*\IEEEeqnarrayrulerow
\end{IEEEeqnarraybox}
 \tikz[overlay,remember picture]{
	\path[cyan,>=stealth,->,thick] (x112e)+(0.2,0) edge [bend left] ($(y112e)+(0.2,-0.3)$);
	\path[cyan,>=stealth,->,thick] (x113e)+(0.2,0) edge [bend left] ($(y113e)+(0.2,-0.3)$);
	\path[cyan,>=stealth,->,thick] (x121e)+(0.2,0) edge [bend left] ($(y121e)+(0.2,-0.3)$);
	\path[cyan,>=stealth,->,thick] (x123e)+(0.2,0) edge [bend left] ($(y123e)+(0.2,-0.3)$);
	\path[cyan,>=stealth,->,thick] (x131e)+(0.05,0) edge [bend left] ($(y131e)+(0.05,-0.3)$);
	\path[cyan,>=stealth,->,thick] (x132e)+(0.05,0) edge [bend left] ($(y132e)+(0.05,-0.3)$);
}
\end{table}

{\bf \textit{4) Desired Function Symbols for Rounds ${\tau>1}$}:} For the following rounds a similar process is repeated in terms of generating auxiliary query sets containing distinct symbols from the desired function evaluation $\vmat{U}^{(v)}=(U^{(v)}_{t,j})$. This is accomplished in lines {16} to {18} by calling the subroutine $\texttt{Desired-Q}$, given in Algorithm~\ref{alg:Desired-setQ_following-rounds}, to generate $Q^{(v)}_j(a_{i,j},\set{D};\tau)$, $i\in[\kappa]$, such that each of these query sets contains $(\alpha_{\tau}-1)-\alpha_{\tau-1}+1=\binom{\mu-1}{\tau-1}\kappa^{\mu-(\tau-1+1)}(\nu-\kappa)^{\tau-1}$ distinct symbols from the desired function evaluation $\vmat{U}^{(v)}$.

\begin{example}[continues=ex:RunningMDSCapCodeex_n3k2]
After successfully generating the queries for $\nu \alpha_1=12$ distinct symbols from the desired function evaluation in the initiation step, for round $\tau=2$ we generate the queries for the following $\nu (\alpha_2-\alpha_1)=12$  symbols. To this end, subroutine $\texttt{Desired-Q}$, given in Algorithm~\ref{alg:Desired-setQ_following-rounds}, generates auxiliary query sets $Q^{(1)}_j(a_{i,j},\set{D};2)$ containing distinct symbols from the desired function evaluation,  following a process similar to Algorithm~\ref{alg:initial-round}, however with a different method for determining the queried indices. 
Table~\ref{tab:axQ-tableA2} shows the output of lines {16} to {18} after calling the subroutine $\texttt{Desired-Q}$ for $u\in[3]$. \QEDA
\end{example}

\begin{algorithm}[htbp]
  \caption{\texttt{Desired-Q}}
  \label{alg:Desired-setQ_following-rounds}
  \SetInd{0.35em}{0.35em}
  \small
  \SetKwInOut{Input}{Input}
  \SetKwInOut{Output}{Output}
  \SetKwComment{Comment}{$\triangleright$\ }{}
  \DontPrintSemicolon	   
  \Input{$u$, $\alpha_\tau$, $j$, $v$, and $\tau$}
  \Output{$\varphi^{(v)}(u,\set{D};\tau)$}  
  $\varphi^{(v)}(u,\set{D};\tau)\leftarrow\emptyset$\;
  \For{$l\in[\alpha_{\tau-1}:\alpha_\tau-1]$}{
    $\varphi^{(v)}(u,\set{D};\tau)\leftarrow\varphi^{(v)}(u,\set{D};\tau)
    \cup\bigl\{U^{(v)}_{l\cdot\nu+{u},j}\bigr\}$\;
  }
\end{algorithm}

\begin{table}[htbp!]  
	\centering
	\caption{Initial auxiliary query sets for the second round. Highlighted in red is the row 
            indicator $u\in[\nu]$ used in determining the indices of the queried symbols.}
	 \label{tab:axQ-tableA2}
	\begin{IEEEeqnarraybox}[
		\IEEEeqnarraystrutmode
		\IEEEeqnarraystrutsizeadd{3pt}{2pt}]{v/c/v/c/v/c/v/c/v}
		\IEEEeqnarrayrulerow\\
		& j && 1  && 2 && 3\\
		\hline\hline
			& \multirow{2}{*}{$Q^{(1)}_j({\r 1},\set{D};2)$}
		&& \phantom{x_{1,1},  x_{2,1} ,  x_{3,1} } && x_{4\cdot3+{\r 1},2},  && x_{4\cdot3+{\r 1},3}, &
		\\
		& && \phantom{x_{16,1},  x_{19,1} ,  x_{22,1}  } && x_{16,2},  x_{19,2} ,  x_{22,2}  && x_{16,3},  x_{19,3} ,  x_{22,3}  &	
		\\*\hline
		& \multirow{2}{*}{$Q^{(1)}_j({\r 2},\set{D};2)$}
		&&    x_{4\cdot3+{\r 2},1},  && \phantom{ y_{1,2},  y_{2,2} ,  y_{3,2}} && x_{4\cdot3+{\r 2},3}  &
		\\
		& && x_{17,1},  x_{20,1} ,  x_{23,1}  &&\phantom{ z_{1,2},  z_{2,2} ,  z_{3,2} }  && x_{17,3},  x_{20,3} ,  x_{23,3} &	
		\\*\hline
		& \multirow{2}{*}{$Q^{(1)}_j({\r 3},\set{D};2)$}
		&&     x_{4\cdot3+{\r 3},1}, &&  x_{4\cdot3+{\r 3},2},   && \phantom{y_{9,3},  y_{10,3} ,  y_{11,3} }&
		\\
		& && x_{18,1},  x_{21,1} ,  x_{24,1} && x_{18,2},  x_{21,2} ,  x_{24,2}   && \phantom{x_{1,3},  x_{21,3} ,  x_{24,3}}&	
		\\*\IEEEeqnarrayrulerow
\end{IEEEeqnarraybox}
	%
	%
\end{table}

{\bf \textit{5) Side Information Exploitation}:} In lines {20} to {22}, we generate the \emph{side information} query sets $Q^{(v)}_j(b_{i',j},\set{U};\tau-1)$, $i'\in [\nu-\kappa]$, from the auxiliary query sets $Q^{(v)}_1(a_{i,1},\set{U};\tau-1),\ldots,Q^{(v)}_n(a_{i,n},\set{U};\tau-1)$, $i\in [\kappa]$, of the previous round $\tau-1$, $\tau\in[2:\mu]$, by applying the subroutine \texttt{Exploit-SI}, given by Algorithm~\ref{alg:Exploit-SI}. This subroutine is extended from \cite{SunJafar19_2} based on our coded storage scenario. These side information query sets will be exploited by the user to ensure the recovery and privacy of the subsequent PC schemes. Note that in Algorithm~\ref{alg:Exploit-SI} the function \texttt{Reproduce}$(j,Q^{(v)}_{j'}(u,\set{U};\tau-1))$, $j'\in [n]\setminus\{j\}$, simply reproduces all the queries in the auxiliary query set $Q^{(v)}_{j'}(u,\set{U};\tau-1)$ with a different coordinate $j$.

\begin{algorithm}[htbp!]
  \caption{\texttt{Exploit-SI}}
  \label{alg:Exploit-SI}
  \SetInd{0.35em}{0.35em}
  \small
  \SetKwFunction{Reproduce}{Reproduce}
  \SetKwInOut{Input}{Input}
  \SetKwInOut{Output}{Output}
  \SetKwComment{Comment}{$\triangleright$\ }{}
  \DontPrintSemicolon	 
  
  \Input{$u$, $Q^{(v)}_1(u,\set{U};\tau-1),\ldots,Q^{(v)}_n(u,\set{U};\tau-1)$, $j$, $v$, and $\tau$} 
  \Output{$\varphi^{(v)}(u,\set{U};\tau-1)$}
  
  $\varphi^{(v)}(u,\set{U};\tau-1)\leftarrow\emptyset$\;  
  \For{$i\in [\kappa]$}{
    \For{$j'\in [n]\setminus\{j\}$}{
      \If{$u=a_{i,j'}$}{
        $\varphi^{(v)}(u,\set{U};\tau-1)\leftarrow\Reproduce(j,Q^{(v)}_{j'}(u,\set{U};\tau-1))$\;
        break\;
      }
    }
  }
\end{algorithm}

Next, we update the desired query sets $Q^{(v)}_j(a_{i,j},\set{D};\tau)$ in lines {25} to {31}. First, the function $\texttt{Partition}\bigl(\tilde{Q}^{(v)}_j(\set{U};\tau-1)\bigr)$ denotes a procedure that divides a set into $\kappa$ disjoint equally-sized subsets. This is viable since based on the subroutine \texttt{Initial-Round} and the following subroutine \texttt{M-Sym}, one can show that $\bigcard{\tilde{Q}^{(v)}_j(\set{U};\tau-1)}=\binom{\mu-1}{\tau-1}\kappa^{\mu-(\tau-1)}(\nu-\kappa)^{(\tau-1)-1}\cdot(\nu-\kappa)$ for each round $\tau\in[2:\mu]$, which is always divisible by $\kappa$. Secondly, we assign the new query set of desired symbols $Q^{(v)}_j(a_{i,j},\set{D};\tau)$ for the current round by using an element-wise set addition $\texttt{SetAddition}(Q_1,Q_2)$. The element-wise set addition is defined as $\bigl\{q_{i_l}+q_{i'_l}\colon q_{i_l}\in Q_1,q_{i'_l}\in Q_2, l\in [\rho]\bigr\}$ with $\card{Q_1}=\card{Q_2}=\rho$, where $\rho$ is an appropriate integer.  In lines {33} to {37}, the subroutine $\texttt{M-Sym}$, given in Algorithm~\ref{alg:message-symmetry}, is invoked to generate the undesired query sets $Q^{(v)}_j(a_{i,j},\set{U};\tau)$ by utilizing message symmetry. This subroutine selects symbols of undesired functions evaluations to generate $\tau$-sums that enforce symmetry in the round queries. The procedure resembles the subroutine \texttt{M-Sym} proposed in \cite{SunJafar19_2}.
In Algorithm~\ref{alg:message-symmetry},
$\Pi_\tau$ denotes the set of all possible selections of $\tau$ distinct indices in $[\mu]$ and $\texttt{Lexico}(\Pi_\tau)$ denotes the corresponding set of ordered selections (the indices $(v_1,\ldots,v_\tau)$ of a selection of $\Pi_\tau$ are ordered in natural lexicographical order).
Further, the notation $U^{(v_x)}_{\ast,j}$ implies that the row index of the symbol can be arbitrary. This is the case since only the function indices $(v_1,\ldots,v_\tau)$ are necessary to determine $i_z$, $\forall\,z\in [\tau]$. As a result, symmetry over the functions is maintained. Moreover, for $Q_j^{(v)}(a_{i,j},\set{U};\tau)$, $i\in [\kappa]$, we obtain for each $\tau\in[2:\mu]$ the remaining $\tau$-sum types, such that each of these query sets contains $\binom{\mu-1}{\tau} \kappa^{\mu-(\tau-1+1)}(\nu-\kappa)^{\tau-1}$ symbols.
\begin{algorithm}[htbp!]
  \caption{\texttt{M-Sym}}
  \label{alg:message-symmetry}
  \SetInd{0.35em}{0.35em}
  \small
  \SetKwFunction{Msym}{M-sym}
  \SetKwFunction{Lexico}{Lexico}
  \SetKwInOut{Input}{Input}
  \SetKwInOut{Output}{Output}
  \SetKwComment{Comment}{$\triangleright$\ }{}
  \DontPrintSemicolon	 
  
  \Input{$Q^{(v)}_j(u,\set{D};\tau)$, $j$, $v$, and $\tau$}
  \Output{$\varphi^{(v)}(u,\set{U};\tau)$}
  
  $\varphi^{(v)}(u,\set{U};\tau)\leftarrow\emptyset$\;
  \For{$(v_1,\ldots,v_\tau)\in\Lexico(\Pi_\tau)$, $v\notin \{v_1,\ldots,v_\tau \}$}{
    $\varphi^{(v)}(u,\set{U};\tau)\leftarrow\varphi^{(v)}(u,\set{U};\tau)\cup
    \bigl\{U^{(v_1)}_{i_1,j}+\ldots+U^{(v_\tau)}_{i_\tau,j}\bigr\}$ $\textnormal{ such that }
    \forall\,z\in [\tau] $, $\exists\,U^{(v)}_{i_z,j}+\sum\limits_{\substack{x\in [\tau]\\ x\neq z}}
    U^{(v_x)}_{\ast,j}\in Q^{(v)}_j(u,\set{D};\tau)$\;
  }
\end{algorithm}

\begin{example}[continues=ex:RunningMDSCapCodeex_n3k2]
		After determining the indices of the desired function evaluations to be queried by each database in round $\tau=2$, we now deploy side information to preserve the privacy for the desired function evaluation. This is accomplished by generating $\tau$-sums of each possible \emph{type} and enforcing index symmetry. To this end, we first identify the side information available from the previous round, queried from the neighboring databases, to be exploited according to the interference matrix $\mat{B}_{1\times 3}$. This process is performed by invoking Algorithm~\ref{alg:Exploit-SI}, which generates \emph{complement} sets for the undesired query sets of the previous round, i.e., $Q^{(1)}_j(a_{i,j},\set{U};1)$. Table~\ref{tab:axQ-tableA3}(a) shows the output of Algorithm~\ref{alg:Exploit-SI} for $u\in[3]$. Next, these side information query sets are then partitioned into $\kappa$ groups to be exploited in different $Q^{(1)}_j(a_{i,j},\set{D};2)$ for $i\in [\kappa]$. 
		%
		%
		The partitioning guarantees that the two sets used in generating the $\tau$-sums in lines {28} to {30} of Algorithm~\ref{alg:generation_QuerySet} have an equal number of elements. 
		Finally, message and index symmetry is guaranteed by passing the generated auxiliary query sets $Q^{(1)}_j(a_{i,j},\set{D};2)$ to the subroutine $\texttt{M-Sym}$, i.e.,~Algorithm~\ref{alg:message-symmetry}, that generates $\tau$-sums of the remaining \emph{types}. Table~\ref{tab:axQ-tableA3}(b) illustrates the final query sets for round $\tau=2$. 
		\QEDA
	\end{example}
\begin{table}[htbp!]  
	\centering
	\caption{Side information generation and exploitation. (a) shows the side information generated by the \texttt{Exploit-SI} algorithm from the auxiliary query sets of the first round. The generated side information shown in (a) is exploited in forming $\tau$-sums of the final auxiliary query sets of the second round shown in (b). The cyan arrows indicate that the \texttt{M-Sym} algorithm is used. Finally, highlighted in red is the row 
            indicator $u\in[\nu]$ used in determining the indices of the queried symbols. 
	}
	 \label{tab:axQ-tableA3}
	\begin{IEEEeqnarraybox}[
		\IEEEeqnarraystrutmode
		\IEEEeqnarraystrutsizeadd{3pt}{2pt}]{v/c/v/c/v/c/v/c/v}
		\IEEEeqnarrayrulerow\\
		& j && 1  && 2 && 3\\
		\hline\hline
			& \multirow{2}{*}{$Q^{(1)}_j({\r 1},\set{U};1)$}
		&&  y_{1,1},  y_{2,1} ,  y_{3,1} ,  y_{4,1}  && \phantom{y_{1,2},  y_{2,2} ,  y_{3,2} ,  y_{4,2}}  && \phantom{ y_{1,3},  y_{2,3} ,  y_{3,3} ,  y_{4,3} }&
		\\
		& && z_{1,1},  z_{2,1} ,  z_{3,1} ,  z_{4,1}  &&  \phantom{z_{1,2},  z_{2,2} ,  z_{3,2} ,  z_{4,2}}   && \phantom{z_{1,3},  z_{2,3} ,  z_{3,3} ,  z_{4,3}}&	
		\\*\hline
		& \multirow{2}{*}{$Q^{(1)}_j({\r 2},\set{U};1)$}
		&&    \phantom{y_{5,1},  y_{6,1} ,  y_{7,1} ,  y_{8,1}}  && y_{5,2},  y_{6,2} ,  y_{7,2} ,  y_{8,2} && \phantom{ y_{5,3},  y_{6,3} ,  y_{7,3} ,  y_{8,3} }&
		\\
		& && \phantom{z_{5,1},  z_{6,1} ,  z_{7,1} ,  z_{8,1}} &&z_{5,2},  z_{6,2} ,  z_{7,2} ,  z_{8,2} && \phantom{ z_{5,3},  z_{6,3} ,  z_{7,3} ,  z_{8,3} } &	
		\\*\hline
		& \multirow{2}{*}{$Q^{(1)}_j({\r 3},\set{U};1)$}
		&&     \phantom{y_{9,1},  y_{10,1} ,  y_{11,1} ,  y_{12,1}}   &&  \phantom{ y_{9,2},  y_{10,2} ,  y_{11,2} ,  y_{12,2} } && y_{9,3},  y_{10,3} ,  y_{11,3} ,  y_{12,3} &
		\\
		& && \phantom{z_{9,1},  z_{10,1} ,  z_{11,1} ,  z_{12,1} }  && \phantom{ z_{9,2},  z_{10,2} ,  z_{11,2} ,  z_{12,2}} && z_{9,3},  z_{10,3} ,  z_{11,3} ,  z_{12,3}&	
		\\*\IEEEeqnarrayrulerow
\end{IEEEeqnarraybox}

 (a)
\\[2mm]

	\begin{IEEEeqnarraybox}[
	\IEEEeqnarraystrutmode
	\IEEEeqnarraystrutsizeadd{3pt}{2pt}]{v/c/v/c/v/c/v/c/v}
	\IEEEeqnarrayrulerow\\
	& j && 1  && 2 && 3\\
	\hline\hline
	& \multirow{4}{*}{$Q^{(1)}_j({\r 1},\set{D};2)$}
	&& \phantom{y_{1,1},  y_{2,1} ,  y_{3,1} ,  y_{4,1}} && x_{13,2} + y_{5,2}  && x_{13,3}+ y_{9,3} &
		\\
	& && \phantom{x_{16,2} + z_{5,2}  } && x_{16,2} + z_{5,2} && x_{16,3} +z_{9,3}   &	
		\\
	& && \phantom{x_{19,2} + y_{7,2} } && x_{19,2} + y_{7,2} &&  x_{19,3} +y_{11,3}  &	
		\\
	& && \phantom{x_{22,2} +  z_{7,2}  } && x_{22,2} +  z_{7,2}  \tikzmark{x212e}&& x_{22,3} +z_{11,3} \tikzmark{x213e}  &	
	\\*\hline
	& \multirow{2}{*}{$Q^{(1)}_j({\r 1},\set{U};2)$}
		&& \phantom{y_{17,1}+z_{ 14,1}} && y_{16,2} + z_{13,2} \tikzmark{y212e} && y_{16,3} +z_{13,3}  \tikzmark{y213e}  &	
	\\
	& && \phantom{y_{23,1}+z_{ 20,1}}&& y_{22,2} +  z_{19,2}  \tikzmark{z212e}  && y_{22,3} +z_{19,3} \tikzmark{z213} &	
	\\*\hline
	& \multirow{4}{*}{$Q^{(1)}_j({\r 2},\set{D};2)$}
	&&    x_{14,1}+y_{ 1,1}  && \phantom{ y_{5,2},  y_{6,2} ,  y_{7,2} ,  y_{8,2}} && x_{14,3}+y_{10,3} &
	\\
	& && x_{17,1}+z_{ 1,1}  && \phantom{ x_{17,3}+z_{11,3}}  && x_{17,3}+z_{10,3} &	
	\\
	& && x_{20,1}+y_{ 3,1} && \phantom{ x_{20,3} +y_{12,3} }  && x_{20,3} +y_{12,3} &	
	\\
	& &&  x_{23,1}+z_{ 3,1} \tikzmark{x221e} &&\phantom{  x_{23,3}+z_{12,3}}  &&  x_{23,3}+z_{12,3} \tikzmark{x223e}&	
	\\*\hline
		& \multirow{2}{*}{$Q^{(1)}_j({\r 2},\set{U};2)$}
	&& y_{17,1}+z_{ 14,1} \tikzmark{y221e} && \phantom{y_{16,2} + z_{13,2} } && y_{17,3}+z_{14,3}  \tikzmark{y223e}&	
	\\
	& && y_{23,1}+z_{ 20,1}  \tikzmark{z221e}  && \phantom{y_{22,2} +  z_{19,2} }  && y_{23,3}+z_{20,3}\tikzmark{z223e} &	
	\\*\hline
	& \multirow{4}{*}{$Q^{(1)}_j({\r 3},\set{D};2)$}
	&&     x_{15,1}+y_{ 2,1}  &&  x_{15,2}+y_{6,2}  && \phantom{y_{9,2},  y_{10,2} ,  y_{11,2} ,  y_{12,2}}&
	\\
	& && x_{18,1}+z_{ 2,1}  && x_{18,2}+z_{ 6,2} && \phantom{x_{1,3},  x_{21,3} ,  x_{24,3}}&	
		\\
	& && x_{21,1}+y_{ 4,1} && x_{21,2}+y_{ 8,2}   && \phantom{x_{1,3},  x_{21,3} ,  x_{24,3}}&	
		\\
	& &&  x_{24,1}+z_{ 4,1} \tikzmark{x231e} &&   x_{24,2}+z_{8 ,2}   \tikzmark{x232e} && \phantom{x_{1,3},  x_{21,3} ,  x_{24,3}}&	
	\\*\hline
	& \multirow{2}{*}{$Q^{(1)}_j({\r 3},\set{U};2)$}
	&& y_{18,1}+z_{ 15,1} \tikzmark{y231e}  && y_{18,2}+z_{ 15,2} \tikzmark{y232e}  && \phantom{y_{17,3}+z_{14,3}} &	
	\\	
	& &&  y_{24,1}+z_{ 21,1} \tikzmark{z231e}  &&y_{24,2}+z_{21 ,2} \tikzmark{z232e}    && \phantom{ y_{23,3}+z_{20,3} }&	
	\\*\IEEEeqnarrayrulerow
\end{IEEEeqnarraybox}

(b)

 \tikz[overlay,remember picture]{
	\path[cyan,>=stealth,->,thick] (x212e)+(0.2,0) edge [bend left] ($(y212e)+(0.2,-0.3)$);
	\path[cyan,>=stealth,->,thick] (x213e)+(0.2,0) edge [bend left] ($(y213e)+(0.2,-0.3)$);
	%
	\path[cyan,>=stealth,->,thick] (x221e)+(0.2,0) edge [bend left] ($(y221e)+(0.2,-0.3)$);
	\path[cyan,>=stealth,->,thick] (x223e)+(0.2,0) edge [bend left] ($(y223e)+(0.2,-0.3)$);
	%
	\path[cyan,>=stealth,->,thick] (x231e)+(0.05,0) edge [bend left] ($(y231e)+(0.05,-0.3)$);
	\path[cyan,>=stealth,->,thick] (x232e)+(0.05,0) edge [bend left] ($(y232e)+(0.05,-0.3)$);
}
\end{table}

{\bf \textit{6) Query Set Assembly}:} Finally, in lines {39} to {48}, we assemble each query set from disjoint query subsets obtained in all $\tau$ rounds. It can be shown that $Q_j^{(v)}(\set{D};\tau) \cup Q_j^{(v)}(\set{U};\tau)$ contains $\kappa^{\mu-(\tau-1)}(\nu-\kappa)^{\tau-1}$ $\tau$-sums for every $\tau$-sum type as follows.

For the initialization round, $\tau=1$, from Step~3) above, the total number of queried symbols is given by
\begin{IEEEeqnarray*}{rCl}
  \bigcard{Q_j^{(v)}(\set{D};1) \cup Q_j^{(v)}(\set{U};1)} &=&  \kappa \left[  \kappa^{\mu-1} +\binom{\mu-1}{1} \kappa^{\mu-1}  \right] =  \binom{\mu}{1} \kappa^{\mu-1+1} (\nu-\kappa)^{1-1}.   
\end{IEEEeqnarray*}
For the following rounds, $\tau\in [2:\mu]$, from Steps 4) and 5) above, we have
\begin{IEEEeqnarray*}{rCl}
\bigcard{Q_j^{(v)}(\set{D};\tau) \cup Q_j^{(v)}(\set{U};\tau)} &=&  \kappa \left[ \binom{\mu-1}{\tau-1} \kappa^{\mu-\tau}(\nu-\kappa)^{\tau-1} +  \binom{\mu-1}{\tau} \kappa^{\mu-\tau}(\nu-\kappa)^{\tau-1}  \right] \\
&=&  \left(\binom{\mu-1}{\tau-1} +\binom{\mu-1}{\tau} \right) \kappa^{\mu-\tau+1}(\nu-\kappa)^{\tau-1} \\*%
&=&   \binom{\mu}{\tau} \kappa^{\mu-\tau+1}(\nu-\kappa)^{\tau-1}.
\end{IEEEeqnarray*}
 In summary, the total number of queries generated by Algorithm~\ref{alg:generation_QuerySet} is
\begin{IEEEeqnarray}{c}
  \sum_{j=1}^{n}\bigcard{Q_j^{(v)}}= n\sum_{\tau=1}^{\mu}{\mu \choose \tau}\kappa^{\mu-\tau+1}(\nu-\kappa)^{\tau-1}.
  \label{eq:query-card_PC}
\end{IEEEeqnarray}

\begin{example}[continues=ex:RunningMDSCapCodeex_n3k2]
		After performing steps 
		4) and 5)  for the following final round, i.e., for $\tau=3$, the queries for the remaining $\nu (\alpha_3-\alpha_2)=3$ distinct symbols of the desired function evaluation are generated. In the final step, i.e., step 6), the auxiliary query subsets are aggregated according to the row 
                indicator
			\begin{table}[htbp!]  
	\centering
	\caption{Generic PC query sets for $v=1$ for the generic PC achievable rate matrix $\mat{\Lambda}^{\textnormal{PC}}_{2,3}$ of \cref{ex:MDSCapCodeex_n3k2}, $f=2$ messages, and $\mu=3$ candidate functions.} 
	 \label{tab:axQ-tableA4}
	\begin{IEEEeqnarraybox}[
		\IEEEeqnarraystrutmode
		\IEEEeqnarraystrutsizeadd{3pt}{2pt}]{v/c/v/c/v/c/v/c/v}
		\IEEEeqnarrayrulerow\\
		& j && 1  && 2 && 3\\
		\hline\hline
			& \multirow{2}{*}{$Q^{(1)}_j(\set{D};1)$}
		&& x_{5,1},  x_{6,1} ,  x_{7,1} ,  x_{8,1} \tikzmark{x121e} && x_{1,2},  x_{2,2} ,  x_{3,2} ,  x_{4,2} \tikzmark{x112e} &&  \tikzmark{x113b} x_{1,3},  x_{2,3} ,  x_{3,3} ,  x_{4,3} \tikzmark{x113e}&
		\\ \cline{3-9}
		& &&  x_{9,1},  x_{10,1} ,  x_{11,1} ,  x_{12,1} \, \tikzmark{x131e}  && \tikzmark{x132b}\, x_{9,2},  x_{10,2} ,  x_{11,2} ,  x_{12,2} \tikzmark{x132e} &&\tikzmark{x123b} x_{5,3},  x_{6,3} ,  x_{7,3} ,  x_{8,3} \tikzmark{x123e}&
		\\*\hline
			& \multirow{4}{*}{$Q^{(1)}_j(\set{U};1)$}
		&& y_{5,1},  y_{6,1} ,  y_{7,1} ,  y_{8,1} \tikzmark{y121e} && y_{1,2},  y_{2,2} ,  y_{3,2} ,  y_{4,2}  \tikzmark{y112e} && \tikzmark{y113b} y_{1,3},  y_{2,3} ,  y_{3,3} ,  y_{4,3} \tikzmark{y113e}&
		\\
		& && z_{5,1},  z_{6,1} ,  z_{7,1} ,  z_{8,1} \tikzmark{z121e}&& z_{1,2},  z_{2,2} ,  z_{3,2} ,  z_{4,2}  \tikzmark{z112e}  && z_{1,3},  z_{2,3} ,  z_{3,3} ,  z_{4,3} \tikzmark{z113}&
		\\ \cline{3-9}
		&  &&  y_{9,1},  y_{10,1} ,  y_{11,1} ,  y_{12,1}   \tikzmark{y131e} && \tikzmark{y132b} \, y_{9,2},  y_{10,2} ,  y_{11,2} ,  y_{12,2} \, \tikzmark{y132e} && \tikzmark{y123b} y_{5,3},  y_{6,3} ,  y_{7,3} ,  y_{8,3} \tikzmark{y123e}&
		\\
		& &&z_{9,1},  z_{10,1} ,  z_{11,1} ,  z_{12,1} \tikzmark{z131e}  && \tikzmark{z132b} z_{9,2},  z_{10,2} ,  z_{11,2} ,  z_{12,2} \, \tikzmark{z132e}   &&  z_{5,3},  z_{6,3} ,  z_{7,3} ,  z_{8,3} \tikzmark{z123e} &	
		\\*\hline
			& \multirow{8}{*}{$Q^{(1)}_j(\set{D};2)$}
		&& x_{14,1}+y_{ 1,1} && x_{13,2} + y_{5,2}  && x_{13,3}+ y_{9,3} &
		\\
		& && x_{17,1}+z_{ 1,1} && x_{16,2} + z_{5,2} && x_{16,3} +z_{9,3}    &	
		\\
		& && x_{20,1}+y_{ 3,1} && x_{19,2} + y_{7,2} &&  x_{19,3} +y_{11,3}  &	
		\\
		& && x_{23,1}+z_{ 3,1} && x_{22,2} +  z_{7,2} && x_{22,3} +z_{11,3}  &	
		\\\cline{3-9}
		& &&     x_{15,1}+y_{ 2,1}   &&  x_{15,2}+y_{6,2} && x_{14,3}+y_{10,3} &
		\\
		& && x_{18,1}+z_{ 2,1} && x_{18,2}+z_{ 6,2} && x_{17,3}+z_{10,3} &	
		\\
		& && x_{21,1}+y_{ 4,1} && x_{21,2}+y_{ 8,2}  && x_{20,3} +y_{12,3} &	
		\\
		& &&  x_{24,1}+z_{ 4,1}&&x_{24,2}+z_{8 ,2}   &&  x_{23,3}+z_{12,3} &	
		\\*\hline
		& \multirow{4}{*}{$Q^{(1)}_j(\set{U};2)$}
		&& y_{17,1}+z_{ 14,1} && y_{16,2} + z_{13,2} && y_{16,3} +z_{13,3}    &	
		\\
		& && y_{23,1}+z_{ 20,1} && y_{22,2} +  z_{19,2} && y_{22,3} +z_{19,3}  &	
		\\\cline{3-9}
		& && y_{18,1}+z_{ 15,1} && y_{18,2}+z_{ 15,2} && y_{17,3}+z_{14,3} &	
		\\	
		& &&  y_{24,1}+z_{ 21,1}&&y_{24,2}+z_{21 ,2}   &&  y_{23,3}+z_{20,3} &
		\\*\hline	
		& \multirow{2}{*}{$Q^{(1)}_j(\set{D};3)$}
		 && x_{26 ,1}+y_{16,1}+z_{ 13,1} &&x_{25,2} +y_{17,2} + z_{14,2} &&x_{ 25,3}+ y_{18,3} +z_{15,3}    &		
		\\\cline{3-9}
		& && x_{ 27,1}+y_{22,1}+z_{ 19,1} && x_{27,2} + y_{23,2}+z_{ 20,2} && x_{ 26,3}+y_{24,3}+z_{21,3} &		
		\\*\IEEEeqnarrayrulerow
\end{IEEEeqnarraybox}
	%
	%
\end{table}
		$u=a_{i,j}$, $i\in[\kappa]$, to form the final query set for each database.
		For example, the query sets for the first database are formed by aggregating the auxiliary query sets $Q_1^{(1)}(2,\set{D};\tau) \cup Q_1^{(1)}(3,\set{D};\tau)$ and $Q_1^{(1)}(2,\set{U};\tau) \cup Q_1^{(1)}(3,\set{U};\tau)$ for all $\tau\in[3]$.   
		Table~\ref{tab:axQ-tableA4} gives the final query sets for Example \ref{ex:RunningMDSCapCodeex_n3k2}. 
		\QEDA
\end{example}

{\bf \textit{7) Privacy}:} 
It is worth mentioning that the queries generated by Algorithm~\ref{alg:generation_QuerySet} inherently satisfy the privacy condition of \eqref{eq:privacy}, which is guaranteed by satisfying the index, message, and database symmetry principles as for all the PIR schemes in \cite{SunJafar17_1,BanawanUlukus18_1,KumarLinRosnesGraellAmat19_1}. We also would like to emphasize that the achievable rates of our proposed PC schemes can be further improved by removing the redundant queries caused from the dependency among the virtual messages. Note that this will not break the privacy condition and will be discussed together with recovery in the following sections for each of the proposed PC schemes.
 
As mentioned at the beginning of this section, the presented generic query generation is, so far, a PIR-like scheme from a linearly-coded DSS with dependent virtual messages representing the evaluations of the candidate functions. 
In contrast to  simple PIR solutions, in %
PC we have the opportunity to exploit the dependencies induced by performing computations over the same set of messages, i.e., the $f$ independent messages $\vmat{W}^{(1)},\ldots,\vmat{W}^{(f)}$, while keeping the requested index $v$ private from each database. As shown in the recent PC literature (e.g., \cite{SunJafar19_2,MirmohseniMaddahAli18_1,ObeadKliewer18_1}), one is able to exploit this dependency to optimize the download cost by trading communication overhead with offline computation performed at the user side. In the following, we exploit the redundancy among the virtual messages ${\vmat{X}^{(v)}}, v\in[\mu]$, to enhance the achievable rate and accordingly tailor the \texttt{Q-Gen} algorithm to the case of PLC in Section~\ref{sec:private-linear-computation_coded-DSSs}, and to the general case of PPC in Sections~\ref{sec:achievable-scheme_PPC} and \ref{sec:achievable-scheme_PPC_systematic}, respectively.

\section {Private Linear Computation From Coded DSSs} 
\label{sec:private-linear-computation_coded-DSSs}

One of the main results of this paper is the derivation of the PLC capacity for a coded DSS where data is encoded and stored using a linear code from the class of MDS-PIR capacity-achieving codes \cite{KumarLinRosnesGraellAmat19_1}. The problem of PLC translates, in the PPC setup, to restricting the candidate function set to polynomials of degree $g=1$. Based on the PLC converse bound of \cref{thm:PLCconverse_MDS-PIRcapacity-achieving-codes}, in this section we represent PLC as a special case of PPC and construct a capacity-achieving scheme using the generic query generation algorithm of \cref{sec:generic-query-generation_PC_coded-DSSs}. In the following \cref{thm:PLCcapacity_MDS-PIRcapacity-achieving-codes}, we settle the PLC capacity for a DSS where data is stored using an MDS-PIR capacity-achieving code.

	\begin{theorem}
		\label{thm:PLCcapacity_MDS-PIRcapacity-achieving-codes}
		Consider a DSS with $n$ noncolluding databases that uses an $[n,k]$ MDS-PIR capacity-achieving code $\code{C}$ to store $f$ messages. Then, the maximum achievable PLC rate over all possible PLC protocols, i.e., the PLC capacity $\const{C}_{\textnormal{PLC}}$, is 
		\begin{IEEEeqnarray*}{c}
		\const{C}_{\textnormal{PLC}}\eqdef\Bigl(1-\frac{k}{n}\Bigr)\inv{\left[1-\Bigl(\frac{k}{n}\Bigr)^r\right]},
		\end{IEEEeqnarray*}
		where $r$ is the rank of the linear mapping from \eqref{eq:linear-mappingV}.
\end{theorem}
We remark  that since all MDS codes are MDS-PIR capacity-achieving codes, it follows that if $\rank{\mat{V}}=f$, then the PLC capacity for an MDS-coded DSS is equal to the MDS-PIR capacity $\const{C}_{\textnormal{MDS-PIR}}$ \cite{ObeadKliewer18_1}.

In PLC, a user wishes to privately compute exactly one \emph{linear} function evaluation from the $\mu$ candidate linear functions evaluations $\vmat{X}^{(1)},\ldots,\vmat{X}^{(\mu)}$ from the coded DSS.  With $\vmat{X}^{(v)}=\bigl(X^{(v)}_1,\ldots,X^{(v)}_\const{L}\bigr)$, the $\mu$-tuple $\trans{\bigl(X^{(1)}_l,\ldots,X^{(\mu)}_l\bigr)}$, $\forall\,l\in [\const{L}]$, is mapped by \eqref{eq:linear-mappingV}. Hence, the user privately generates an index $v\in[\mu]$ and wishes to compute the $v$-th linear function while keeping the index $v$ private from each database. The capacity-achieving PLC scheme is provided in the following subsections.
%

\subsection{Query Generation for PLC}
\label{sec:query-generation_coded-PLC}

We use the generic query generation algorithm of Section~\ref{sec:generic-query-generation_PC_coded-DSSs} (see \cref{alg:generation_QuerySet}). 
Given that the messages are stored using an $[n,k]$ MDS-PIR capacity-achieving code $\code{C}$, we can construct a $\nu\times n$ MDS-PIR capacity-achieving matrix $\mat{\Lambda}^{\textnormal{PIR}}_{\kappa,\nu}$. This matrix is used as the generic PC achievable rate matrix $\mat{\Lambda}^{\textnormal{PC}}_{\kappa,\nu}$ for PLC, and we can obtain the PC interference matrices $\mat{A}_{\kappa\times n}$ and $\mat{B}_{(\nu-\kappa)\times n}$ as defined in Section~\ref{sec:pc-achievable-rate} (see \cref{def:PCinterference-matrices}). In other words, for PLC we impose the additional condition that for each row $\vect{\lambda}_i$ of the generic PC achievable rate matrix the support $\chi(\vect{\lambda}_i)$ contains an information set. This condition is required for the recovery of the desired function evaluation. As a result, we require the size of the messages to be $\const{L}= \nu^{\mu}\cdot k $ (i.e., ${\beta=\nu^\mu}$). %
In PLC, \eqref{eq:IndexPrep} can simply be written as 
\begin{IEEEeqnarray}{c}
	\label{eq:IndexPrepPLC}
U^{(v')}_{t,j}\triangleq \vect{v}_{v'}{\vect{C}}_{\pi(t),j}, \,t\in[\beta], \, j\in[n], \, v'\in[\mu], 
\end{IEEEeqnarray}
where $\vect{v}_{v'}$ represents the $v'$-th row vector of the matrix $\mat{V}_{\mu\times f}=({v}_{i,j})$. For the desired linear function indexed with $v\in[\mu]$, the queries $Q^{(v)}_j$ are generated by invoking \cref{alg:generation_QuerySet} from \cref{sec:query-generation} as follows: 
\begin{IEEEeqnarray*}{c}
  \{Q^{(v)}_1,\dots,Q^{(v)}_n\} \leftarrow \texttt{Q-Gen}(v, \mu, \kappa,\nu,n, \mat{A}_{\kappa\times n}, \mat{B}_{(\nu-\kappa)\times n}).
\end{IEEEeqnarray*}
%
%
To illustrate the key concepts of the coded PLC scheme, we use the following Example~\ref{ex:PLCex_n4k2f4mu4} as a running example for this section. 
\begin{example}
  \label{ex:PLCex_n4k2f4mu4}
  Consider four messages $\vmat{W}^{(1)}$, $\vmat{W}^{(2)}$, $\vmat{W}^{(3)}$, and $\vmat{W}^{(4)}$ that are stored in a DSS using the $[4,2]$ MDS-PIR capacity-achieving code $\code{C}$ given in Example~\ref{ex:MDSCapCodeex_n4k2}
for which  
 \begin{IEEEeqnarray*}{c} 
    \mat{\Lambda}^{\textnormal{PIR}}_{1,2}=
    \begin{pmatrix}
      1 & 0 &1 & 0
      \\
      0 & 1 &0 & 1
    \end{pmatrix}, \quad 
\mat{A}_{1\times 4}=\begin{pmatrix} 1 & 2 &1 & 2 \end{pmatrix}, \quad 
\mat{B}_{1\times 4} =\begin{pmatrix} 2 & 1 &2 & 1 \end{pmatrix}
  \end{IEEEeqnarray*}  
  are a generic PC achievable rate matrix and the corresponding PC interference matrices $\mat{A}_{1\times 4}$ and $\mat{B}_{1\times 4}$, respectively. Suppose that the user wishes to obtain a linear function evaluation $\vmat{X}^{(v)}$ from a set of $\mu=4$ candidate linear functions evaluations, whose $\mat{V}_{\mu\times f}$ from \eqref{eq:linear-mappingV} is given by
  \begin{IEEEeqnarray*}{rCl}
    \mat{V}_{4\times 4} = \begin{pmatrix}
      1 & 0 &0 & 1 \\ 
      1 & 1 &0 & 0 \\
      2 & 1 &0 & 1 \\
      4 & 1 &0 & 3  
    \end{pmatrix}.
  \end{IEEEeqnarray*}
  We simplify notation by letting $x_{t,j}=U^{(1)}_{t,j}$, $y_{t,j}=U^{(2)}_{t,j}$, $z_{t,j}=U^{(3)}_{t,j}$, and  $w_{t,j}=U^{(4)}_{t,j}$ for all ${t\in[\beta]}$, $j\in[n]$, where $\beta=\nu^\mu=16.$  First, let the desired linear function index be $v=1$. For this example, the construction of the query sets is briefly presented in the following steps.\footnote{With some abuse of notation for the sake of simplicity, the generated queries are sets containing their answers.}
  
  {\bf \textit{Initialization (Round ${\tau=1}$)}:} Algorithm~\ref{alg:generation_QuerySet} starts with ${\tau=1}$ to generate auxiliary query sets for each database holding $\kappa^{\mu-1}=1$ distinct instances of $x_{t,j}$. By message symmetry this also applies to $y_{t,j}$, $z_{t,j}$, and $w_{t,j}$. The auxiliary query sets for the first round are shown in Table~\ref{tab:axQ-table}(a). Note that the queries for $z_{t,j}$ and $w_{t,j}$ can be generated offline by the user and thus are later removed from the query sets. Moreover, in Table~\ref{tab:axQ-table}(a), we highlight in red the row indicator $u\in[\nu]$ as specified by the interference matrix $\mat{A}_{1\times 4}$, i.e.,~$u=a_{1,j}$. Using this indicator, we determine the indices of the queried symbols as seen in the algorithm \texttt{Initial-Round}, i.e.,~Algorithm~\ref{alg:initial-round}.

    \begin{table}[htbp!]  
  \centering
  \caption{Auxiliary query sets for each round. Highlighted in red is the {\lin row} 
    indicator $u\in[\nu]$ used in determining the indices of the queried symbols. The magenta dashed arrows and the cyan arrows indicate that the \texttt{Exploit-SI} algorithm and the \texttt{M-Sym} algorithm are used, respectively.} 
  \label{tab:axQ-table}
  \vskip -1mm 
  \Resize[0.7\columnwidth]{ 
    \begin{IEEEeqnarraybox}[
      \IEEEeqnarraystrutmode
      \IEEEeqnarraystrutsizeadd{3pt}{2pt}]{v/c/v/c/v/c/v/c/v/c/v}
      \IEEEeqnarrayrulerow\\
      & j && 1  && 2 && 3 && 4\\
      \hline\hline
      & Q^{(1)}_j(a_{1,j},\set{D};1)
      && x_{({\r 1}-1)\cdot 1+1,1} &&  x_{({\r 2}-1)\cdot 1+1,2} && x_{({\r 1}-1)\cdot 1+1,3} && x_{({\r 2}-1)\cdot 1+1,4} &
      \\*\hline
      & \multirow{3}{*}{$Q^{(1)}_j(a_{1,j},\set{U};1)$}
      && y_{({\r 1}-1)\cdot 1+1,1}\tikzmark{y11} &&  y_{({\r 2}-1)\cdot 1+1,2} && y_{({\r 1}-1)\cdot 1+1,3}\tikzmark{y13}
      && y_{({\r 2}-1)\cdot 1+1,4} &
      \\
      & && z_{({\r 1}-1)\cdot 1+1,1}\tikzmark{z11} &&  z_{({\r 2}-1)\cdot 1+1,2} && z_{({\r 1}-1)\cdot 1+1,3}\tikzmark{z13}
      && z_{({\r 2}-1)\cdot 1+1,4} &
      \\
      & && w_{({\r 1}-1)\cdot 1+1,1}\tikzmark{w11} &&  w_{({\r 2}-1)\cdot 1+1,2} && w_{({\r 1}-1)\cdot 1+1,3}\tikzmark{w13}
      && w_{({\r 2}-1)\cdot 1+1,4} &
      \\*\IEEEeqnarrayrulerow
    \end{IEEEeqnarraybox}
}

  (a)
  \\[2mm]
  \Resize[0.8\columnwidth]{ 
    \begin{IEEEeqnarraybox}[
      \IEEEeqnarraystrutmode
      \IEEEeqnarraystrutsizeadd{3pt}{2pt}]{v/c/v/c/v/c/v/c/v/c/v}
      \IEEEeqnarrayrulerow\\
      & j && 1  && 2 && 3 && 4\\
      \hline\hline
      & \multirow{3}{*}{$Q^{(1)}_j(a_{1,j},\set{D};2)$}
      && x_{1\cdot 2+{\r 1},1}+y_{2,1} &&  x_{1\cdot 2+{\r 2},2}+y_{1,2}\tikzmark{y22} && x_{1\cdot 2+{\r 1},3}+y_{2,3}
      && x_{1\cdot 2+{\r 2},4}+y_{1,4}\tikzmark{y24} &
      \\
      & && x_{2\cdot 2+{\r 1},1}+z_{2,1} && x_{2\cdot 2+{\r 2},2}+z_{1,2}\tikzmark{z22} && x_{2\cdot 2+{\r 1},3}+z_{2,3}
      && x_{2\cdot 2+{\r 2},4}+z_{1,4}\tikzmark{z24} &
      \\
      & && x_{3\cdot 2+{\r 1},1}+w_{2,1} && x_{3\cdot 2+{\r 2},2}+w_{1,2}\tikzmark{w22} && x_{3\cdot 2+{\r 1},3}+w_{2,3}
      && x_{3\cdot 2+{\r 2},4}+w_{1,4}\tikzmark{w24} &
      \\*\hline
      & \multirow{3}{*}{$Q^{(1)}_j(a_{1,j},\set{U};2)$}
      && y_{4+{\r 1},1}+z_{2+{\r 1},1} && y_{4+{\r 2},2}+z_{2+{\r 2},2}\tikzmark{yz22}
      && y_{4+{\r 1},3}+z_{2+{\r 1},3} && y_{4+{\r 2},4}+z_{2+{\r 2},4}\tikzmark{yz24} &
      \\
      & && y_{6+{\r 1},1}+w_{2+{\r 1},1} && y_{6+{\r 2},2}+w_{2+{\r 2},2}\tikzmark{yw22}
      && y_{6+{\r 1},3}+w_{2+{\r 1},3} && y_{6+{\r 2},4}+w_{2+{\r 2},4}\tikzmark{yw24} &
      \\
      & && z_{6+{\r 1},1}+w_{4+{\r 1},1} && z_{6+{\r 2},2}+w_{4+{\r 2},2}\tikzmark{zw22}
      && z_{6+{\r 1},3}+w_{4+{\r 1},3} && z_{6+{\r 2},4}+w_{4+{\r 2},4}\tikzmark{zw24} &
      \\*\IEEEeqnarrayrulerow
    \end{IEEEeqnarraybox}
}

  (b)
  \\[2mm]
  \Resize[\columnwidth]{
    \begin{IEEEeqnarraybox}[
      \IEEEeqnarraystrutmode
      \IEEEeqnarraystrutsizeadd{3pt}{2pt}]{v/c/v/c/v/c/v/c/v/c/v}
      \IEEEeqnarrayrulerow\\
      & j && 1  && 2 && 3 && 4\\
      \hline\hline
      & \multirow{3}{*}{$Q^{(1)}_j(a_{1,j},\set{D};3)$} 
      && x_{4\cdot 2+{\r 1},1}+y_{6,1}+z_{4,1}\tikzmark{yz31} && x_{4\cdot 2+{\r 2},2}+y_{5,2}+z_{3,2}
      && x_{4\cdot 2+{\r 1},3}+y_{6,3}+z_{4,3}\tikzmark{yz33} && x_{4\cdot 2+{\r 2},4}+y_{5,4}+z_{3,4} &
      \\
      & && x_{5\cdot 2+{\r 1},1}+y_{8,1}+w_{4,1}\tikzmark{yw31} && x_{5\cdot 2+{\r 2},2}+y_{7,2}+w_{3,2}
      && x_{5\cdot 2+{\r 1},3}+y_{8,3}+w_{4,3}\tikzmark{yw33} && x_{5\cdot 2+{\r 2},4}+y_{7,4}+w_{3,4} &
       \\
      & && x_{6\cdot 2+{\r 1},1}+z_{8,1}+w_{6,1}\tikzmark{zw31} && x_{6\cdot 2+{\r 2},2}+z_{7,2}+w_{5,2}
      && x_{6\cdot 2+{\r 1},3}+z_{8,3}+w_{6,3}\tikzmark{zw33} && x_{6\cdot 2+{\r 2},4}+z_{7,4}+w_{5,4} &
      \\*\hline
      & Q^{(1)}_j(a_{1,j},\set{U};3)
      && y_{12+{\r 1},1}+z_{10+{\r 1},1}+w_{8+{\r 1},1}\tikzmark{yzw31} && y_{12+{\r 2},2}+z_{10+{\r 2},2}+w_{8+{\r 2},2}
      && y_{12+{\r 1},3}+z_{10+{\r 1},3}+w_{8+{\r 1},3}\tikzmark{yzw33} && y_{12+{\r 2},4}+z_{10+{\r 2},4}+w_{8+{\r 2},4} &
      \\*\IEEEeqnarrayrulerow
    \end{IEEEeqnarraybox}}
  (c)
  \\[2mm]
  \Resize[\columnwidth]{
  	\begin{IEEEeqnarraybox}[
  		\IEEEeqnarraystrutmode
  		\IEEEeqnarraystrutsizeadd{3pt}{2pt}]{v/c/v/c/v/c/v/c/v/c/v}
  		\IEEEeqnarrayrulerow\\
  		& j && 1  && 2 && 3 && 4\\
  		\hline\hline
  		& Q^{(1)}_j(a_{1,j},\set{D};4)
  		&& x_{7\cdot 2+{\r 1},1}+y_{14,1}+z_{12,1}+w_{10,1} && x_{7\cdot 2+{\r 2},2}+y_{13,2}+z_{11,2} +w_{9,2} \tikzmark{yzw42}
  		&& x_{7\cdot 2+{\r 1},3}+y_{14,3}+z_{12,3}+w_{10,3} && \tikzmark{yzw44} x_{7\cdot 2+{\r 2},4}+y_{13,4}+z_{11,4} +w_{9,4}  &
  		\\*\IEEEeqnarrayrulerow
  \end{IEEEeqnarraybox}}
  (d)
  \tikz[overlay,remember picture]{
    \draw[magenta,dashed,>=stealth,->,thick] ($(z11)+(-0.1,-0.80)$) to ($(y22)+(0.65,0.7)$);
    \draw[magenta,dashed,>=stealth,->,thick] ($(z13)+(0.75,-0.80)$) to ($(y22)+(0.65,0.7)$);
    \draw[magenta,dashed,>=stealth,->,thick] ($(z11)+(-0.1,-0.80)$) to ($(y24)+(0.95,0.7)$);
    \draw[magenta,dashed,>=stealth,->,thick] ($(z13)+(0.75,-0.80)$) to ($(y24)+(0.95,0.7)$);
    \path[cyan,>=stealth,->,thick] (w22)+(1.25,0.5) edge [bend left] ($(yz22)+(1.23,0.3)$);
    \path[cyan,>=stealth,->,thick] (w24)+(2.03,0.5) edge [bend left] ($(yz24)+(2.01,0.3)$);
    \draw[magenta,dashed,>=stealth,->,thick] ($(zw22)+(-0.3,-0.2)$) to ($(yz31)+(-0.6,0.2)$);
    \draw[magenta,dashed,>=stealth,->,thick] ($(zw22)+(-0.3,-0.2)$) to ($(yz33)+(-1.2,0.2)$);
    \draw[magenta,dashed,>=stealth,->,thick] ($(zw24)+(-0,-0.2)$) to ($(yz31)+(-0.6,0.2)$);
    \draw[magenta,dashed,>=stealth,->,thick] ($(zw24)+(-0,-0.2)$) to ($(yz33)+(-1.2,0.2)$);
    \path[cyan,>=stealth,->,thick] (zw31)+(-0.01,0.4) edge [bend left] ($(yzw31)+(-0.31,0.2)$);
    \path[cyan,>=stealth,->,thick] (zw33)+(-0.25,0.4) edge [bend left] ($(yzw33)+(-0.55,0.2)$);
    \draw[magenta,dashed,>=stealth,->,thick] ($(yzw31)+(-2.4,-0.2)$) to ($(yzw42)+(-3.3,0.25)$);
    \draw[magenta,dashed,>=stealth,->,thick] ($(yzw33)+(-3.2,-0.2)$) to ($(yzw42)+(-3.3,0.25)$);
    \draw[magenta,dashed,>=stealth,->,thick] ($(yzw31)+(-2.4,-0.2)$) to ($(yzw44)+(-1.35,0.25)$);
    \draw[magenta,dashed,>=stealth,->,thick] ($(yzw33)+(-3.2,-0.2)$) to ($(yzw44)+(-1.35,0.25)$);
  }
  \vskip -2mm 
\end{table}

  {\bf \textit{Following Rounds (${\tau>1}$)}:} As can be seen from Table~\ref{tab:axQ-table}(b)--(d), using the PC interference matrices $\mat{A}_{1\times 4}$ and $\mat{B}_{1\times 4}$, Algorithm~\ref{alg:generation_QuerySet} generates auxiliary query sets $Q^{(1)}_{j}(a_{1,j},\set{D};\tau)$ containing desired linear function evaluations to be decoded by exploiting side information. In particular, the algorithm generated $(\tau-1)$-sums of side information containing symbols from undesired linear functions evaluations based on $\mat{A}_{\kappa\times n}$ in the previous round. In the current round, it generates desired symbols as sums of a single symbol from the desired linear function evaluation and side information based on $\mat{B}_{(\nu-\kappa)\times n}$. Similar to Table~\ref{tab:axQ-table}(a), in Table~\ref{tab:axQ-table}(b)--(d), we highlight with red the row indicator $u=a_{1,j} \in[\nu]$. Here, this indicator is used in determining the indices of desired linear function evaluations in $Q^{(1)}_{j}(a_{1,j},\set{D};\tau)$ following the algorithm \texttt{Desired-Q}, i.e.,~\cref{alg:Desired-setQ_following-rounds}. In addition, we illustrate with magenta dashed arrows the side information exploitation following the algorithm \texttt{Exploit-SI}, i.e., \cref{alg:Exploit-SI}. Note that, by utilizing the code coordinates forming an information set in the code array, it can be shown that the side information based on $\mat{B}_{(\nu-\kappa)\times n}$ can be decoded. For example, in round $3$, since $\{2,4\}$ is an information set of the storage code $\code{C}$, the code symbols $y_{6,1}+z_{4,1}$ and $y_{6,3}+z_{4,3}$ can be obtained by knowing $y_{6,2}+z_{4,2}$ and $y_{6,4}+z_{4,4}$, from which the corresponding symbols $x_{6,1}$ and $x_{6,3}$ can be obtained by canceling the side information. Hence, the symbols from the desired linear function can be obtained. After generating the desired auxiliary query sets $Q^{(1)}_j(a_{1,j},\set{D};\tau)$, the undesired auxiliary query sets $Q^{(1)}_j(a_{1,j},\set{U};\tau)$ are generated by enforcing message symmetry. In Table~\ref{tab:axQ-table}(b)--(d), we indicate with cyan arrows the message symmetry enforcement procedure following the algorithm \texttt{M-Sym}, i.e.,~\cref{alg:message-symmetry}, and with red the resulting index symmetry in $Q^{(1)}_j(a_{1,j},\set{U};\tau)$ based on the desired linear function indices. \QEDA 
    \end{example}

 \subsection{Recovery of Desired Function Evaluation} 
Given that we construct the capacity-achieving PLC scheme using the generic query generation algorithm presented in Section~\ref{sec:generic-query-generation_PC_coded-DSSs}, so far the PLC scheme is a PIR-like scheme that privately retrieve a virtual message from a linearly-coded DSS. This virtual message represents the evaluation of the desired function over coded symbols, however, the user wishes to privately retrieve the evaluation of the desired function over the original information symbols. 
As a result, due to the fact that we are performing computation over coded storage, the coded PLC scheme includes two extra steps over other uncoded PC schemes. Namely, decoding the desired function evaluation symbols and decoding and canceling the side information. Thus, the correct decoding of the desired function evaluation relies on the correct decoding of the queried symbols from all virtual messages. To this end, in the following, we show that we can reliably recover the desired function evaluation from the queried symbols. \\ 
 The main argument behind the reliable recovery of the desired function evaluation is the fact that the candidate linear functions and linear coding commute, i.e., evaluating a function over coded symbols is equal to encoding the symbols of the function evaluation. To see that, let $\hat{t}=\pi(t)$ where $t, \hat{t} \in [\beta]$ be the private permutation selected by the user and let $\vect{g}_j= \trans{\bigl(g_{1,j}, g_{2,j},\dots, g_{k,j}\bigr)}$ be the $j$-th column of the generator matrix $\mat{G}^\code{C}$ for  the $[n,k]$ linear storage code.  One can verify, from \eqref{eq:IndexPrepPLC}, that for all $v'\in[\mu]$, we have
 	 \begin{IEEEeqnarray}{rCl}
 			U^{(v')}_{\hat{t},j}	& = & \vect{v}_{v'}{\vect{C}}_{\hat{t},j} 
                =\sum_{i=1}^{f}  v_{v',i} C_{\hat{t},j}^{(i)} \nonumber\\
 		& = &  \sum_{i=1}^{f}  v_{v',i} \sum_{h=1}^{k} W_{\hat{t},h}^{(i)} g_{h,j} 
                =\sum_{h=1}^{k}  g_{h,j}  \sum_{i=1}^{f}  v_{v',i} W_{\hat{t},h}^{(i)} \nonumber\\	
  	 & = &  \sum_{h=1}^{k} X_{\hat{t},h}^{(v')} g_{h,j}, 
 		\label{eq:PLCRecov}
 	\end{IEEEeqnarray}
 where $(X^{(v')}_{1,1},\dots,X^{(v')}_{1,k},X^{(v')}_{2,1},\dots, X^{(v')}_{\beta,k} )=(X_1^{(v')},\dots,X^{(v')}_{k},X^{(v')}_{k+1},\dots,X_{\const{L}}^{(v')})=\vmat{X}^{(v')}$. Note that \eqref{eq:PLCRecov} resembles the process of encoding the segment $\bigl(X^{(v')}_{\hat{t},1},\dots,X^{(v')}_{\hat{t},k}\bigr)$ of the candidate linear function evaluation $\vmat{X}^{(v')}$ using the $[n,k]$ storage code. Thus, one can consider the construction of our PLC scheme, so far, as a coded PIR scheme 
 over a virtual coded DSS storing the evaluations of the candidate functions. As a result, using the same $[n,k]$ linear code for decoding the symbols obtained from the answer sets guarantees the reliable retrieval of the desired function evaluation.

\subsection{Sign Assignment and Redundancy Elimination} 
\label{sec:SignAssignment-RedundancyElimination_PLC}
Our proposed PLC scheme is further constructed with two additional procedures: \emph{sign assignment} and \emph{redundancy elimination}. After running Algorithm~\ref{alg:generation_QuerySet}, the user will know which row indices of the stored code symbols he/she is going to request. To reduce the total number of downloaded symbols, the linear dependency among the candidate linear functions evaluations is exploited. To this end, an initial sign $\sigma^{(v)}_t$ is first privately generated by the user with a uniform distribution over $\{-1,+1\}$ for all $t\in [\beta]$, i.e., the same selected sign is identically applied to all symbols from different function evaluations with the same index. Next, depending on the desired linear function index $v\in[\mu]$, we  apply a deterministic sign assignment procedure that carefully scales each pre-signed symbol in the query sets, i.e.,~$\sigma^{(v)}_t U^{(v')}_{t,j}$,  $v'\in[\mu]$, by $\{+1,-1\}$. 
The intuition behind the sign assignment is to introduce a uniquely solvable equation system from the different $\tau$-sum types given the side information available from all other databases. By obtaining such a system of equations in each round, the user can determine some of the queries offline to decode the desired linear function evaluations and/or interference, thus reducing the download rate.
On the other hand, the privately selected initial sign for $\sigma^{(v)}_t, t\in[\beta]$, acts as a one-time pad that randomizes over the deterministic sign assignment procedure. 
Here, we adopt a similar sign assignment process over each symbol in the query sets, as introduced in \cite[Sec.~IV-B]{SunJafar19_2} and presented in Appendix~\ref{sec:sign_assignment}. 
Moreover, we remark that after sign assignment, the recovery condition of the scheme is inherently maintained since it can be seen as a coded PIR scheme as Protocol~1 in~\cite{KumarLinRosnesGraellAmat19_1}. 
  The key idea of redundancy elimination is illustrated with \cref{ex:PLCex_n4k2f4mu4} below.
  \begin{example} [continues=ex:PLCex_n4k2f4mu4]
    First, without loss of generality, we assume the initial sign assignment $\sigma^{(v)}_{t}=+1$ is privately selected by the user  for all $t\in{[\beta]}$. Next, we apply the sign assignment process to the query sets for $v=1$ (refer to Appendix~\ref{sec:sign_assignment} for more details). The resulting queries after sign assignment are shown in Table~\ref{tab:answers-table_Ex1}. 
    In the following, we show that we can remove some redundant queries from each database and the desired linear function evaluation $\vmat{X}^{(1)}$ can still be recovered.  For example, in the first round ($\tau=1$), it can be easily seen from $\mat{V}_{\mu\times f}$ that the queried symbols of $z_{t,j}$ and $w_{t,j}$ can be generated offline by the user as functions of $x_{t,j}$ and $y_{t,j}$, i.e., $z_{t,j}=x_{t,j}+y_{t,j}$ and $w_{t,j}=3x_{t,j}+y_{t,j}$ for all $t\in[\beta]$ and $j\in [n]$. Moreover, the coefficient vectors associated with $x_{t,j}$ and $y_{t,j}$ are the two row basis vectors of the coefficient matrix  $\mat{V}_{\mu\times f}$ ($r=\rank{\mat{V}}=2$).  Thus, we can represent the candidate functions evaluations in terms of this  basis  with a deterministic linear mapping $\mat{\hat{V}}_{\mu\times r}=(\hat{v}_{i,l})$ of size $\mu\times r$ as follows:
    \begin{IEEEeqnarray}{rCl}
      \label{eq:Alt-linear-mappingV}
      \begin{pmatrix}
        x_{t,j}\\
        y_{t,j}\\
        z_{t,j}\\
        w_{t,j}
      \end{pmatrix}=\underbrace{\begin{pmatrix}
  	1 & 0\\
  	0 & 1\\
  	1 & 1\\
  	3 & 1
  \end{pmatrix}}_{\mat{\hat{V}}_{\mu\times r}}
  	\begin{pmatrix}
  		x_{t,j}\\
  		y_{t,j}
  	\end{pmatrix}.
  \end{IEEEeqnarray}
  That is true due to the commutativity of the performed linear functions, i.e., the storage code and the candidate functions, and given that the coefficient matrix $\mat{V}_{\mu\times f}$ of the candidate functions  is available to the user.
  Thus, the queries for these symbols, i.e., $z_{t,j}$ and $w_{t,j}$, are redundant and can be removed from the query sets regardless of which function evaluation is desired by the user.   
  Next, in round $\tau=2$ and for the $1$st database, from the deterministic linear mapping $\mat{\hat{V}}_{\mu\times r}=(\hat{v}_{i,l})$ of \eqref{eq:Alt-linear-mappingV}, one can verify that
  \begin{IEEEeqnarray}{rCl}
    \IEEEeqnarraymulticol{3}{l}{%
      \hat{v}_{3,2}(y_{7,1}-w_{3,1})-\hat{v}_{4,2}(y_{5,1}-z_{3,1})-(\hat{v}_{3,1}\cdot \hat{v}_{4,2}-\hat{v}_{4,1}\cdot \hat{v}_{3,2})x_{3,1}-\hat{v}_{4,1}x_{5,1}+\hat{v}_{3,1}x_{7,1}}\nonumber\\*\quad%
    & = &
    1(y_{7,1}-w_{3,1})-1(y_{5,1}-z_{3,1})-(1\cdot 1-3\cdot 1)x_{3,1}-3x_{5,1}+1x_{7,1}  \nonumber\\
	& = &
    1(y_{7,1}-3x_{3,1}-1y_{3,1})-(y_{5,1}-x_{3,1}-y_{3,1})+2x_{3,1}-3x_{5,1}+x_{7,1}
    \nonumber\\
    & = &(x_{7,1}+y_{7,1})-(3x_{5,1}+y_{5,1})=z_{7,1}-w_{5,1},
    \label{eq:Eq_v=1_Ex1}
  \end{IEEEeqnarray}
  and hence we do not need to download the $2$-sum $z_{7,1}-w_{5,1}$. Similarly, we can do the same exercise for the other databases. The redundant queries are marked in blue in Table~\ref{tab:answers-table_Ex1} and the indices $t\in [\beta]$ of the desired linear function evaluations are marked in red. This completes the recovery part. The resulting PLC rate becomes $\frac{\nu^\mu\cdot k}{\const{D}}=\frac{16\cdot 2}{12\cdot 4}=\frac{2}{3}$, which is equal to the PLC capacity in \cref{thm:PLCcapacity_MDS-PIRcapacity-achieving-codes} with $r=\rank{\mat{V}}=2$. This demonstrates the optimality of the PLC scheme. \QEDA 
     \begin{table}[thbp!]
  \centering
  \caption{PLC query sets for $v=1$ after sign assignment for rounds one to four for the $[4,2]$ code of \cref{ex:PLCex_n4k2f4mu4},  $f=4$ messages, and $\mu=4$ candidate linear functions. Red subscripts indicate the indices of the desired linear function evaluations. The redundant queries are marked in blue.}
  \label{tab:answers-table_Ex1}
  \vskip -1mm
  \Resize[\columnwidth]{
    \begin{IEEEeqnarraybox}[
      \IEEEeqnarraystrutmode
      \IEEEeqnarraystrutsizeadd{3pt}{2pt}]{v/c/v/c/v/c/v/c/v/c/v}
      \IEEEeqnarrayrulerow\\
      & j && 1 && 2 && 3 && 4\\
      \hline\hline
      &  Q^{(v)}_j(\set{D};1)
      && x_{{\r 1},1} &&  x_{{\r 2},2} && x_{{\r 1},3} && x_{{\r 2},4} &
      \\*\cline{1-11}      
      & Q^{(v)}_j(\set{U};1)
      && y_{1,1}, {\b z_{1,1},w_{1,1}} &&  y_{2,2},  {\b z_{2,2},  w_{2,2}} && y_{1,3},  {\b z_{1,3},  w_{1,3}} && y_{2,4}, {\b z_{2,4} , w_{2,4}}  &
      \\*\cline{1-11}      
      & \multirow{3}{*}{$Q^{(v)}_j(\set{D};2)$}
      && x_{{\r 3},1}-y_{2,1}&&  x_{{\r 4},2}-y_{1,2} && x_{{\r 3},3}-y_{2,3} && x_{{\r 4},4}-y_{1,4} &
      \\ 
      & 
      && x_{{\r 5},1}-z_{2,1} &&  x_{{\r 6},2}-z_{1,2} && x_{{\r 5},3}-z_{2,3} && x_{{\r 6},4}-z_{1,4} &
      \\ 
      & 
      && x_{{\r 7},1}-w_{2,1}&&  x_{{\r 8},2}-w_{1,2} && x_{{\r 7},3}-w_{2,3} && x_{{\r 8},4}-w_{1,4} &
      \\*\cline{1-11}      
      & \multirow{3}{*}{$Q^{(v)}_j(\set{U};2)$}  
      && y_{5,1}-z_{3,1}&&  y_{6,2}-z_{4,2} && y_{5,3}-z_{3,3} && y_{6,4}-z_{4,4} &
      \\ 
      &  && y_{7,1}-w_{3,1} &&  y_{8,2}-w_{4,2} && y_{7,3}-w_{3,3} && y_{8,4}-w_{4,4} &
      \\ 
      &  && {\b z_{7,1}-w_{5,1}} &&  {\b z_{8,2}-w_{6,2}} && {\b z_{7,3}-w_{5,3}} && {\b z_{8,4}-w_{6,4}} &
      \\*\cline{1-11}      
      & \multirow{3}{*}{$Q^{(v)}_j(\set{D};3)$}
      && x_{{\r 9},1}-y_{6,1}+z_{4,1}  && x_{{\r 10},2}-y_{5,2}+z_{3,2}
      && x_{{\r 9},3}-y_{6,3}+z_{4,3} && x_{{\r 10},4}-y_{5,4}+z_{3,4} &
      \\ 
      & && x_{{\r 11},1}-y_{8,1}+w_{4,1} && x_{{\r 12},2}-y_{7,2}+w_{3,2}
      && x_{{\r 11},3}-y_{8,3}+w_{4,3} && x_{{\r 12},4}-y_{7,4}+w_{3,4} &
      \\ 
      & && x_{{\r 13},1}-z_{8,1}+w_{6,1}  && x_{{\r 14},2}-z_{7,2}+w_{5,2}
      && x_{{\r 13},3}-z_{8,3}+w_{6,3} && x_{{\r 14},4}-z_{7,4}+w_{5,4} &
      \\*\cline{1-11}      
      & Q^{(v)}_j(\set{U};3)
      && y_{13,1}-z_{11,1}+w_{9,1}   && y_{14,2}-z_{12,2}+w_{10,2}
      && y_{13,3}-z_{11,3}+w_{9,3} && y_{14,4}-z_{12,4}+w_{10,4} &   
       \\*\cline{1-11}      
      & Q^{(v)}_j(\set{D};4)
      && x_{{\r 15},1}-y_{14,1}+z_{12,1}-w_{10,1} 
      && x_{{\r 16},2}-y_{13,2}+z_{11,2} -w_{9,2} && x_{{\r 15},3}-y_{14,3}+z_{12,3}-w_{10,3}
      && x_{{\r 16},4}-y_{13,4}+z_{11,4} -w_{9,4}&   
      \\*\IEEEeqnarrayrulerow
    \end{IEEEeqnarraybox}}
\end{table}

\end{example}
From the above example we note the following.
\begin{itemize}
	\item There is a deterministic linear mapping, i.e., $\mat{\hat{V}}_{\mu\times r}$, that captures the dependencies among the candidate linear functions evaluations. 
	\item We maintain the same characteristics of the query construction that facilitate the exploitation of the linear dependencies among the candidate functions evaluations as for the uncoded PLC scheme in \cite{SunJafar19_2}. These characteristics include index assignment, sign assignment, and lexicographic ordering of the elements of $\tau$-sums. As a result, some of the queries become redundant and can be removed from the query sets while maintaining the decodability of the desired function evaluation. 
	\item  The candidate functions are computed over the coded symbols stored in each database individually. Consequently, from the perspective of the queries of each database, the linear dependency among the symbols of the candidate functions evaluations is present, i.e., the fact that the computation is performed over coded storage is transparent to the redundancy elimination process. This can be seen from \eqref{eq:Eq_v=1_Ex1}. 
	\item The number of redundant queries depends on the rank of the coefficient matrix $\mat{V}_{\mu\times f}$, i.e., $r=\rank{\mat{V}}$. This can be clearly observed for the $1$-sum symbols where out of the $\mu$ symbols, $\mu-r$ can be computed offline given the symbols of the functions evaluations associated with the $r$ row basis vectors of $\mat{V}_{\mu\times f}$ are available.   
\end{itemize}
Based on this insight we can state the following lemma for redundancy elimination.
\begin{lemma}
	\label{lem:redundancy}
	For all {$v\in[\mu]$}, each database $j\in[n]$, and based on the side information available from the databases, any $\mu-r\choose \tau$ $\tau$-sum types out of all possible $\mu\choose \tau$ types in each round {${\tau\in[\mu-r]}$} of the query sets are redundant.
\end{lemma}
The proof of Lemma~\ref{lem:redundancy} is presented in Appendix~\ref{sec:proof_PLCredundancy}. The proof  is based on the insight that the redundancy resulting from the linear dependencies between virtual messages is also present with MDS-PIR capacity-achieving codes.  
Since both repetition and MDS codes are MDS-PIR capacity-achieving codes, Lemma~\ref{lem:redundancy} generalizes both \cite[Lem.~1]{SunJafar19_2} and \cite[Lem.~1]{ObeadKliewer18_1}. 
We now make the final modification to our PLC query sets by first directly applying the sign assignment over $\sigma^{(v)}_t U^{(v')}_{t,j}$,  $v'\in[\mu]$, and then remove the $\tau$-sums corresponding to the redundant $\tau$-sum types from every round  $\tau\in{[\mu-r]}$. Note that the amount of redundancy is dependent on the rank of the functions matrix, $\rank{\mat{V}}=r\leq \min\{\mu,f\}$, thus generalizing the MDS-coded PLC case. Finally, we generate the queries $Q^{(v)}_{[n]}$. 

\subsection{Privacy} 
 As mentioned in Section~\ref{sec:query-generation}, the queries generated by Algorithm~\ref{alg:generation_QuerySet} inherently satisfy the privacy condition of \eqref{eq:privacy}, which is guaranteed by satisfying the index, message, and database symmetry principles as for all the PIR schemes in \cite{SunJafar17_1,BanawanUlukus18_1,KumarLinRosnesGraellAmat19_1}. That is, given the fixed and symmetric construction of the queries, there always exists a one-to-one mapping between the queries, $Q^{(v)}_j\leftrightarrow Q^{(v')}_j, \forall\,j\in[n]$, in terms of the queried symbols indices $t\in[\beta]$, where  $v,v' \in[\mu]$ and $v\neq v'$. Given this one-to-one mapping along with a permutation $\pi(t)$ over these indices  privately selected uniformly at random by the user, the queries are indistinguishable and equally likely. 
 Moreover, after the sign assignment process a one-to-one mapping between the assigned signs is found following a simple sign flipping rule for $\sigma_t$. The rule states that, to map the queries of $Q^{(v')}_j$ to $Q^{(v)}_j$, one should only consider the desired queries, i.e., queries that contain symbols associated with $\vmat{X}^{(v')}$. For such queries in each round $\tau$, we replace $\sigma_*$ with $-\sigma_*$ for each element to the right of the desired function evaluation symbol $U^{(v')}_*$ in the lexicographically ordered query if the query is sorted in a subgroup indexed with an odd $S$ (see Appendix~\ref{sec:sign_assignment}). Next, we flip the sign of elements to the left of the desired function evaluation symbol $U^{(v')}_*$ if the query is sorted in a subgroup indexed with an even $S$.
 The proof of the correctness of this rule and thus the privacy after sign assignment follows directly from \cite[Sec.~VI-B]{SunJafar19_2}.
 For completeness, we also show with \cref{ex:PLCex_n4k2f4mu4} that the user's privacy is still maintained after the sign assignment process and the removal of redundant queries. 
  \begin{example} [continues=ex:PLCex_n4k2f4mu4]
Here, to show that the queries are identically distributed regardless of the desired function evaluation index $v\in[4]$ we show that there exists a one-to-one mapping from the queries for $v=1$ to the queries for $v=3$ for all databases. Without loss of generality, we again assume the initial sign assignment $\sigma^{(3)}_{t}=+1$ privately selected by the user for all $t\in{[\beta]}$. In Table~\ref{tab:answers-table_Ex1_3}, the queries for $v=3$ are presented following \cref{alg:generation_QuerySet} and the sign assignment process of Appendix~\ref{sec:sign_assignment}. From Tables~\ref{tab:answers-table_Ex1} and~\ref{tab:answers-table_Ex1_3} one can verify that the index and sign mapping
\begin{IEEEeqnarray}{rCl} \label{eq:privacy_1to1map}
 \text{Databases 1 and 3:}  && \, (3,2,5,9,6,4,11,8,13,\sigma_{13},15,14,12, \sigma_{10}) \nonumber \\
 \xrightarrow[]{v=3}&& \, (5,3,2,6,4,9,13,11,8,-\sigma_8,14,12,15, -\sigma_{10}) \IEEEyesnumber\IEEEyessubnumber\IEEEeqnarraynumspace \label{eq:privacy_map_1n3}  \\
  \text{Databases 2 and 4:}&& \,  (4,1,6,10,5,3,12,7,14,\sigma_{14},16,13,11,\sigma_9) \nonumber\\
   \xrightarrow[]{v=3} && \, (6,4,1,5,3,10,14,12,7,-\sigma_7,13,11,16,-\sigma_9)  \IEEEyessubnumber \label{eq:privacy_map_2n4} 
\end{IEEEeqnarray}
converts the queries for $v=1$ to the queries for $v=3$. To see this mapping, compare the $\tau$-sums $x_{t_1,1}-y_{t_2,1}$ and $x_{t'_1,1}-y_{t'_2,1}$ from the queries of the first database of Tables~\ref{tab:answers-table_Ex1} and~\ref{tab:answers-table_Ex1_3}, respectively. It can be seen that the indices $t_1=3$ and $t_2=2$ of the queries for $v=1$ convert into the indices $t'_1=5$ and $t'_2=3$ of the queries for $v=3$, respectively. Thus, we have the mapping $(t_1,t_2)\rightarrow(t'_1,t'_2)=(3,2)\rightarrow(5,3)$ and due to the index symmetry of the query construction this mapping is fixed for all symbols with the corresponding indices. A similar comparison between the remaining $\tau$-sums results in the index and sign mapping of \eqref{eq:privacy_map_1n3} and \eqref{eq:privacy_map_2n4}. 
One can similarly verify that there exists a mapping from the queries for $v=1$ to the queries for $v=2$ or that for $v=4$, i.e., $Q^{(1)}_{[n]} \leftrightarrow Q^{(2)}_{[n]}$ and $Q^{(1)}_{[n]} \leftrightarrow Q^{(4)}_{[n]}$. Since a  permutation over these indices, i.e., $\pi(t)$ and an initial sign $\sigma^{(v)}_t$ are uniformly and privately selected by the user independently of the desired function evaluation index $v$, these queries are equally likely and indistinguishable. 
Next, to verify the correctness of the sign flipping rule stated above, consider the desired queries of the third round ($\tau=3$) for the query sets for $v=3$ in \cref{tab:answers-table_Ex1_3}. For database $1$, one can verify that the query $x_{6,1}-y_{4,1}+z_{9,1}$ is sorted in the subgroup indexed by $S=1$. As $S$ is odd and no element is placed to the right of $z_
{9,1}$ the signs are left unchanged. However, for the query $-x_{8,1}-z_{11,1}+w_{4,1}$ which falls in the subgroup indexed by $S=2$, the sign of the element to the left of $z_{11,1
}$, i.e., $x_{8,1}$, is flipped. That is, we change $\sigma_8$ to $-\sigma_8$ and that matches the sign mapping in \eqref{eq:privacy_map_1n3} for this query. Moreover, due to index symmetry, this mapping also matches the sign assignment for $\sigma_8$ for the query $-y_{8,1}-z_{13,1}+w_{6,1}$.
Finally, for redundancy elimination,  we only need to show that for any desired index $v\in[4]$, the removed redundant $\tau$-sums can be chosen to be of the same type. For instance, let us consider the $1$st database. In the $2$nd round, see Table~\ref{tab:answers-table_Ex1_3}, it can be shown that the queries for desired index $v=3$ satisfy the equation
\begin{IEEEeqnarray*}{rCl}
	\IEEEeqnarraymulticol{3}{l}{%
		(1\cdot 1-3\cdot 1)(x_{5,1}-y_{3,1})-1(x_{7,1}-w_{3,1})-3z_{3,1}-1z_{5,1}+1z_{7,1}}\nonumber\\*\quad%
	& = &
	-2(x_{5,1}-y_{3,1})-(x_{7,1}-3x_{3,1}-y_{3,1})-3(x_{3,1}+y_{3,1})-(x_{5,1}+y_{5,1})+(x_{7,1}+y_{7,1})
	\nonumber\\
	& = &1(y_{7,1}-(3x_{5,1}+y_{5,1}))=y_{7,1}-w_{5,1},
	\label{eq:Eq_v=3_Ex1}
\end{IEEEeqnarray*}
which implies that the $2$-sum $z_{7,1}-w_{2,1}$ can be removed from the download, since $z_{7,1}$ can be obtained from downloading $x_{5,1}-y_{3,1}$, $x_{7,1}-w_{3,1}$, $x_{2,1}-z_{3,1}$, $y_{2,1}-z_{5,1}$, and $y_{7,1}-w_{5,1}$. Hence, the redundant $\tau$-sum type for $v=3$ can be chosen to be equal to the redundant $\tau$-sum type for $v=1$ (see \eqref{eq:Eq_v=1_Ex1}). A similar argument can be made for $v=2$ and $v=4$, which ensures that the privacy of the scheme is not effected by redundancy elimination.
\begin{table}[thbp!]
  \centering
  \caption{PLC query sets for $v=3$ after sign assignment for rounds one to four for the $[4,2]$ code of \cref{ex:PLCex_n4k2f4mu4},  $f=4$ messages, and $\mu=4$ candidate linear functions. Red subscripts indicate the indices of the desired linear function evaluations. The redundant queries are marked in blue. }
  \label{tab:answers-table_Ex1_3}
  \vskip -1mm
  \Resize[\columnwidth]{
    \begin{IEEEeqnarraybox}[
      \IEEEeqnarraystrutmode
      \IEEEeqnarraystrutsizeadd{3pt}{2pt}]{v/c/v/c/v/c/v/c/v/c/v}
      \IEEEeqnarrayrulerow\\
      & j && 1 && 2 && 3 && 4\\
      \hline\hline 
      &  Q^{(v)}_j(\set{D};1)
      && {\b z_{{\r 1},1}}&&  {\b z_{{\r 2},2}} && {\b z_{{\r 1},3}} && {\b z_{{\r 2},4}} &
      \\*\cline{1-11}      
      & Q^{(v)}_j(\set{U};1)
      && x_{1,1}, y_{1,1}, {\b w_{1,1}}  &&  x_{2,2},  y_{2,2}, {\b w_{2,2}} && x_{1,3},  y_{1,3}, {\b w_{1,3}} && x_{2,4}, y_{2,4} ,{\b w_{2,4}}  &
      \\*\cline{1-11}      
      & \multirow{3}{*}{$Q^{(v)}_j(\set{D};2)$} 
      && x_{2,1}-z_{{\r 3},1} &&  x_{1,2}-z_{{\r 4},2} && x_{2,3}-z_{{\r 3},3} && x_{1,4}-z_{{\r 4},4} &
      \\ 
      & 
      && y_{2,1}-z_{{\r 5},1} &&  y_{1,2}-z_{{\r 6},2} && y_{2,3}-z_{{\r 5},3} && y_{1,4}-z_{{\r 6},4} &
      \\ 
      & 
      && {\b z_{{\r 7},1}-w_{2,1}} &&  {\b z_{{\r 8},2}-w_{1,2}} && {\b z_{{\r 7},3}-w_{2,3}} && {\b z_{{\r 8},4}-w_{1,4}} &
      \\*\cline{1-11}      
      & \multirow{3}{*}{$Q^{(v)}_j(\set{U};2)$}   
      &&  x_{5,1}-y_{3,1} &&  x_{6,2}-y_{4,2} && x_{5,3}-y_{3,3} && x_{6,4}-y_{4,4} &
      \\ 
      &  &&  x_{7,1}-w_{3,1} &&  x_{8,2}-w_{4,2} && x_{7,3}-w_{3,3} && x_{8,4}-w_{4,4} &
      \\ 
      &  && y_{7,1}-w_{5,1} &&  y_{8,2}-w_{6,2} && y_{7,3}-w_{5,3} &&  y_{8,4}-w_{6,4} &
      \\*\cline{1-11}      
      & \multirow{3}{*}{$Q^{(v)}_j(\set{D};3)$} 
      && x_{6,1}-y_{4,1}+z_{{\r 9},1} && x_{5,2}-y_{3,2}+z_{{\r 10},2}
      && x_{6,3}-y_{4,3}+z_{{\r 9},3} && x_{5,4}-y_{3,4}+z_{{\r10},4} &
      \\ 
      & && -x_{8,1}-z_{{\r 11},1}+w_{4,1} && -x_{7,2}-z_{{\r 12},2}+w_{3,2}
      && -x_{8,3}-z_{{\r 11},3}+w_{4,3} && -x_{7,4}-z_{{\r 12},4}+w_{3,4} &
      \\ 
      & && -y_{8,1}-z_{{\r 13},1}+w_{6,1}  && -y_{7,2}-z_{{\r 14},2}+w_{5,2}
      && -y_{8,3}-z_{{\r 13},3}+w_{6,3} &&- y_{7,4}-z_{{\r 14},4}+w_{5,4} &
      \\*\cline{1-11}      
      & Q^{(v)}_j(\set{U};3) 
      && x_{13,1}-y_{11,1}+w_{9,1}  && x_{14,2}-y_{12,2}+w_{10,2}
      && x_{13,3}-y_{11,3}+w_{9,3} && x_{14,4}-y_{12,4}+w_{10,4} &   
       \\*\cline{1-11}      
      & Q^{(v)}_j(\set{D};4) 
      && x_{14,1}-y_{12,1}+z_{{\r 15},1}+w_{10,1}
      && x_{13,2}-y_{11,2}+z_{{\r 16},2} +w_{9,2} && x_{14,3}-y_{12,3}+z_{{\r 15},3}+w_{10,3}
      && x_{13,4}-y_{11,4}+z_{{\r 16},4} +w_{9,4}&   
      \\*\IEEEeqnarrayrulerow
    \end{IEEEeqnarraybox}}
\end{table}
 \QEDA
\end{example}

\subsection{Achievable PLC Rate} 
\label{sec:achievable-rate_PLC}

The resulting achievable PLC rate of Algorithm~\ref{alg:generation_QuerySet} after removing redundant $\tau$-sums according to Lemma~\ref{lem:redundancy} becomes
\begin{IEEEeqnarray}{rCl}
  \label{eq:Cap_proof}
  \const{R} & \overset{(a)}{=} &\frac{k \nu^\mu}{n\sum_{\tau=1}^{\mu}
    \Bigl({\mu\choose\tau}-{\mu-r\choose\tau}\Bigr)\kappa^{\mu-(\tau-1)}(\nu-\kappa)^{\tau-1}} \nonumber
  \\
  & \overset{(b)}{=} &\frac{\kappa\nu^\mu}{\nu\sum_{\tau=1}^{\mu}
    \Bigl({\mu\choose\tau}-{\mu-r\choose\tau}\Bigr)\kappa^{\mu-(\tau-1)}(\nu-\kappa)^{\tau-1}} \nonumber
  \\
  & = &\frac{\nu^\mu\bigl(\frac{\nu-\kappa}{\nu}\bigr)}{\sum_{\tau=1}^{\mu}
    \Bigl({\mu\choose\tau}-{\mu-r\choose\tau}\Bigr)\kappa^{\mu-\tau}(\nu-\kappa)^\tau} \nonumber
  \\
  &  & \qquad\quad \vdots \nonumber
  \\[1.5mm]
  & \overset{(c)}{=}  &\frac{\nu^\mu\big(1-\frac{\kappa}{\nu}\big)}{\nu^\mu\!-\kappa^r \nu^{\mu-r}} \nonumber
  \\* [1mm]
  & = &\Bigl(1-\frac{\kappa}{\nu}\Bigr)\inv{\left[1-\Bigl(\frac{\kappa}{\nu}\Bigr)^r\right]},  
\end{IEEEeqnarray}
where we recall that $\binom{m}{n} =  0$ if $m<n$; $(a)$ follows from the PLC rate in Definition~\ref{def:def_info-PCrate}, \eqref{eq:query-card_PC}, and \cref{lem:redundancy}; $(b)$ follows from Definition~\ref{def:MDS-PIRcapacity-achieving-codes}; and $(c)$ follows by adapting similar steps as in the proof given in \cite{ObeadKliewer18_1} (see also the proof of the achievable PPC rate of \cref{thm:PMCrate_LagrangeCoded-DSS} in \cref{sec:PPCrate_codes}). Note that the rate in \eqref{eq:Cap_proof} matches the converse in \cref{thm:PLCconverse_MDS-PIRcapacity-achieving-codes}, which proves \cref{thm:PLCcapacity_MDS-PIRcapacity-achieving-codes}. 

\section{A General PPC Scheme for RS-Coded DSSs With Lagrange Encoding}
\label{sec:achievable-scheme_PPC}

In the following, we build a PPC scheme based on Lagrange encoding and our PLC scheme in Section~\ref{sec:private-linear-computation_coded-DSSs}. Note that a higher degree polynomial, i.e., $g>1$, can be written as a linear combination of monomials, and therefore any private monomial computation (PMC) scheme is a special case of PPC. Thus, a PPC scheme can be obtained from a PLC scheme by replacing independent messages with a monomial basis. We first discuss the PPC case in general and then provide an example for the special case of PMC.

In RS-coded DSSs, each message is encoded using an $[n,k]$ RS code as follows. Each $\vect{W}^{(m)}_i$ is encoded by an RS code ${\cal RS}_{k}(\vect{\alpha})$ with evaluation vector $\vect{\alpha}=(\alpha_1,\ldots,\alpha_n)$ over $\Field_q$ into a length-$n$ codeword $\vect{C}^{(m)}_i$ where $\vect{C}^{(m)}_i= \vect{W}^{(m)}_i \mat{G}_{\mathcal{RS}_k}(\vect{\alpha},\vect{\gamma})=\bigl(C^{(m)}_{i,1},\ldots,C^{(m)}_{i,n}\bigr)$ and $C^{(m)}_{i,j} = \ell(\alpha_j)$, $j\in[n]$. Consider an RS-coded DSS with $n$ noncolluding databases storing $f$ messages. The user wishes to retrieve the evaluation of the $v$-th polynomial function $\vmat{X}^{(v)}$, $v\in[\mu]$, from the available information from queries $Q^{(v)}_j$ and answer strings $A^{(v)}_j$, $j\in[n]$, satisfying the conditions of \eqref{eq:privacy} and \eqref{eq:recovery}. 

\subsection{Lagrange Coded Computation}
\label{sec:lagrange-computation}

Lagrange coded computation \cite{YuLiRavivKalanSoltanolkotabiAvestimehr19_1} is a framework that can be applied to any function computation when the function of interest is a multivariate polynomial of the messages. We extend the application of this framework to PMC and PPC by utilizing the following argument.

Let $\ell_t^{(m)}(z)$ be the Lagrange interpolation polynomial associated with the length-$k$ message segment $\vect{W}^{(m)}_t$ for some $t\in[\beta]$ and $m\in[f]$. Recall that $\ell_t^{(m)}(z)$ evaluated at $\gamma_j$ results in an information symbol $W^{(m)}_{t,j}$ and when evaluated at $\alpha_j$ we obtain a code symbol $C^{(m)}_{t,j}$. Let $\vect{\ell}_t(z)=(\ell_t^{(1)}(z), \ldots, \ell_t^{(f)}(z))$ be a vector of $f$ Lagrange interpolation polynomials associated with the messages $\vect{W}^{(1)}_t,\ldots,\vect{W}^{(f)}_t$. Now, given a multivariate polynomial function $\phi(\vect{W}_{t,j})$ of degree at most $g$, we introduce the composition function $\psi_t(z) =\phi(\vect{\ell}_t(z))$. Accordingly, evaluating $\psi_t(z)$ at any $\gamma_j$, $j\in[k]$, is equal to evaluating the polynomial function over the uncoded information symbols, i.e., $\phi(\vect{W}_{t,j})$ and similarly, evaluating $\psi_t(z)$ at $\alpha_j$, $j \in [n]$, will result in the evaluation of the polynomial function over the coded symbols, i.e.,  $\phi(\vect{C}_{t,j})$. Since each Lagrange interpolation polynomial of $\vect{\ell}_t(z)$ is a polynomial of degree at most $k-1$, it follows that $\deg{\psi_t(z)}\leq g(k-1)$ and we require up to $g(k-1)+1$ coefficients to interpolate and determine the polynomial $\psi_t(z)$. 

Note that $\psi_t(z)$ is a linear combination of monomials $z^{i}\in\Field_q[z]$, $i\leq g(k-1)$, and the underlying code  $\tilde{\code{C}}$ for  $(\psi_t(\alpha_1),\ldots,\psi_t(\alpha_n))$, referred to as the \emph{polynomial decoding code}, is given by the $g$-fold star-product ${\cal RS}_k^{\star g}(\vect{\alpha})$ of the storage code ${\cal RS}_k(\vect{\alpha})$ according to \cite[Lem.~6]{RavivKarpuk19_2}. This is due to the fact that the span of ${\cal RS}_k^{\star g}(\vect{\alpha})$ is given by linear combinations of codewords in ${\cal RS}_k^{\star g}(\vect{\alpha})$ where each code symbol represents a monomial. With other words, to construct coded PPC schemes that retrieve polynomials of degree at most $g$, we require $g(k-1)+1 \leq n$ and $d_{\textnormal{min}}^{\tilde{\code{C}}} \geq n-(g(k-1)+1)+1$, where $d_{\textnormal{min}}^{\tilde{\code{C}}}$ denotes the minimum distance of $\tilde{\code{C}}$, to be able to decode the computation correctly. It follows from Proposition~\ref{prop:RSstarProduct} that $\tilde{\code{C}} = {\cal RS}_{{\tilde{k}}}(\vect{\alpha})$ with dimension $\tilde{k} = \min\{g(k-1)+1,n\} = g(k-1)+1$ and $d_{\textnormal{min}}^{\tilde{\code{C}}} = n-\tilde{k}+1 = n-(g(k-1)+1)+1$.

\subsection{PPC Achievable Rate Matrix}  
\label{sec:ppc-achievable-rate}

We now specialize the definition of a generic PC achievable rate matrix from \cref{def:generic-PCachievable-rate} to the coded PPC problem as follows.
\begin{definition}
  \label{def:PPCachievable-rate-matrix}
  Let $\code{C}$ be an arbitrary $[n,k]$ code and denote by $\tilde{\code{C}}=\code{C}^{\star g}$ the $\tilde{k}$-dimensional code generated by the $g$-fold star-product of $\code{C}$ with itself. A $\nu\times n$ binary matrix ${\mat{\Lambda}}_{\kappa,\nu}^{\textnormal{PPC}}$ is called a \emph{PPC achievable rate matrix} for $(\code{C},\tilde{\code{C}})$, if it is a generic PC achievable rate matrix with $\frac{\kappa}{\nu}=\frac{\tilde{k}}{n}$, and for each row $\vect{\lambda}_i$, $\chi(\vect{\lambda}_i)$ is always an information set for $\tilde{\code{C}}$, $i\in [\nu]$.
\end{definition}

Similar to the PLC scheme presented in \cref{sec:query-generation_coded-PLC}, the resulting PPC scheme requires the length of each message to be $\const{L}=\nu^{\mu}\cdot k$. The queries $Q^{(v)}_j$ are generated by setting $(\kappa, \nu)=(\tilde{k}, n)$ and invoking \cref{alg:generation_QuerySet} from Section~\ref{sec:generic-query-generation_PC_coded-DSSs} as follows:
\begin{IEEEeqnarray*}{c}
  \{Q^{(v)}_1,\dots,Q^{(v)}_n\} \leftarrow \texttt{Q-Gen}(v, \mu, \tilde{k},n,n, \mat{A}_{\tilde{k}\times n}, \mat{B}_{(n-\tilde{k})\times n}).
\end{IEEEeqnarray*}%

\subsection{Sign Assignment and Redundancy Elimination}
\label{sec:SignAssignment-RedundancyElimination_PPC}

Here, we generalize the coded PLC scheme of Section~\ref{sec:private-linear-computation_coded-DSSs} in terms of exploiting the dependency between the virtual messages. Since any polynomial is a linear function of the monomial basis of size $\const{M}_g(f)$, a PPC scheme can be seen as a PLC scheme performed over a set of $\const{M}_g(f)$ messages. Hence, the redundancy resulting from the linear dependencies between the virtual messages is also present for PPC and we can extend Lemma~\ref{lem:redundancy} and \cite[Lem.~1]{SunJafar19_2} to this scheme. To exploit the dependency between the virtual messages we adopt a similar sign assignment process to each queried symbol of the virtual monomial messages as mentioned in Section~\ref{sec:SignAssignment-RedundancyElimination_PLC}. Using Lagrange interpolation, we will show that it results in a uniquely solvable equation system from the different $\tau$-sum types given the side information available from all other databases. By obtaining such a system of equations in each round $\tau\in[\mu]$ of the protocol, the user can determine some of the answers offline.

Now, consider $\tau$-sum types for $\tau=1$, where we download individual segments of each virtual message including $f$ independent messages. For this type, the user can determine any polynomial from the $f$ obtained  message segments. Based on this insight we can state the following lemma.

\begin{lemma}
  \label{lem:PPCredundancy}
  Let $\mu \in [f:\upmu_g(f)]$ be the number of candidate polynomial functions evaluations, including the $f$ independent messages. For each query set, for all $v\in[\mu]$, each database $j\in[n]$, and based on the queried segments from the $f$ independent messages, any $\mu-f\choose 1$ $1$-sum types out of all possible types $\mu\choose 1$ are redundant. On the other hand, for $\tau\in[2:\mu]$, any $\max\{\mu-{\const M}_g(f),0\}\choose\tau$ $\tau$-sum types out of $\mu\choose\tau$ types are redundant. Thus, the number of nonredundant $\tau$-sum types with $\tau>1$ is given by $\rho(\mu,\tau)\triangleq{\mu \choose \tau}-{\max\{\mu-{\const M}_g(f), 0\}\choose \tau}$.
\end{lemma}
The proof of Lemma~\ref{lem:PPCredundancy} is presented in Appendix~\ref{sec:proof_PPCredundancy}.
In the next subsection, we show that the privacy and recovery conditions of our proposed PPC scheme are satisfied.

\subsection{Recovery and Privacy}
\label{sec:decoding-privacy_ppc}

The scheme works as the PLC scheme in \cref{sec:private-linear-computation_coded-DSSs} using the code $\tilde{\code{C}}$ instead of the storage code $\code{C}$. This is the case since for \emph{any} polynomial evaluation code $\code{D}$, $\code{D}^{\ast i} \subseteq \code{D}^{\ast j}$ for all $i \in [j]$, $j \in \Naturals$,  since the all-ones codeword is in $\code{D}$ (see also \cite[Lem.~6]{RavivKarpuk19_2}). Moreover, since the definition of the PPC achievable rate matrix in \cref{def:PPCachievable-rate-matrix} is analogous to the corresponding definition of a PIR achievable rate matrix in \cref{def:PIRachievable-rate-matrix} (by using $\tilde{\code{C}}$ instead of $\code{C}$), it can directly be seen that the arguments in the proof  of \cite[Thm.~1]{KumarLinRosnesGraellAmat19_1} (see \cite[App.~B]{KumarLinRosnesGraellAmat19_1}) can be applied. Hence, it follows that $\tilde{k}$ distinct evaluations of $\psi_t(z) = \phi(\vect{\ell}_t(z))$ for each segment $t$ can be recovered. Since $\deg{\psi_t(z)} \leq \tilde{k}-1$, it follows that the polynomial $\psi_t(z)$ can be reconstructed via polynomial interpolation and then the desired polynomial functions evaluations can be recovered by evaluating $\psi_t(z)$ at $\gamma_j$, $j\in[k]$, which is equal to evaluating the desired polynomial $\phi(\cdot)$ over the uncoded information symbols, i.e., $\phi(\vect{W}_{t,j})$ due to Lagrange encoding. 

As for the privacy of the PPC scheme, using an argumentation similar to the PLC scheme, it can be seen that for any desired index $v\in[\mu]$, the redundant $\tau$-sum types according to Lemma~\ref{lem:PPCredundancy} can be fixed, i.e.,~the same $\tau$-sum types are redundant for all $v\in[\mu]$, and hence the queries satisfy the privacy condition. See also Example~\ref{ex:PMCex_n4k2f2mu3} below which illustrates  that the privacy and recovery conditions are indeed satisfied. 

\subsection{Achievable PPC Rate}
\label{sec:PPCrate_codes}  

Since $\tilde{\code{C}}$ is an $[n,\tilde{k}]$ MDS code ($\code{C}$ is an RS code), there always exists a PPC achievable rate matrix $\mat{\Lambda}^{\textnormal{PPC}}_{\kappa,\nu}$ with $\frac{\kappa}{\nu}=\frac{\tilde{k}}{n}$. Hence, using \cref{lem:PPCredundancy} we can prove the following theorem.

\begin{theorem}
  \label{thm:PMCrate_LagrangeCoded-DSS}
  Consider a DSS that uses an $[n,k]$ RS code $\code{C}$ to store $f$ messages over $n$	noncolluding databases using Lagrange encoding. Let $\mu \in [f:\upmu_g(f)]$ be the number of candidate polynomial functions evaluations of degree at most $g$,  including the $f$ independent messages. Then, the PPC rate 
  \begin{IEEEeqnarray}{rCl}
    \const{R}_{\textnormal{PPC}}
    & = & \begin{cases} \frac{1}{f} \HH_{\textnormal{min}}  & \text{if $n \leq g(k-1)+1$}, \\ 
    \frac{\frac{k}{\tilde{k}}\bigl(1-\frac{\tilde{k}}{n} \bigr)\HH_{\textnormal{min}}}
    {1-  \bigl(\frac{\tilde{k}}{n}\bigr)^{\min\{\mu,{\const M}_g(f)\}} - (\min\{\mu,{\const M}_g(f)\}-f)\bigl(1-\frac{\tilde{k}}{n} \bigr)\bigl(\frac{\tilde{k}}{n} \bigr)^{\mu-1}} & \text{otherwise} \end{cases}
    \label{eq:PPCrate_LagrangeCoded-DSS}
  \end{IEEEeqnarray}
  is achievable.
\end{theorem}

\begin{IEEEproof}
From \eqref{eq:query-card_PC} and Lemma~\ref{lem:PPCredundancy}, the achievable PPC rate after removing redundant $\tau$-sums becomes
\begin{IEEEeqnarray}{rCl}
  \label{eq:Cap_proof_first}
  \const{R} & \overset{(a)}{=} &\frac{k\nu^\mu\HH_\textnormal{min}}{n\Bigl({\mu \choose 1}-{\mu-f\choose 1}\Bigr){\kappa}^{\mu} + n\sum_{\tau=2}^{\mu}\rho(\mu,\tau) {\kappa}^{\mu-\tau+1}{(\nu-\kappa)}^{\tau-1}}
  \nonumber\\
  & = & \frac{k\nu^{\mu}\HH_\textnormal{min}}{n\Bigl[f{\kappa}^{\mu}+\sum_{\tau=2}^{\mu}\rho(\mu,\tau) {\kappa}^{\mu-\tau+1}{(\nu-\kappa)}^{\tau-1}\Bigr]},
  \end{IEEEeqnarray}
 where $(a)$ follows from the PPC rate in Definition~\ref{def:def_info-PCrate}, \eqref{eq:query-card_PC}, and Lemma~\ref{lem:PPCredundancy}.  %
Now, if $\nu = \kappa$, or equivalently (from Definition~\ref{def:PPCachievable-rate-matrix}) $n = \tilde{k} \overset{(b)}{=} \min\{g(k-1)+1,n\}$, i.e., $n = g(k-1)+1$  (since $n$ cannot be strictly smaller than $g(k-1)+1$ by assumption and $(b)$ is from Proposition~\ref{prop:RSstarProduct}), then it follows directly from \eqref{eq:Cap_proof_first} that $\const{R} = \frac{k}{n f} \HH_{\textnormal{min}}$. Moreover, it can be seen in this case that the proposed scheme reduces to the trivial scheme where the $f$ independent  files are downloaded and then the desired function evaluation is performed offline. However, the proposed scheme requires an unnecessarily high redundancy to decode the $f$ files, i.e., $\tilde{k}=n$ instead of $\tilde{k}=k$. As a result, for the case of $n\leq g(k-1)+1$, we opt out of any other achievable scheme and achieve the PPC rate $\frac{1}{f} \HH_{\textnormal{min}}$ by simply downloading all $f$ files and performing the desired function evaluation offline. Otherwise, i.e., $\nu >\kappa$, or equivalently (from Definition~\ref{def:PPCachievable-rate-matrix}), $n > \tilde{k}=\min\{g(k-1)+1,n\}$,  i.e., $n  > g(k-1)+1$, then from \eqref{eq:Cap_proof_first} we have
  \begin{IEEEeqnarray}{rCl}
  \const{R}  
  & \overset{(c)}{=}  & \frac{k(\nu-\kappa)\HH_\textnormal{min}}{n\kappa}\inv{\left[\frac{f(\nu-\kappa)}{\nu}\Bigl(\frac{\kappa}{\nu}\Bigr)^{\mu-1}+\frac{1}{\nu^{\mu}}\sum_{\tau=2}^{\mu}\rho(\mu,\tau)\kappa^{\mu-\tau}{(\nu-\kappa)}^{\tau}\right]} \nonumber \\
  & \overset{(d)}{=} &\frac{k\HH_\textnormal{min}}{n}\Bigl(\frac{n}{\tilde{k}}-1\Bigr)\inv{\left[f \Bigl(1-\frac{\tilde{k}}{n}\Bigr)  \Bigl( \frac{{\tilde{k}}}{n}\Bigr)^{\mu-1}  + \frac{1}{n^{\mu}}\sum_{\tau=2}^{\mu}\left({\mu \choose \tau}-{\max\{\mu-{\const M}_g(f), 0\}\choose \tau}\right) {\tilde{k}}^{\mu-\tau}{(n-\tilde{k})}^{\tau} \right]}
  \nonumber\\
  & = & \frac{k\HH_\textnormal{min}}{\tilde{k}} \Bigl(1-\frac{\tilde{k}}{n}\Bigr) \inv{\left[ f \Bigl(1-\frac{\tilde{k}}{n}\Bigr)  \Bigl( \frac{{\tilde{k}}}{n}\Bigr)^{\mu-1}  + \frac{1}{n^{\mu}}  \left( \sum_{\tau=0}^{\mu}{\mu \choose \tau} {\tilde{k}}^{\mu-\tau}(n-\tilde{k})^{\tau} - \mu\tilde{k}^{\mu-1}(n-\tilde{k}) -{\tilde{k}}^{\mu} \right)  \right.
    \nonumber\\ 
    & & \hspace{20ex}\left.-\frac{1}{n^{\mu}}\sum_{\tau=2}^{\mu}{\max\{\mu-{\const M}_g(f), 0\}\choose \tau} {\tilde{k}}^{\mu-\tau}(n-\tilde{k})^{\tau}  \right]} \nonumber
  \\
  & \overset{(e)}{=} & \frac{k\HH_\textnormal{min}}{\tilde{k}} \Bigl(1-\frac{\tilde{k}}{n}\Bigr) \inv{\left[ f \Bigl(1-\frac{\tilde{k}}{n}\Bigr)  \Bigl( \frac{{\tilde{k}}}{n}\Bigr)^{\mu-1}  + \frac{1}{n^{\mu}}  \left(n^{\mu} - \mu\tilde{k}^{\mu-1}(n-\tilde{k}) -{\tilde{k}}^{\mu} \right)\right. \nonumber
    \\ 
    & & \hspace{20ex}\left.-\frac{1}{n^{\mu}} \left( \sum_{\tau=0}^{\eta}
        {\eta \choose \tau} {\tilde{k}}^{\mu-\tau}(n-\tilde{k})^{\tau} - \eta {\tilde{k}}^{\mu-1}(n-\tilde{k}) - {\tilde{k}}^{\mu} \right)   \right]} \nonumber
  \\
  & = & \frac{k\HH_\textnormal{min}}{\tilde{k}} \Bigl(1-\frac{\tilde{k}}{n}\Bigr) \inv{\left[ f \Bigl(1-\frac{\tilde{k}}{n}\Bigr) \Bigl( \frac{{\tilde{k}}}{n}\Bigr)^{\mu-1}  
      +   1 - \mu \Bigl(1-\frac{\tilde{k}}{n}\Bigr) \Bigl( \frac{{\tilde{k}}}{n}\Bigr)^{\mu-1}  - \Bigl( \frac{{\tilde{k}}}{n}\Bigr)^{\mu}   \right. \nonumber
    \\ 
    & & \hspace{20ex}\left.-\frac{1}{n^{\mu}} \left( {\tilde{k}}^{\mu-\eta} \sum_{\tau=0}^{\eta}{\eta \choose \tau} {\tilde{k}}^{\eta-\tau}(n-\tilde{k})^{\tau} \right) + \eta \Bigl(1-\frac{\tilde{k}}{n}\Bigr) \Bigl( \frac{{\tilde{k}}}{n}\Bigr)^{\mu-1}  + \Bigl( \frac{{\tilde{k}}}{n}\Bigr)^{\mu} \right]} \nonumber
  \\
  & = & \frac{k\HH_\textnormal{min}}{\tilde{k}} \Bigl(1-\frac{\tilde{k}}{n}\Bigr) \inv{\left[1+  (f-\mu+\eta) \Bigl(1-\frac{\tilde{k}}{n}\Bigr) \Bigl( \frac{{\tilde{k}}}{n}\Bigr)^{\mu-1}  
      -   \frac{1}{n^{\mu}} \left( {\tilde{k}}^{\mu-\eta} {n}^{\eta} \right)   \right]} \nonumber
  \\
  & = & \frac{k\HH_\textnormal{min}}{\tilde{k}} \Bigl(1-\frac{\tilde{k}}{n}\Bigr) \inv{\left[1-  \big(\mu-\eta -f\big) \Bigl(1-\frac{\tilde{k}}{n}\Bigr) \Bigl( \frac{{\tilde{k}}}{n}\Bigr)^{\mu-1}  
      -  \Bigl( \frac{{\tilde{k}}}{n}\Bigr)^{\mu-\eta}    \right]} \nonumber
  \\
  &=& \frac{\frac{k}{\tilde{k}}\bigl(1-\frac{\tilde{k}}{n}\bigr)\HH_\textnormal{min}}{1-  \bigl(\frac{\tilde{k}}{n}\bigr)^{\min\{\mu,{\const M}_g(f)\}} - (\min\{\mu,{\const M}_g(f)\}-f)\bigl(1-\frac{\tilde{k}}{n}\bigr)\bigl(\frac{\tilde{k}}{n} \bigr)^{\mu-1}},  \nonumber
\end{IEEEeqnarray}
where $(c)$ follows since $\nu > \kappa$; $(d)$ holds since we have $\frac{\kappa}{\nu}=\frac{\tilde{k}}{n}$ from Definition~\ref{def:PPCachievable-rate-matrix}; and $(e)$ follows by defining $\eta\triangleq \max\{\mu-{\const M}_g(f), 0\}$ and the fact that $\binom{m}{n} = 0$ if $m<n$.
\end{IEEEproof}

\begin{corollary}
  \label{cor:PMCrate_LagrangeCoded-DSS}
  Consider a DSS that uses an $[n,k]$ RS code $\code{C}$ to store $f$ messages over $n$ noncolluding databases using Lagrange encoding. Let $\mu \in [f:\upmu_g(f)]$ be the number of candidate polynomial functions evaluations of degree at most $g$,  including the $f$ independent messages. Then, the PPC rate 
    \begin{IEEEeqnarray}{rCl}
    \const{R}_{\textnormal{PPC},\infty}
    & = & \frac{k}{n} \left( \frac{\max\{n-g(k-1)-1,0\}}{g(k-1)+1} \right) \HH_{\textnormal{min}}
    \label{eq:PPCrate_LagrangeCoded-DSS_asymp}
  \end{IEEEeqnarray}
  is achievable as $f \to \infty$.
\end{corollary}

\begin{IEEEproof}
If $n \leq g(k-1)+1$, then it follows from \eqref{eq:PPCrate_LagrangeCoded-DSS} that the PPC rate approaches zero as $f \to \infty$, which is in accordance with \eqref{eq:PPCrate_LagrangeCoded-DSS_asymp}. Otherwise, if $n > g(k-1)+1$, the result follows directly from \eqref{eq:PPCrate_LagrangeCoded-DSS} by taking the limit $f \to \infty$ and using the fact that $\tilde{k} \overset{(a)}{=} \min\{g(k-1)+1,n\} = g(k-1)+1 < n$, where $(a)$ follows from Proposition~\ref{prop:RSstarProduct}.
\end{IEEEproof}

Note that the asymptotic PPC rate in \eqref{eq:PPCrate_LagrangeCoded-DSS_asymp} is equal to the rate of the general scheme from \cite{RavivKarpuk19_2} when $\HH_{\textnormal{min}}=1$. This difference is due to the simplified rate definition used in \cite{RavivKarpuk19_2}. Moreover, our proposed scheme cannot readily be obtained using the concept of refinement and lifting of so-called one-shot schemes as introduced for PIR in \cite{DOliveiraElRouayheb18_1}, since this concept cannot readily be applied to the function computation case.

\subsection{Special Case: PMC Scheme}
\label{sec:PIRrate_codes}

As the rate of PMC is a decreasing function of the number of candidate monomial functions, we can increase the PMC rate by limiting ourselves to the set of monomials excluding \emph{parallel} monomials. To this end, we define a parallel monomial as a monomial resulting from raising another monomial to a positive integer power, i.e., to $\{{\vect{W}}^{\vect{i}}: \vect{i} \in \Naturals_0^{f},\, 1\leq \textsf{wt}(\vect{i})\leq g,\,\vect{i} \mid p,\,p\in\set{P}_g\}$. Here,  $\set{P}_g$ denotes the set of prime numbers less or equal to $g$ and  $\vect{i}=(i_1,\ldots,i_f) \mid p$ means that all nonzero $i_j$, $j \in [f]$, are divisors of $p$. For example, for a bivariate monomial over the variables $x$ and $y$ of degree at most $g=2$ the set of possible monomials is $\{x,y, xy, x^2,y^2\}$. Note that $x^2$ is a parallel monomial as it can be obtained by raising the monomial $x$ to the power of $2$. Thus, $x^2$ and $y^2$ are parallel monomials and can be excluded from the set of candidate monomials. Denote by $\set{P}=\{ p_{1},\ldots,p_{|\set{P}|}\}$ an arbitrary nonempty subset of $\set{P}_g$. By applying the Legendre formula for counting the prime numbers less or equal to $g$, we obtain the number of nonparallel monomials as
 
\begin{IEEEeqnarray*}{rCl}
  \widetilde{\const{M}}_g(f)& = & \const{M}_g(f) +\sum_{\substack{\forall\set{P}\subseteq \set{P}_g: \set{P} \neq \emptyset, \\ p_1  \cdots p_{|\set{P}|} \leq g}} (-1)^{{|\set{P}|}} \left[ {\left( \genfrac{}{}{0pt}{0}{\left\lfloor{\frac{g}{p_{1}\cdots p_{{|\set{P}|}}}}\right\rfloor +f}{\left\lfloor {\frac{g}{p_{1}\cdots p_{{|\set{P}|}}}}\right\rfloor}\right)}-1\right].
\end{IEEEeqnarray*}%

We illustrate the key concept of our proposed scheme in Theorem \ref{thm:PMCrate_LagrangeCoded-DSS} with an example.

\begin{example}
  \label{ex:PMCex_n4k2f2mu3}
  Consider two messages $\vmat{W}^{(1)}$ and $\vmat{W}^{(2)}$ that are stored in a noncolluding DSS using a $[4,2]$ RS code $\code{C}$. Suppose that the user wishes to obtain a monomial function evaluation $\vmat{X}^{(v)}$ from the  set of nonparallel monomial functions of degree at most $g=2$. We have $\mu=\widetilde{\const{M}}_2(2)=3$, $v\in [3]$, and the candidate set of monomial functions evaluations is $\{\vmat{W}^{(1)},\vmat{W}^{(2)},\vmat{W}^{(1)}\!\star\vmat{W}^{(2)}\}$, where $\star$ denotes element-wise multiplication. Let the desired monomial function index be $v=1$, i.e.,~the user wishes to obtain the function evaluation $\vmat{X}^{(1)}=\vmat{W}^{(1)}$. We have $\tilde{k}=g(k-1)+1=3$ and 
  \begin{IEEEeqnarray*}{rCl}
    \mat{\Lambda}^{\textnormal{PPC}}_{3,4}=
    \begin{pmatrix}
      1 & 1 & 1 & 0 
      \\
      1 & 1 & 0 & 1 
      \\
      1 & 0 & 1 & 1
      \\
      0 & 1 & 1 & 1
    \end{pmatrix}
  \end{IEEEeqnarray*}
  is a valid PPC achievable rate matrix for $(\code{C},\tilde{\code{C}})$. From $\mat{\Lambda}^{\textnormal{PPC}}_{3,4}$ we further obtain the PC interference matrices
  \begin{IEEEeqnarray*}{c}
    \mat{A}_{3\times 4} = 
    \begin{pmatrix}
      1 &1 &1 &2
      \\
      2 &2 &3 &3 
      \\
      3 &4 &4 &4 
    \end{pmatrix}
    \text{ and }
    \mat{B}_{1\times 4} = 
    \begin{pmatrix}
      4 &3 &2 &1
    \end{pmatrix}
    \vspace{-1ex}
  \end{IEEEeqnarray*}
from \cref{def:PCinterference-matrices}.

  We simplify notation by letting $x_{t,j}=C^{(1)}_{t,j}$, $y_{t,j}=C^{(2)}_{t,j}$, and $z_{t,j}=C^{(1)}_{t,j}\cdot C^{(2)}_{t,j}$ for all ${t\in[\beta]}$, $j\in[n]$, where $\beta=\nu^{\mu}=64$. Since the desired function evaluation is $\vmat{X}^{(1)}$, the goal is to privately obtain $x_{t,j}$, $t\in[\beta]$, and successfully decode $\vmat{X}^{(1)}$. The construction of the query sets is briefly presented in the following steps.\footnote{With some abuse of notation for the sake of simplicity, the generated queries are sets containing their answers.}
	
  {\bf \textit{Initialization (Round ${\tau=1}$)}:} We start with ${\tau=1}$ to generate query sets for each database $j$ holding $\kappa^{\mu}=27$ distinct instances of $x_{t,j}$. By message symmetry this also applies to $y_{t,j}$ and $z_{t,j}$.

  {\bf \textit{Following Rounds ($\tau \in  [2:3]$)}:} Using the PC interference matrices $\mat{A}_{3\times 4}$ and $\mat{B}_{1\times 4}$ for the exploitation of side information for the $j$-th database, $j\in [n]$, we generate the desired query sets $Q^{(1)}_j(\set{D};\tau)$ by querying a number of new symbols of the desired monomial jointly combined with symbols from other monomials queried in the previous round from database $i\neq j$. Next, the undesired query sets $Q^{(1)}_j(\set{U};\tau)$ (if $\tau=2)$ are generated by enforcing message symmetry.
  
  In the end, we we apply the sign assignment procedure to the query sets for $v=1$ and make the final modification to the queries by removing all the $1$-sums corresponding to the redundant $1$-sum types from the first round (see Lemma~\ref{lem:PPCredundancy}). This translates to removing the queries for $z_{t,j}$, since they can be generated offline by the user given $x_{t,j}$ and $y_{t,j}$. The resulting query sets are shown in Table~\ref{tab:answers-table}, where $u_{a:b,j} \eqdef \{u_{a,j},\ldots,u_{b,j}\}$ for $u=x,y,z$, and the side information is highlighted with blue and red for rounds $\tau=2$ and $\tau=3$, respectively. 
  Similar to \cref{ex:PLCex_n4k2f4mu4}, by using Lagrange interpolation, it can be shown that the side information based on $\mat{B}_{(\nu-\kappa)\times n}$ can be decoded. For instance, in round $2$, since $y_{1:3,1}$, $y_{1:3,2}$, $y_{1:3,3}$ obtained from round $1$ are enough to reconstruct the associated Lagrange interpolation polynomial, the side information $y_{1:3,4}$ can be obtained, from which the desired polynomial function evaluations $x_{43:45,4}$ can be decoded by side information cancellation. Note that in this example, it is clear that $z_{t,j}$ is redundant, no matter which $v$ is requested, and hence privacy is ensured.  The PMC rate of the scheme is equal to $\frac{k\nu^{\mu} \HH_{\textnormal{min}}}{\const{D}}=\frac{2\times4^3}{3\times4\times28} \HH_{\textnormal{min}}=0.3810 \cdot  \HH_{\textnormal{min}}$, where the value of $\HH_{\textnormal{min}} = \HH(\vmat{X}^{(3)})$ depends on the underlying field.
\QEDA
\end{example}

\begin{table}[t]
  \centering
  \caption{The PMC query sets for $v=1$ after sign assignment and removal of redundant queries for a $[4,2]$ RS-coded DSS with Lagrange encoding storing $f=2$ messages, where the $\mu=3$ candidate monomial functions evaluations are $\{\vmat{X}^{(1)}=\vmat{W}^{(1)},\vmat{X}^{(2)}=\vmat{W}^{(2)},\vmat{X}^{(3)}=\vmat{W}^{(1)}\star\vmat{W}^{(2)}\}$. Blue and red subscripts indicate side information exploitation in rounds $\tau=2$ and $\tau=3$, respectively.}
  \label{tab:answers-table}
  \vskip -1mm
  \Resize[\columnwidth]{
    \begin{IEEEeqnarraybox}[
      \IEEEeqnarraystrutmode
      \IEEEeqnarraystrutsizeadd{4pt}{2pt}]{v/c/v/c/v/c/v/c/v/c/v}
      \IEEEeqnarrayrulerow\\
      & j && 1 && 2 && 3 && 4\\
      \hline\hline
      & Q^{(1)}_j(\set{D};1)
      && x_{1:9,1},\, x_{10:18,1},\, x_{19:27,1} &&  x_{1:9,2},\, x_{10:18,2}, \, x_{28:36,2} && x_{1:9,3},\, x_{19:27,3},\, \,x_{28:36,3} && x_{10:18,4},\, x_{19:27,4},\,x_{28:36,4} &
      \\*\cline{1-11}      
      & Q^{(1)}_j(\set{U};1)
      && y_{1:9,1},\, y_{10:18,1},\, y_{19:27,1} &&  y_{1:9,2},\, y_{10:18,2}, \, y_{28:36,2} && y_{1:9,3},\, y_{19:27,3},\, \,y_{28:36,3} && y_{10:18,2},\, y_{19:27,3},\,y_{28:36,4} &
      \\*\cline{1-11}      
      & \multirow{6}{*}{$Q^{(1)}_j(\set{D};2)$}
      && x_{37:39,1}-y_{{\b 28:30},1} &&  x_{37:39,2}-y_{{\b 19:21},2} && x_{37:39,3}-y_{{\b 10:12},3} && x_{43:45,4}-y_{{\b 1:3},4} &
      \\ 
      & 
      && x_{40:42,1}-z_{{\b 28:30},1} &&  x_{40:42,2}-z_{{\b 19:21},2} && x_{40:42,3}-z_{{\b 10:12},3} && x_{46:48,4}-z_{{\b 1:3},4} &
      \\
      &
      && x_{43:45,1}-y_{{\b 31:33},1} &&  x_{43:45,2}-y_{{\b 22:24},2} && x_{49:51,3}-y_{{\b 13:15},3} && x_{49:51,4}-y_{{\b 4:6},4} &
      \\ 
       & 
      && x_{46:48,1}-z_{{\b 31:33},1} &&  x_{46:48,2}-z_{{\b 22:24},2} && x_{52:54,3}-z_{{\b 13:15},3} && x_{52:54,4}-z_{{\b 4:6},4} &
      \\
      &
      && x_{49:51,1}-y_{{\b 34:36},1} &&  x_{55:57,2}-y_{{\b 25:27},2} && x_{55:57,3}-y_{{\b 16:18},3} && x_{55:57,4}-y_{{\b 7:9},4} &
      \\
       & 
      && x_{52:54,1}-z_{{\b 34:36},1} &&  x_{58:60,2}-z_{{\b 25:27},2} && x_{58:60,3}-z_{{\b 16:18},3} && x_{58:60,4}-z_{{\b 7:9},4} &
      \\*\cline{1-11}      
      & \multirow{3}{*}{$Q^{(1)}_j(\set{U};2)$}
      && y_{40:42,1}-z_{37:39,1} &&   y_{40:42,2}-z_{37:39,2} &&  y_{40:42,3}-z_{37:39,3} && y_{46:48,4}-z_{43:45,4} &
      \\
      &
      && y_{46:48,1}-z_{43:45,1} &&  y_{46:48,2}-z_{43:45,2} && y_{52:54,3}-z_{49:51,3} && y_{52:54,4}-z_{49:51,4} &
      \\
      &
      && y_{52:54,1}-z_{49:51,1} &&  y_{58:60,2}-z_{55:57,2} && y_{58:60,3}-z_{55:57,3} && y_{58:60,4}-z_{55:57,4} &
      \\*\cline{1-11}      
      & \multirow{3}{*}{$Q^{(1)}_j(\set{D};3)$}
      && x_{61,1}-y_{{\r  58},1}+z_{{\r  55},1} && x_{61,2}-y_{{\r  52},2}+z_{{\r  49},2}
      && x_{61,3}-y_{{\r  46},3}+z_{{\r  43},3} && x_{62,4}-y_{{\r  40},4}+z_{{\r  37},4} &
      \\
      &
      && x_{62,1}-y_{{\r  59},1}+z_{{\r  56},1} && x_{62,2}-y_{{\r  53},2}+z_{{\r  50},2}
      && x_{63,3}-y_{{\r  47},3}+z_{{\r  44},3} && x_{63,4}-y_{{\r  41},4}+z_{{\r  38},4} &
      \\
      &
      && x_{63,1}-y_{{\r  60},1}+z_{{\r  57},1} && x_{64,2}-y_{{\r  54},2}+z_{{\r  51},2}
      && x_{64,3}-y_{{\r  48},3}+z_{{\r  45},3} && x_{64,4}-y_{{\r  42},4}+z_{{\r  39},4} &
      \\*\IEEEeqnarrayrulerow
    \end{IEEEeqnarraybox}}
\end{table}

\section{PPC Scheme for RS-Coded DSSs With Systematic Lagrange Encoding}  
\label{sec:achievable-scheme_PPC_systematic}

In this section, we consider the case of RS-coded DSSs with systematic Lagrange encoding and first specialize the definition of a generic PC achievable rate matrix from \cref{def:generic-PCachievable-rate} to this scenario.

\subsection{PPC Systematic Achievable Rate Matrix}

In contrast to the PPC scheme in \cref{sec:achievable-scheme_PPC}, the basic idea is to utilize the systematic part of the RS code to recover the requested function evaluation directly, i.e.,~we do not need to interpolate the systematic downloaded symbols to determine the requested function evaluation. Thus, we can further enhance the download rate. However, due to the generic PC query design principles, namely, message symmetry and side information exploitation, we are restricted in how to exploit side information obtained from the systematic nodes. Specifically, for decodability (side information cancellation) to be possible, the side information obtained from the systematic nodes must be utilized in an isolated manner within an information set of the \emph{polynomial decoding code} (see \cref{sec:lagrange-computation}), such that we can reverse the order of the decoding procedure (i.e.,~unlike our RS-coded PPC scheme, we interpolate first and then cancel the side information).  This restriction is further illustrated by a careful construction of a PPC systematic achievable rate matrix (\cref{def:SysPPCachievable-rate-matrix} below) and the corresponding interference matrices. Moreover, we modify the general PPC scheme to utilize only the necessary number of nodes, denoted by $\hat{n}$, that guarantee the isolated use of systematic side information. %
Accordingly, we specialize \cref{def:generic-PCachievable-rate} as follows.

\begin{definition} 
  \label{def:SysPPCachievable-rate-matrix}
  Let $\code{C}$ be an arbitrary $[n,k]$ code and denote by $\tilde{\code{C}}=\code{C}^{\star g}$ the $\tilde{k}$-dimensional code generated by the $g$-fold star-product of $\code{C}$ with itself. Moreover, let\footnote{Note that the first requirement of the final case of \eqref{eq:systematic_n} is unnecessary as $\bigl\lfloor \frac{n}{\tilde{k}} \bigr\rfloor \geq 1$ always. However, it is included for symmetry reasons.}
  \begin{IEEEeqnarray}{rCl}
    \label{eq:systematic_n}
    \hat{n} \triangleq \begin{cases}  n  & \text{if } \bigl\lfloor \frac{n}{\tilde{k}} \bigr\rfloor = 1 \text{ and } n-\bigl\lfloor \frac{n}{\tilde{k}}\bigr\rfloor \tilde{k} <k,\\
      k+(\bigl\lfloor \frac{n}{\tilde{k}}\bigr\rfloor -1)\tilde{k} & \text{if } \bigl\lfloor \frac{n}{\tilde{k}} \bigr\rfloor > 1 \text{ and } n-\bigl\lfloor \frac{n}{\tilde{k}}\bigr\rfloor \tilde{k} <k,\\
      k+\bigl\lfloor \frac{n}{\tilde{k}}\bigr\rfloor  \tilde{k} & \text{if }  \bigl\lfloor \frac{n}{\tilde{k}} \bigr\rfloor  \geq  1 \text{ and } n-\bigl\lfloor \frac{n}{\tilde{k}}\bigr\rfloor \tilde{k} \geq k. 
    \end{cases} \end{IEEEeqnarray} 
  Then, a $\nu\times \hat{n}$ binary matrix $\mat{\Lambda}^{\textnormal{S,PPC}}_{\kappa,\nu}$ is called a \emph{PPC systematic achievable rate matrix} for $(\code{C},\tilde{\code{C}})$
  if the following conditions are satisfied.
  \begin{enumerate}
  \item \label{item:3} $\mat{\Lambda}^{\textnormal{S,PPC}}_{\kappa,\nu}$ is a $\kappa$-column regular matrix, and
  \item \label{item:4} there are exactly $\varrho \triangleq \bigl\lfloor\frac{\hat{n}}{\tilde{k}} \bigr\rfloor \kappa$ rows $\{\vect{\lambda}_i\}_{i\in[\varrho]}$ and $\nu-\varrho$ rows
    $\{\vect{\lambda}_{i+\varrho}\}_{i\in [\nu-\varrho]}$ of $\mat{\Lambda}^{\textnormal{S,PPC}}_{\kappa,\nu}$  such that
    $\forall\,i\in [\varrho]$, $\chi(\vect{\lambda}_i)$ contains an information set for $\tilde{\code{C}}$ and
    $\forall\,i\in [\nu-\varrho]$, $\chi(\vect{\lambda}_{i+\varrho})=[k]$.
  \end{enumerate}
\end{definition}
The following lemma shows how to construct a PPC systematic achievable rate matrix with $(\kappa,\nu)=\bigl(k,\hat{n}- \bigl\lfloor\frac{\hat{n}}{\tilde{k}} \bigr\rfloor (\tilde{k}-k)\bigr)$.
\begin{lemma}
  \label{lem:S-PPC}
  Let $\code{C}$ be an arbitrary $[n,k]$ code and $\tilde{\code{C}}= \code{C}^{\star g}$. Then, there exists a PPC systematic achievable rate matrix $\mat{\Lambda}^{\textnormal{S,PPC}}_{\kappa,\nu}$ for $(\code{C},\tilde{\code{C}})$ with $(\kappa,\nu)=\bigl(k,\hat{n}- \bigl\lfloor\frac{\hat{n}}{\tilde{k}} \bigr\rfloor (\tilde{k}-k)\bigr)$, where $\tilde{k}$ is the dimension of $\tilde{\code{C}}$.
\end{lemma}
\begin{IEEEproof} 
  Let $\hat{\delta}\eqdef\bigl\lfloor\frac{\hat{n}}{\tilde{k}}\bigr\rfloor$ and $\Gamma\eqdef\hat{n}-\hat{\delta}\tilde{k}$. From our choices of $\hat{n}$ in \eqref{eq:systematic_n}, one can verify that $\Gamma\leq k$ and $\Gamma$ is well-defined. Accordingly, construct a matrix $\mat{A}_{k\times\hat{n}}$ as in Definition \ref{def:PCinterference-matrices} with
  \begin{IEEEeqnarray}{c}
    a_{i,j}=\hat{\delta} k+i, \text{ if } j\in [k],\,i\in [\Gamma].
    \label{eq:entries_A1}
  \end{IEEEeqnarray}
  In this way, $k\Gamma$ entries of $\mat{A}_{k\times \hat{n}}$ are filled. Next, let $\{a_{i^{(j)}_1,j},\ldots,a_{i^{(j)}_{u(j)},j} \}$, $j\in [\hat{n}]$, denote the remaining empty entries in column $j$ of $\mat{A}_{k\times \hat{n}}$, where $u(j)\leq k$ is the number of empty entries in column $j$. Hence, the $k\hat{n}-k \Gamma =k(\hat{n}-\Gamma)$ entries 
  \begin{IEEEeqnarray}{c}
    \Bigl\{a_{i^{(1)}_1,1},\ldots,a_{i^{(1)}_{u(1)},1},\ldots,a_{i^{(\hat{n})}_1,\hat{n}},\ldots,a_{i^{(\hat{n})}_{u(\hat{n})},\hat{n}}\Bigr\}
    \label{eq:remaining-entries_A}
  \end{IEEEeqnarray}
  are empty. Now, observe that $(\hat{n}-\Gamma)\inv{\hat{\delta}}=\bigl(\hat{n}-(\hat{n}-\hat{\delta}\tilde{k})\bigr)\inv{\hat{\delta}} = \tilde{k}\in\Naturals$. By consecutively assigning $\{1,\ldots,\hat{\delta} k\}$ to the entries of $\mat{A}_{k\times \hat{n}}$ in \eqref{eq:remaining-entries_A} and repeating this process $\tilde{k}$ times, the remaining $\hat{\delta} k\cdot\frac{\hat{n}-\Gamma}{\hat{\delta}}=k(\hat{n}-\Gamma)$ empty entries of $\mat{A}_{k\times \hat{n}}$ are filled. Note that since values of $[\hat{\delta} k]$ are consecutively assigned, the largest number of empty entries of each column of $\mat{A}_{k\times \hat{n}}$ is $k$, and $\hat{\delta}=\bigl\lfloor \frac{\hat{n}}{\tilde{k}}\bigr\rfloor \geq 1$, there are no repeated values of $[\hat{\delta} k]$ in any column of $\mat{A}_{k\times \hat{n}}$, which implies that condition 1) in Definition~\ref{def:SysPPCachievable-rate-matrix} is satisfied. From \eqref{eq:entries_A1} and \eqref{eq:remaining-entries_A}, it can be seen that each $a\in [\hat{\delta} k]=[\varrho]$ occurs in $\tilde{k}$ columns of $\mat{A}_{k\times \hat{n}}$ and each $a\in [\hat{\delta} k+1: \hat{\delta} k+\Gamma]$ occurs in $k$ columns of $\mat{A}_{k\times \hat{n}}$. This implies that condition 2) in Definition~\ref{def:SysPPCachievable-rate-matrix} is satisfied with $\kappa=k$, $\varrho=\hat{\delta}k$, and $\nu=\Gamma+\hat{\delta} k$, which completes the proof.  
\end{IEEEproof}

\begin{lemma}
  \label{lem:nu_formula}
  It holds that 
  \begin{IEEEeqnarray}{rCl}
    \label{eq:nu_cases1}
    \nu=\begin{cases}
      n-\tilde{k}+k  & \text{if }   \bigl\lfloor \frac{n}{\tilde{k}} \bigr\rfloor = 1 \text{ and } n-\bigl\lfloor \frac{n}{\tilde{k}}\bigr\rfloor \tilde{k} <k,\\ 
      \bigl\lfloor \frac{n}{\tilde{k}} \bigr\rfloor k  & \text{if }   \bigl\lfloor \frac{n}{\tilde{k}} \bigr\rfloor > 1 \text{ and } n-\bigl\lfloor \frac{n}{\tilde{k}} \bigr\rfloor\tilde{k} < k,\\
      \bigl\lfloor \frac{n}{\tilde{k}} \bigr\rfloor k+k & \text{if }   \bigl\lfloor \frac{n}{\tilde{k}} \bigr\rfloor \geq 1 \text{ and } n-\bigl\lfloor \frac{n}{\tilde{k}} \bigr\rfloor \tilde{k}\geq k.
    \end{cases} 	
  \end{IEEEeqnarray}
\end{lemma}
\begin{IEEEproof}
  To prove the results, we use \cref{def:SysPPCachievable-rate-matrix} and the fact that $\nu=\hat{n}- \bigl\lfloor\frac{\hat{n}}{\tilde{k}} \bigr\rfloor (\tilde{k}-k)$. 
  Now, if $ \bigl\lfloor \frac{n}{\tilde{k}} \bigr\rfloor = 1$ and $n-\bigl\lfloor \frac{n}{\tilde{k}}\bigr\rfloor \tilde{k} <k$ (the first case from \cref{def:SysPPCachievable-rate-matrix}), then it follows directly that $\nu = \hat{n} - \bigl\lfloor \frac{\hat{n}}{\tilde{k}} \bigr\rfloor (\tilde{k}-k) = n - \bigl\lfloor \frac{n}{\tilde{k}} \bigr\rfloor (\tilde{k}-k) = n-\tilde{k}+k$.
  On the other hand, if $ \bigl\lfloor \frac{n}{\tilde{k}} \bigr\rfloor > 1$ and $n-\bigl\lfloor \frac{n}{\tilde{k}} \bigr\rfloor\tilde{k} < k$ (the second case from \cref{def:SysPPCachievable-rate-matrix}), then after inserting $\hat{n} = k + \bigl( \bigl\lfloor \frac{n}{\tilde{k}} \bigr\rfloor -1\bigr) \tilde{k}$ into the expression for $\nu$, $\nu = k \bigl\lfloor \frac{n}{\tilde{k}}\bigr\rfloor - \bigl\lfloor \frac{k}{\tilde{k}}  \bigr\rfloor (\tilde{k}-k) = k \bigl\lfloor \frac{n}{\tilde{k}} \bigr\rfloor $, since $\bigl\lfloor \frac{k}{\tilde{k}}  \bigr\rfloor (\tilde{k}-k)=0$. In a similar manner, the  remaining case in \eqref{eq:nu_cases1} can be shown.
\end{IEEEproof}

In the following lemma, we show a lower bound to the fraction $\frac{\kappa}{\nu}$. 
\begin{lemma}
	\label{lem:PIRrate_upper-bound}
	If a matrix $\mat{\Lambda}^{\textnormal{S,PPC}}_{\kappa,\nu}(\code{C},\tilde{\code{C}})$ exists for an $[n,k]$  code $\code{C}$ and the $[n, \tilde{k}]$ code $\tilde{\code{C}}$, then we have
	\begin{IEEEeqnarray*}{rCl}
		\frac{\kappa}{\nu}\geq\frac{k}{\hat{n}-  \bigl\lfloor \frac{\hat{n}}{\tilde{k}} \bigr\rfloor (\tilde{k}-k)}.
	\end{IEEEeqnarray*}
\end{lemma} 
\begin{IEEEproof}
  Since by definition each row $\vect{\lambda}_{i}$ of $\mat{\Lambda}^{\textnormal{S,PPC}}_{\kappa,\nu}$ contains an information set for $\tilde{\code{C}}$, $i\in[\varrho]$, $\varrho = \bigl\lfloor\frac{\hat{n}}{\tilde{k}} \bigr\rfloor \kappa$, and each row $\vect{\lambda}_{i+\varrho}=[k]$, $i\in[\nu-\varrho]$, we have $\Hwt{\vect{\lambda}_i}\geq \tilde{k}$, $i\in[\varrho]$, and $\Hwt{\vect{\lambda}_{i+\varrho}}=k$, $i\in[\nu-\varrho]$. Let $\vect{v}_j$, $j\in [\hat{n}]$, be the $j$-th column of $\mat{\Lambda}^{\textnormal{S,PPC}}_{\kappa,\nu}$. If we look at $\mat{\Lambda}^{\textnormal{S,PPC}}_{\kappa,\nu}$ from both a row-wise and a column-wise point of view, we obtain
  \begin{IEEEeqnarray*}{c}
    \varrho\tilde{k}+(\nu-\varrho)k\leq\sum_{i=1}^{\varrho}\Hwt{\vect{\lambda}_i}+\sum_{i=1}^{\nu-\varrho}\Hwt{\vect{\lambda}_{i+\varrho}}=\sum_{j=1}^{\hat{n}}\Hwt{\vect{v}_j}=\kappa \hat{n}.
  \end{IEEEeqnarray*}
  Thus, we have
  \begin{IEEEeqnarray*}{rCl}
    \varrho\tilde{k}-\varrho k+\nu k& = &\varrho(\tilde{k}-k)+\nu k\leq \kappa \hat{n},
  \end{IEEEeqnarray*}
  from which the result follows.
\end{IEEEproof} 
Now, similar to the PLC scheme presented in \cref{sec:query-generation_coded-PLC}, the systematic PPC scheme requires the length of each message to be $\const{L}=\nu^{\mu}\cdot k$. The queries $Q^{(v)}_j$ are generated by setting $(\kappa,\nu)=(k, \hat{n}- \bigl\lfloor \frac{\hat{n}}{\tilde{k}}\bigr\rfloor (\tilde{k}-k))$ and invoking \cref{alg:generation_QuerySet} from Section~\ref{sec:generic-query-generation_PC_coded-DSSs} as follows: 
\begin{IEEEeqnarray*}{c}
  \{Q^{(v)}_1,\dots,Q^{(v)}_{\hat{n}}\} \leftarrow \texttt{Q-Gen}(v, \mu, \kappa,\nu,\hat{n}, \mat{A}_{\kappa\times \hat{n}}, \mat{B}_{(\nu-\kappa) \times \hat{n}}).
\end{IEEEeqnarray*}%
Note that we utilize $\hat{n} \leq n$ databases, including the systematic nodes, in constructing the scheme, while the remaining $n-\hat{n}$ databases are not queried.

\subsection{Sign Assignment and Redundancy Elimination}
Since this scheme is a modified version of the general PPC scheme where we utilize the systematic part of the RS code to recover the requested function evaluation directly, the scheme inherently extend the same redundancy and sign assignment arguments stated in \cref{sec:SignAssignment-RedundancyElimination_PPC}. The only difference between the general PPC scheme and the systematic PPC scheme lies  within the following recovery argument.

\subsection{Recovery and Privacy} 
\label{sec:dec_ppc}

The scheme works as the PPC scheme in \cref{sec:private-linear-computation_coded-DSSs}, however by mixing between the code $\tilde{\code{C}}$ and the storage code $\code{C}$. Due to this mixture, we require a more complicated decoding process. The key idea of the scheme is illustrated in \cref{ex:PMCex_n4k2f2mu3_sys} below.

\begin{example}
  \label{ex:PMCex_n4k2f2mu3_sys}
  Consider the same scenario as in \cref{ex:PMCex_n4k2f2mu3} where $n=4$, $k=2$, and $\tilde{k}=3$. It follows that  $\hat{n}=n=4$, $\nu= \hat{n}-\bigl\lfloor \frac{\hat{n}}{\tilde{k}}\bigr\rfloor(\tilde{k}-k)=3$, $\kappa=k=2$, $\varrho=\bigl\lfloor \frac{\hat{n}}{\tilde{k}}\bigr\rfloor \kappa = 2$,
  and \begin{IEEEeqnarray*}{rCl}
    \mat{\Lambda}_{2,3}^{\textnormal{S,PPC}}=
    \begin{pmatrix}
      1 & 0 & 1 & 1
      \\
      0 & 1 & 1 & 1
      \\
      1 & 1 & 0 & 0
    \end{pmatrix}
  \end{IEEEeqnarray*}
  is a valid PPC systematic achievable rate matrix (see \cref{lem:S-PPC}). We further obtain the PC interference matrices
  \begin{IEEEeqnarray*}{rCl}
    \mat{A}_{2\times 4} = &
    \begin{pmatrix}
      1 &2 &1 &1
      \\
      3 &3 &2 &2
    \end{pmatrix}
    \text{ and } \mat{B}_{1\times 4} = &
    \begin{pmatrix}
      2 &1 &3 &3
    \end{pmatrix}
  \end{IEEEeqnarray*}
  from $\mat{\Lambda}_{2,3}^{\textnormal{S,PPC}}$ using \cref{def:PCinterference-matrices}. For the desired function evaluation $\vmat{X}^{(1)}$, i.e.,~$v=1$, and $\mu=3$ candidate monomial functions evaluations, the resulting query sets are shown in Table~\ref{tab:answers-table2}. Here, similar to \cref{ex:PMCex_n4k2f2mu3}, we deploy the simplified notation $x_{t,j}=C^{(1)}_{t,j}$, $y_{t,j}=C^{(2)}_{t,j}$, and $z_{t,j}=C^{(1)}_{t,j}\cdot C^{(2)}_{t,j}$ for all ${t\in[64]}$, $j\in[5]$, where $u_{a:b,j} \eqdef \{u_{a,j},\ldots,u_{b,j}\}$ for $u =x,y,z$. The PMC rate $\frac{k\nu^{\mu} \HH_{\textnormal{min}}}{\const{D}}=\frac{2\times 3^3}{2\times 4\times 15}  \HH_{\textnormal{min}}=0.45 \cdot \HH_{\textnormal{min}}$  is achievable, where the value of $\HH_{\textnormal{min}} = \HH(\vmat{X}^{(3)})$ depends on the underlying field.\footnote{With some abuse of notation for the sake of simplicity, the generated queries are sets containing their answers.}
	 
\begin{table}[t]
  \centering
  \caption{PMC query sets for $v=1$ after sign assignment and removal of redundant queries for a $[4,2]$ RS-coded DSS with systematic Lagrange encoding storing $f=2$ messages, where the $\mu=3$ candidate monomial functions evaluations are $\{\vmat{X}^{(1)}=\vmat{W}^{(1)},\vmat{X}^{(2)}=\vmat{W}^{(2)},\vmat{X}^{(3)}=\vmat{W}^{(1)}\star\vmat{W}^{(2)}\}$. Blue and red subscripts indicate side information exploitation in rounds $\tau=2$ and $\tau=3$, respectively.}
  \label{tab:answers-table2}
  \vskip -1mm
  \Resize[0.85\columnwidth]{
    \begin{IEEEeqnarraybox}[
      \IEEEeqnarraystrutmode
      \IEEEeqnarraystrutsizeadd{4pt}{2pt}]{v/c/v/c/v/c/v/c/v/c/v}
      \IEEEeqnarrayrulerow\\
      & j && 1 && 2 && 3 && 4\\
      \hline\hline
      & Q^{(1)}_j(\set{D};1)
      && x_{1:4,1},\, x_{9:12,1} &&  x_{5:8,2},\, x_{9:12,2} && x_{1:4,3}, \,x_{5:8,3} && x_{1:4,4},\, x_{5:8,4} &
      \\*\cline{1-11}      
      & Q^{(1)}_j(\set{U};1)
      && y_{1:4,1},\, y_{9:12,1} &&  y_{5:8,2},\, y_{9:12,2} && y_{1:4,3}, \,y_{5:8,3} && y_{1:4,4},\, y_{5:8,4} &
      \\*\cline{1-11}      
      & \multirow{4}{*}{$Q^{(1)}_j(\set{D};2)$}
      && x_{13:14,1}-y_{{\b 5:6},1} &&  x_{17:18,2}-y_{{\b 1:2},2} && x_{13:14,3}-y_{{\b 9:10},3} && x_{13:14,4}-y_{{\b 9:10},4} &
      \\ 
      & 
      && x_{15:16,1}-z_{{\b 5:6},1} &&  x_{19:20,2}-z_{{\b 1:2},2} && x_{15:16,3}-z_{{\b 9:10},3} && x_{15:16,4}-z_{{\b 9:10},4} &
      \\
      &
      && x_{21:22,1}-y_{{\b 7:8},1} &&  x_{21:22,2}-y_{{\b 3:4},2} && x_{17:18,3}-y_{{\b 11:12},3} && x_{17:18,4}-y_{{\b 11:12},4} &
      \\ 
      & 
      && x_{23:24,1}-z_{{\b 7:8},1} &&  x_{23:24,2}-z_{{\b 3:4},2} && x_{19:20,3}-z_{{\b 11:12},3} && x_{19:20,4}-z_{{\b 11:12},4} &
      \\*\cline{1-11}      
      & \multirow{2}{*}{$Q^{(1)}_j(\set{U};2)$}  
      && y_{15:16,1}-z_{13:14,1} &&  y_{19:20,2}-z_{17:18,2} &&  y_{15:16,3}-z_{13:14,3}&&  y_{15:16,4}-z_{13:14,4} &
      \\
      &
      && y_{23:24,1}-z_{21:22,1} &&  y_{23:24,2}-z_{21:22,2} && y_{19:20,3}-z_{17:18,3} && y_{19:20,4}-z_{17:18,4} &
      \\*\cline{1-11}      
     & \multirow{2}{*}{$Q^{(1)}_j(\set{D};3)$}
      && x_{25,1}-y_{{\r 19},1}+z_{{\r 17},1} && x_{26,2}-y_{{\r 15},2}+z_{{\r 13},2}
      && x_{25,3}-y_{{\r 23},3}+z_{{\r 21},3} && x_{25,4}-y_{{\r 23},4}+z_{{\r 21},4} &
      \\
      &
      && x_{27,1}-y_{{\r 20},1}+z_{{\r 18},1} && x_{27,2}-y_{{\r 16},2}+z_{{\r 14},2}
      && x_{26,3}-y_{{\r 24},3}+z_{{\r 22},3} && x_{26,4}-y_{{\r 24},4}+z_{{\r 22},4} &
      \\*\IEEEeqnarrayrulerow
    \end{IEEEeqnarraybox}}
\end{table}

	Now we show that the $\const{L}=k\nu^{\mu}=54$ symbols of the desired function evaluation can be reliably decoded. Note that here we assume that the nodes $j\in\{1,2\}$ are systematic. The goal is to obtain all the desired function evaluation symbols, i.e., the function evaluation symbols for $j\in\{1,2\}$. \\
	{\indent \bf Initialization (Round $\tau=1$):} The following steps are taken.
	\begin{enumerate}
	\item Obtain the desired symbols: 
	From the answers retrieved for the query sets $Q^{(1)}_j({\mathcal D}, 1)$, utilize the information sets $\tilde{\mathcal{I}}_1=\{1,3,4\}$ and $\tilde{\mathcal{I}}_2=\{2,3,4\}$ of $\tilde{\code{C}}$ to decode the symbols of the desired function evaluation $\vmat{X}^{(1)}$ for $j\in \{1,2\}$. In other words, from $x_{1:4,1}$, $x_{1:4,3}$, and $x_{1:4,4}$ we use Lagrange interpolation to obtain $x_{1:4,2}$. Similarly, from $x_{5:8,2}$, $x_{5:8,3}$, and $x_{5:8,4}$ we obtain $x_{5:8,1}$. Finally, from  the information set $\mathcal{I}=\{1,2\}$ of $\code{C}$ we readily have $x_{9:12,1}$ and $x_{9:12,2}$. By the end of this round, we have obtained $k\nu( \kappa^{\mu-1})= 24$ symbols from the desired function evaluation $\vmat{X}^{(1)}$.
	\item Prepare the side information: 
	We prepare the side information symbols retrieved in this round to be used in the next round by the following steps.
	First, for the answers of the query sets $Q^{(1)}_j({\mathcal U}, 1)$, repeat the previous step to decode the undesired symbols $y_{5:8,1}$ and $y_{1:4,2}$. 
	Next, since in this round, due to redundancy elimination, we retrieve symbols of polynomials of degree one, i.e., symbols from the $f=2$ independent files, we can use Lagrange interpolation with $k=2$ symbols from the systematic nodes to obtain coded symbols for $j\notin\{1,2\}$. Accordingly, from $x_{9:12,1}$ and $x_{9:12,2}$ we obtain $x_{9:12,3}$ and $x_{9:12,4}$. Similarly for $y_{9:12,3}$ and $y_{9:12,4}$.
	Finally, using the dependency between $x$, $y$, and $z$ and the available symbols, compute $z_{5:8,1}$, $z_{1:4,2}$, $z_{9:12,3}$, and $z_{9:12,4}$. The obtained symbols are shown in \cref{tab:decoding-table1}--(a).
\end{enumerate}
\begin{table}[h]
  \centering
  \caption{Decoded and computed symbols from the PMC query sets for $v=1$ from \cref{tab:answers-table2}.}
  \label{tab:decoding-table1}
  \vskip -1mm
  \Resize[0.675\columnwidth]{
    \begin{IEEEeqnarraybox}[
      \IEEEeqnarraystrutmode
      \IEEEeqnarraystrutsizeadd{4pt}{2pt}]{v/c/v/c/v/c/v/c/v/c/v}
      \IEEEeqnarrayrulerow\\
      & j && 1 && 2 && 3 && 4\\
      \hline\hline
      & \tilde{Q}^{(1)}_j(\set{D};1)
      && x_{5:8,1} &&  x_{1:4,2} && x_{9:12,3}  && x_{9:12,4} &
      \\*\cline{1-11}      
      & \tilde{Q}^{(1)}_j(\set{U};1)
      && y_{5:8,1},\, z_{5:8,1} && y_{1:4,2},\, z_{1:4,2} &&  y_{9:12,3}, \, z_{9:12,3} &&  y_{9:12,4},\,  z_{9:12,4} &
      \\*\IEEEeqnarrayrulerow
    \end{IEEEeqnarraybox}}
   
(a)
\\[2mm]
\Resize[0.475\columnwidth]{
	\begin{IEEEeqnarraybox}[
		\IEEEeqnarraystrutmode
		\IEEEeqnarraystrutsizeadd{4pt}{2pt}]{v/c/v/c/v/c/v}
		\IEEEeqnarrayrulerow\\
		& j && 1 && 2 \\
		\hline\hline
		& \tilde{Q}^{(1)}_j(\set{D};2)
		&& x_{17:18,1}, \,x_{19:20,1} &&  x_{13:14,2}, \,x_{15:16,2} &
		\\*\cline{1-7}      
		& \tilde{Q}^{(1)}_j(\set{U};2)
		&&  y_{19:20,1}-z_{17:18,1} &&  y_{15:16,2}-z_{13:14,2} & 
		\\*\IEEEeqnarrayrulerow
\end{IEEEeqnarraybox}}

(b)
\\[2mm]
\Resize[0.5\columnwidth]{
	\begin{IEEEeqnarraybox}[
		\IEEEeqnarraystrutmode
		\IEEEeqnarraystrutsizeadd{4pt}{2pt}]{v/c/v/c/v/c/v}
		\IEEEeqnarrayrulerow\\
		& j && 1 && 2 \\
		\hline\hline
		& \tilde{Q}^{(1)}_j(\set{D};3)
		&& {\b x_{25,1}}, \,x_{27,1} &&  {\b x_{26,2}}, \,x_{27,2} &
		\\*\cline{1-7}      
		& \tilde{Q}^{(1)}_j(\set{U};3)
		&&  {\b x_{25,1}}+y_{23,1}-z_{21,1} &&  {\b x_{26,2}}+y_{24,2}-z_{22,2} & 
		\\*\IEEEeqnarrayrulerow
\end{IEEEeqnarraybox}}

(c)
\end{table}

	{\indent \bf Second Round ($\tau=2$):}  The decoding procedure is as follows. 
\begin{enumerate}
	\item Interference cancellation: Utilize the decoded symbols from the set $\tilde{Q}^{(1)}_j({\mathcal U}, 1)$ of \cref{tab:decoding-table1}--(a) to cancel the side information, marked in blue in \cref{tab:answers-table2}, from the answers of the query sets $Q^{(1)}_j({\mathcal D}, 2)$. 
	\item Obtain the desired symbols:
	Similar to the first round, utilize the information sets $\tilde{\mathcal{I}}_1=\{1,3,4\}$ and $\tilde{\mathcal{I}}_2=\{2,3,4\}$ of $\tilde{\code{C}}$ to decode the symbols of the desired function evaluation $\vmat{X}^{(1)}$ for $j\in \{1,2\}$ shown in  $\tilde{Q}^{(1)}_j({\mathcal D}, 2)$ of \cref{tab:decoding-table1}--(b). Together with the symbols directly obtained from $j\in \{1,2\}$, by the end of this round, we have obtained an additional $k\nu ({\mu-1 \choose \tau-1} \kappa^{\mu-\tau}(\nu-\kappa)^{\tau-1})=24$ symbols from the desired function evaluation.
	\item Prepare the side information:
	We prepare the side information $\tau$-sums retrieved in this round to be used in the next round by repeating the previous step to decode the undesired $\tau$-sums $y_{19:20,1}-z_{17:18,1}$ and $y_{15:16,2}-z_{13:14,2}$ of the query sets $\tilde{Q}^{(1)}_j({\mathcal U},2)$. 
	Note that, unlike in the previous round, we do not have enough symbols to utilize Lagrange interpolation to re-encode the $\tau$-sums $y_{19:20,3}-z_{17:18,3}$ and $y_{19:20,4}-z_{17:18,4}$ as they represent polynomials of degree strictly larger than one.
	\end{enumerate}

	{\indent \bf Final Round ($\tau=3$):}  The decoding procedure is as follows. 
	\begin{enumerate}
	\item Interference cancellation: Utilize the decoded $\tau$-sums from the set $\tilde{Q}^{(1)}_j({\mathcal U}, 2)$ of \cref{tab:decoding-table1}--(b) to cancel the side information, marked in red in \cref{tab:answers-table2}, from the query sets $Q^{(1)}_j({\mathcal D}, 3)$ for $j\in\{1,2\}$. As a result we obtain the desired symbols of the set $\tilde{Q}^{(1)}_j({\mathcal D}, 3)$ shown in \cref{tab:decoding-table1}--(c).
	\item Generate new symbols: This step is only required when $\hat{n}-\bigl\lfloor \frac{\hat{n}}{\tilde{k}}\bigr\rfloor \tilde{k} <k$ due to the construction of the interference matrix in the proof of Lemma~\ref{lem:S-PPC}. In particular, the condition is equivalent to $\Gamma < k$. Using the obtained symbols from the previous step, colored in \cref{tab:decoding-table1} for $\tilde{Q}^{(1)}_j({\mathcal D}, 3)$ with blue, along with the side information downloaded in the previous round in $Q^{(1)}_j({\mathcal U}, 2)$, generate $\bigl\lfloor \frac{\hat{n}}{\tilde{k}}\bigr\rfloor\tilde{k}-(n-k)=1$ new $\tau$-sums with identical indices to the $\tau$-sums retrieved from the nonsystematic nodes. These newly generated symbols are shown in $\tilde{Q}^{(1)}_j({\mathcal U}, 3)$.
	\item Obtain the desired symbols: Here, we reverse the order of operation of the previous rounds where we use Lagrange interpolation first and then cancel the side information. First, utilize the information sets $\tilde{\mathcal{I}}_1=\{1,3,4\}$ and $\tilde{\mathcal{I}}_2=\{2,3,4\}$ of $\tilde{\mathcal{C}}$ to decode the $\tau$-sums containing the desired function evaluation for $j\in \{1,2\}$. As a result, we obtain ${\b x_{26,1}}+y_{24,1}-z_{22,1}$ and ${\b x_{25,2}}+y_{23,2}-z_{21,2}$. Next, cancel the side information from the $\tau$-sums directly obtained from $Q^{(1)}_j({\mathcal U}, 2)$ for $j\in \{1,2\}$. Finally, by the end of this round, we have obtained the final $k\nu ({\mu-1 \choose \tau-1} \kappa^{\mu-\tau}(\nu-\kappa)^{\tau-1})=6$ symbols from the desired function evaluation $\vmat{X}^{(1)}$.
	\end{enumerate}
	In summary, the total number of desired function evaluation symbols obtained from this decoding process is 
	$k \nu \sum_{\tau=1}^{\mu}{\mu-1 \choose \tau-1}\kappa^{\mu-\tau}(\nu-\kappa)^{\tau-1} = k\nu^{\mu} = 54$.
	\QEDA
\end{example}

\begin{remark} 
  \label{rem:1}
  The systematic scheme above reduces to the systematic PPC scheme presented in \cite{ObeadLinRosnesKliewer19_1} if and only if $n - \tilde{k} \leq k$. In particular, this happens if and only if  the storage code rate $k / n \geq k /(k+g(k-1)+1)$. Otherwise, $\hat{n}$ is smaller than $n$ and the PPC rate becomes larger than the one for the systematic scheme in  \cite{ObeadLinRosnesKliewer19_1}.
\end{remark}

  \cref{rem:1} can be easily verified with the following argument. The two schemes are equivalent \emph{if and only if} $n = \hat{n}$ and $\nu = k + \min\{k,n-\tilde{k}\}$ (see \cite[Thm.~2]{ObeadLinRosnesKliewer19_1}). Assume that $n - \tilde{k} \leq k$. Then, $1 \leq \bigl \lfloor \frac{n}{\tilde{k}} \bigr\rfloor \leq \bigl \lfloor 1 + \frac{k}{\tilde{k}} \bigr\rfloor \leq 2$. If $\bigl\lfloor \frac{n}{\tilde{k}}  \bigr\rfloor=1$, then it follows directly from \eqref{eq:systematic_n} and Lemma~\ref{lem:nu_formula} that $n = \hat{n}$ and $\nu = k+ n - \tilde{k} = k + \min\{k,n-\tilde{k}\}$. 
  Otherwise, if $\bigl\lfloor \frac{n}{\tilde{k}}  \bigr\rfloor=2$, then $k = \tilde{k}$, $3k> n\geq 2k$, and from \eqref{eq:systematic_n}, we have $\hat{n} = k + \tilde{k} = 2k$. 
  Since, by assumption, we have $n-\tilde{k} \leq k$, it follows that $n\leq k+\tilde{k}= 2k$. Combining the two inequalities over $n$, specifically, $ 3k>n\geq 2k$ and $n\leq 2k$, we conclude that $n=2k$ and  
   it holds that $n = \hat{n}$. Now, from Lemma~\ref{lem:nu_formula}, $\nu = 2k = k + \min\{k,n-\tilde{k}\}$, and the equivalence of the two schemes follows. The ``only-if'' part follows in a similar manner. Finally, the lower bound on the storage code rate follows directly from the condition $n - \tilde{k} \leq k$.

\subsection{Achievable PPC Rate} 

Using \cref{lem:PPCredundancy,lem:S-PPC}, the following theorem follows.

\begin{theorem}
  \label{thm:RS_PPC_rate}
  Consider a DSS that uses an $[n,k]$ RS code $\code{C}$ to store $f$ messages over $n$ noncolluding databases using systematic Lagrange encoding. Let $\mu \in [f:\mu_g(f)]$ be the number of candidate polynomial functions evaluations of degree at most $g$, including the $f$ independent messages. Then, the PPC rate
  \begin{IEEEeqnarray}{rCl}
    \const{R}^{\textnormal{S}}_\textnormal{PPC}&=& \begin{cases}  \frac{1}{ f} \HH_{\textnormal{min}} & \text{if $n \leq g(k-1)+1$}, \\
      \frac{\frac{k}{\hat{n}}(\frac{\nu-\kappa}{\kappa}) \HH_{\textnormal{min}}}
      {1-  \bigl(\frac{\kappa}{\nu} \bigr)^{\min\{\mu,{\const M}_g(f)\}} - (\min\{\mu,{\const M}_g(f)\}-f)\bigl(1-\frac{\kappa}{\nu} \bigr)\bigl(\frac{\kappa}{\nu} \bigr)^{\mu-1}} & \text{otherwise}
    \end{cases}
    \label{eq:PPCrate_SysLagrangeCoded-DSS-2}
  \end{IEEEeqnarray} 
  with $\frac{\kappa}{\nu} = \frac{k}{\hat{n}- \bigl\lfloor\frac{\hat{n}}{\tilde{k}} \bigr\rfloor (\tilde{k}-k)}$ and $\hat{n}$ as defined in \eqref{eq:systematic_n},  is achievable.
\end{theorem}
\begin{IEEEproof}
  From \eqref{eq:query-card_PC}  and by removing redundant $\tau$-sums from the query sets according to Lemma~\ref{lem:PPCredundancy}, the  achievable PPC rate becomes
  \begin{IEEEeqnarray}{rCl}
    \label{eq:Cap_proof_syst_first}
    \const{R} & \overset{(a)}{=} &\frac{k \nu^\mu \HH_{\textnormal{min}}}{\hat{n}\Bigl({\mu \choose 1}-{\mu-f\choose 1}\Bigr){\kappa}^{\mu} + {\hat{n}\sum_{\tau=2}^{\mu}\rho(\mu,\tau) {\kappa}^{\mu-\tau+1}{(\nu-\kappa)}^{\tau-1}}}
      \nonumber \\
      & = & \frac{ k \nu^{\mu} \HH_{\textnormal{min}}}{\hat{n} \kappa \Bigl[ f{\kappa}^{\mu-1} + \sum_{\tau=2}^{\mu}
      \rho(\mu,\tau) {\kappa}^{\mu-\tau}{(\nu-\kappa)}^{\tau-1} \Bigr]}, 
  \end{IEEEeqnarray}
  where $(a)$ follows from the PPC rate in Definition~\ref{def:def_info-PCrate}, \eqref{eq:query-card_PC}, and Lemma~\ref{lem:PPCredundancy}. 
  
  Now, we first consider the case where $\nu=\kappa$ and show that it is equivalent to $n \leq g(k-1)+1$.  Assume that $\nu = \kappa = k$. Then, for the first case of \eqref{eq:nu_cases1} it follows that $\tilde{k}=n$. For the second and third cases of \eqref{eq:nu_cases1}, to obtain $\nu=k$, we must have $\bigl \lfloor \frac{n}{\tilde{k}} \bigr\rfloor = 1$ or $\bigl \lfloor \frac{n}{\tilde{k}} \bigr\rfloor = 0$, respectively, which violates the condition of the second case and is never true for the third case.
  Since, by Proposition~\ref{prop:RSstarProduct},  $\tilde{k} = \min\{g(k-1)+1,n\} = n$, it follows that $n \leq g(k-1)+1$. Conversely, if $n \leq g(k-1)+1$, then $\tilde{k}=\min\{g(k-1)+1,n\} = n$, and it follows from  \eqref{eq:nu_cases1} (the first case) that $\nu = \kappa$. Hence, in summary, we have shown that $\nu = \kappa$ is equivalent to $n \leq g(k-1)+1$.
 As a result, for $n \leq g(k-1)+1$, it  follows directly from \eqref{eq:Cap_proof_syst_first} that $\const{R} = \frac{k}{\hat{n} f} \HH_{\textnormal{min}}$. Moreover, it can be seen in this case that the proposed systematic PPC scheme reduces  to the trivial scheme for which all  the $f$ independent files are downloaded and the desired function evaluation is performed offline. However, similar to the general PPC scheme, the proposed systematic PPC scheme requires an unnecessarily high redundancy to decode the $f$ files, i.e., $\tilde{k}=\hat{n}$ instead of $\tilde{k}=k$. As a result, for the case of $n\leq g(k-1)+1$, we again opt out of any other achievable scheme and achieve the PPC rate $\frac{1}{f} \HH_{\textnormal{min}}$ by simply downloading all $f$ files and performing the desired function evaluation offline. 
  
  On the other hand, if $\nu >\kappa$, or equivalently, $n > g(k-1)+1$, then from \eqref{eq:Cap_proof_syst_first} we have
  \begin{IEEEeqnarray}{rCl}
    \const{R} &  \overset{(b)}{=} & 
    \frac{k\HH_{\textnormal{min}}}{\hat{n}\kappa}   \inv{\left[ \frac{f {\kappa}^{\mu-1}}{\nu^\mu}  + \frac{1}{\nu^{\mu}(\nu-\kappa)}\sum_{\tau=2}^{\mu}
        \rho(\mu,\tau) {\kappa}^{\mu-\tau}{(\nu-\kappa)}^{\tau} \right]}   \nonumber
    \\
    & = &  \frac{k(\nu-\kappa)\HH_{\textnormal{min}}}{\hat{n}\kappa}  \inv{\left[\frac{f(\nu-\kappa)}{\nu} \Bigl( \frac{\kappa}{\nu}\Bigr)^{\mu-1}   + \frac{1}{\nu^{\mu}}\sum_{\tau=2}^{\mu}
        \rho(\mu,\tau) {\kappa}^{\mu-\tau}{(\nu-\kappa)}^{\tau} \right]} \nonumber
    \\
    &  & \qquad\quad \vdots \nonumber
    \\[2mm]
    & \overset{(c)}{=}& \frac{ \frac{k}{\hat{n}}(\frac{\nu-\kappa}{\kappa}) \HH_{\textnormal{min}}}
    {1-  \bigl(\frac{\kappa}{\nu}\bigr)^{\min\{\mu,{\const M}_g(f)\}} - (\min\{\mu,{\const M}_g(f)\}-f)\bigl(1-\frac{\kappa}{\nu} \bigr)\bigl(\frac{\kappa}{\nu} \bigr)^{\mu-1}}, \nonumber
  \end{IEEEeqnarray} 
  where $(b)$ follows since $\nu > k$ and $(c)$ results from following similar steps as in the proof of the achievable PPC rate of \cref{thm:PMCrate_LagrangeCoded-DSS} in \cref{sec:PPCrate_codes}. 
\end{IEEEproof}
\begin{corollary} 
\label{cor:RS_PPC_rate}
Consider a DSS that uses an $[n,k]$ RS code $\code{C}$ to store $f$ messages over $n$ noncolluding databases using systematic Lagrange encoding. Let $\mu \in [f:\upmu_g(f)]$ be the number of candidate polynomial functions evaluations of degree at most $g$, including the $f$ independent messages. Then, the PPC rate
\begin{IEEEeqnarray}{rCl}
  \const{R}^{\textnormal{S}}_{\textnormal{PPC},\infty}& = &\begin{cases} \frac{1}{n}\bigl({\max\{n-g(k-1)-1, 0\}}\bigr) \HH_{\textnormal{min}}  & \text{if }   \bigl\lfloor \frac{n}{\tilde{k}} \bigr\rfloor = 1 \text{ and } n-\bigl\lfloor \frac{n}{\tilde{k}} \bigr\rfloor \tilde{k} < k,
    \\
    \frac{1}{\hat{n}}\bigl(\bigl\lfloor \frac{n}{g(k-1)+1} \bigr\rfloor k -k \bigr) \HH_{\textnormal{min}} & \text{if }   \bigl\lfloor \frac{n}{\tilde{k}} \bigr\rfloor > 1 \text{ and } n-\bigl\lfloor \frac{n}{g(k-1)+1} \bigr\rfloor(g(k-1)+1) < k,
    \\
    \frac{1}{\hat{n}}\bigl(\lfloor \frac{n}{g(k-1)+1} \bigr\rfloor k \bigr) \HH_{\textnormal{min}} & \text{if }   \bigl\lfloor \frac{n}{\tilde{k}} \bigr\rfloor  \geq  1 \text{ and }n-\bigl\lfloor \frac{n}{g(k-1)+1} \bigr\rfloor(g(k-1)+1)\geq k,
  \end{cases} 	
  \label{eq:PPCrate_SysLagrangeCoded-DSS-2_asymp}
\end{IEEEeqnarray}
with $\hat{n}$ as defined in \eqref{eq:systematic_n}, is asymptotically achievable for $f \to \infty$.
\end{corollary}
\begin{IEEEproof}
  If $n \leq g(k-1)+1$, then it follows from \eqref{eq:PPCrate_SysLagrangeCoded-DSS-2}  that the PPC rate approaches zero as $f \to \infty$, which is in accordance with \eqref{eq:PPCrate_SysLagrangeCoded-DSS-2_asymp} (first case, since $ \bigl\lfloor \frac{n}{\tilde{k}} \bigr\rfloor = 1$ and $n-\bigl\lfloor \frac{n}{\tilde{k}} \bigr\rfloor \tilde{k}=0 < k$). Otherwise, if $n > g(k-1)+1$, the result follows directly from \eqref{eq:PPCrate_SysLagrangeCoded-DSS-2} by taking the limit $f \to \infty$ and using \eqref{eq:nu_cases1} and the fact  (see Proposition~\ref{prop:RSstarProduct}) that $\tilde{k} = \min\{g(k-1)+1,n\} = g(k-1)+1$.
\end{IEEEproof}
Note that when
  $n-\tilde{k} \leq k$, the asymptotic PPC rate in \eqref{eq:PPCrate_SysLagrangeCoded-DSS-2_asymp} is equal to the rate of the systematic scheme from \cite[Thm.~3]{Karpuk18_1}, \cite{RavivKarpuk19_2} when $\HH_{\textnormal{min}}=1$. This difference is due to the simplified rate definition used in \cite{Karpuk18_1,RavivKarpuk19_2}. However, for the case when $n-\tilde{k} > k$, with the simplified rate definition, i.e., for $\HH_{\textnormal{min}}=1$, the asymptotic PPC rate in \eqref{eq:PPCrate_SysLagrangeCoded-DSS-2_asymp} is larger compared to the PPC rate of the systematic scheme from \cite[Thm.~3]{Karpuk18_1}, \cite{RavivKarpuk19_2}. See also Remark~\ref{rem:1}.

\section{Numerical Results} 
\label{sec:numerical-results}

\begin{figure}[t]
	\centering
	\begin{subfigure}{.49\textwidth}
		\centering
		\resizebox{0.95\textwidth}{!} 
		{
%
%
\definecolor{mycolor1}{rgb}{1.00000,0.00000,1.00000}%
\begin{tikzpicture}

\begin{axis}[%
width=4.967in, 
height=4.329in,
at={(1.95in,0.779in)},
scale only axis,
separate axis lines,
every outer x axis line/.append style={black},
every x tick label/.append style={font=\color{black}},
every x tick/.append style={black},
xmin=1,
xmax=8,
xtick={1, 2, 3, 4, 5, 6, 7, 8},
xlabel={Number of messages $f$},
every outer y axis line/.append style={black},
every y tick label/.append style={font=\color{black}},
every y tick/.append style={black},
ymin=0.1,
ymax=0.6,
ytick={0, 0.1, 0.2, 0.3, 0.4, 0.5, 0.6, 0.7, 0.8},
ylabel={PMC rate $\const{R}$},
axis background/.style={fill=white},
xmajorgrids,
ymajorgrids,
grid style={gray,opacity=0.75,dotted},
grid style={gray,opacity=0.75,dotted},
legend style={legend cell align=left, align=left, draw=white!15!black, fill=white}
]

\addplot [color=mycolor1, line width=1.5pt, mark=square*, mark options={mycolor1,fill=white}]
table[row sep=crcr]{%
	1	0.579380164285695\\
	2	0.450629016666652\\
	3	0.423725791791031\\
	4	0.416619279559735\\
	5	0.414632421594621\\
	6	0.414068223886962\\
	7	0.41390730644965\\
	8	0.41386135300907\\
};
\addlegendentry{\small Converse bound (Thm.~\ref{thm:converse_bound})}

\addplot [color=red, mark=*,  line width=1.5pt,mark options={red,fill=white}]
  table[row sep=crcr]{%
1	0.38625344285713\\
2	0.237915473653148\\
3	0.22168944039731\\
4	0.220738565673707\\
5	0.220716455604713\\
6	0.220716253803995\\
7	0.220716253062335\\
8	0.220716253061218\\
};
\addlegendentry{\small RS-L (Thm.~\ref{thm:PMCrate_LagrangeCoded-DSS})} 

\addplot [color=black, line width=1.5pt, mark options={black,fill=white}]
  table[row sep=crcr]{%
1	0.220716253061217\\
2	0.220716253061217\\
3	0.220716253061217\\
4	0.220716253061217\\
5	0.220716253061217\\
6	0.220716253061217\\
7	0.220716253061217\\
8	0.220716253061217\\
};
\addlegendentry{\small PPC scheme \cite{RavivKarpuk19_2}}

\addplot [color=blue, line width=1.5pt, mark=diamond*, mark options={blue,fill=white}]
table[row sep=crcr]{%
	1	0.413842974489782\\
	2	0.26312498378038\\
	3	0.248958414239293\\
	4	0.248316449804179\\
	5	0.248305848852434\\
	6	0.248305784839244\\
	7	0.248305784693995\\
	8	0.248305784693869\\
};
\addlegendentry{\small Sys.~RS-L (Thm.~\ref{thm:RS_PPC_rate})} 

\addplot [color=green, line width=1.5pt, mark options={green,fill=white}]
table[row sep=crcr]{%
	1	0.165537189795913\\
	2	0.165537189795913\\
	3	0.165537189795913\\
	4	0.165537189795913\\
	5	0.165537189795913\\
	6	0.165537189795913\\
	7	0.165537189795913\\
	8	0.165537189795913\\
};
\addlegendentry{\small Sys.~PPC scheme \cite{Karpuk18_1}} 

\end{axis}
\end{tikzpicture}%
}
		\vspace{-0.5ex}
		\caption{For $\mu={\const{M}}_2(f)$.}
		\label{fig:PMC_all}
	\end{subfigure}%
	\begin{subfigure}{.49\textwidth}
		\centering
		\resizebox{0.95\textwidth}{!} 
		{
%
%
\definecolor{mycolor1}{rgb}{1.00000,0.00000,1.00000}%
\begin{tikzpicture}

\begin{axis}[%
width=4.967in, 
height=4.329in,
at={(1.95in,0.779in)},
scale only axis,
separate axis lines,
every outer x axis line/.append style={black},
every x tick label/.append style={font=\color{black}},
every x tick/.append style={black},
xmin=1,
xmax=8,
xtick={1, 2, 3, 4, 5, 6, 7},
xlabel={Number of messages $f$},
every outer y axis line/.append style={black},
every y tick label/.append style={font=\color{black}},
every y tick/.append style={black},
ymin=0.2,
ymax=1,
ylabel={PMC rate $\const{R}$},
axis background/.style={fill=white},
xmajorgrids,
ymajorgrids,
grid style={gray,opacity=0.75,dotted},
legend style={legend cell align=left, align=left, draw=white!15!black, fill=white}
]
\addplot [color=mycolor1, line width=1.5pt, mark=square*, mark options={mycolor1,fill=white}]
table[row sep=crcr]{%
	1	1\\
	2	0.70444313178854\\
	3	0.662386825413105\\
	4	0.651277612408273\\
	5	0.64817166850409\\
	6	0.647289689791154\\
	7	0.647038136563762\\
	8	0.64696630012092\\
};
\addlegendentry{\small Converse bound (Thm.~\ref{thm:converse_bound})}

\addplot [color=red, mark=*,  line width=1.5pt,mark options={red,fill=white}]
table[row sep=crcr]{%
	1	0.666666666666666\\
	2	0.422665879073124\\
	3	0.356064971634178\\
	4	0.345683730333475\\
	5	0.345048318405106\\
	6	0.34503350636826\\
	7	0.345033371169571\\
	8	0.345033370672687\\
};
\addlegendentry{\small RS-L (Thm.~\ref{thm:PMCrate_LagrangeCoded-DSS})} 

\addplot [color=black, line width=1.5pt, mark options={black,fill=white}]
  table[row sep=crcr]{%
1	0.345033370671938\\
2	0.345033370671938\\
3	0.345033370671938\\
4	0.345033370671938\\
5	0.345033370671938\\
6	0.345033370671938\\
7	0.345033370671938\\
8	0.345033370671938\\
};
\addlegendentry{\small PPC scheme \cite{RavivKarpuk19_2}}

\addplot [color=blue, line width=1.5pt, mark=diamond*, mark options={blue,fill=white}]
table[row sep=crcr]{%
	1	0.714285714\\
	2	0.462098264292774\\
	3	0.39710860465152\\
	4	0.388569987168796\\
	5	0.388169210702191\\
	6	0.388162582124122\\
	7	0.388162542096833\\
	8	0.388162542006009\\
};
\addlegendentry{\small Sys.~RS-L (Thm.~\ref{thm:RS_PPC_rate})} 

\addplot [color=green, line width=1.5pt, mark options={green,fill=white}]
table[row sep=crcr]{%
	1	0.258775028003953\\
	2	0.258775028003953\\
	3	0.258775028003953\\
	4	0.258775028003953\\
	5	0.258775028003953\\
	6	0.258775028003953\\
	7	0.258775028003953\\
	8	0.258775028003953\\
};
\addlegendentry{\small Sys.~PPC scheme \cite{Karpuk18_1}}

\end{axis}
\end{tikzpicture}%
}
		\vspace{-0.5ex}
		\caption{For $\mu=\widetilde{\const{M}}_2(f)$.}
		\label{fig:PMC_unparl}
	\end{subfigure}%
	\caption{Achievable PMC rates as a function of the number of messages $f$ for $n=7$, $k=2$, and $g=2$.}
	\label{fig:PMC}
\end{figure}
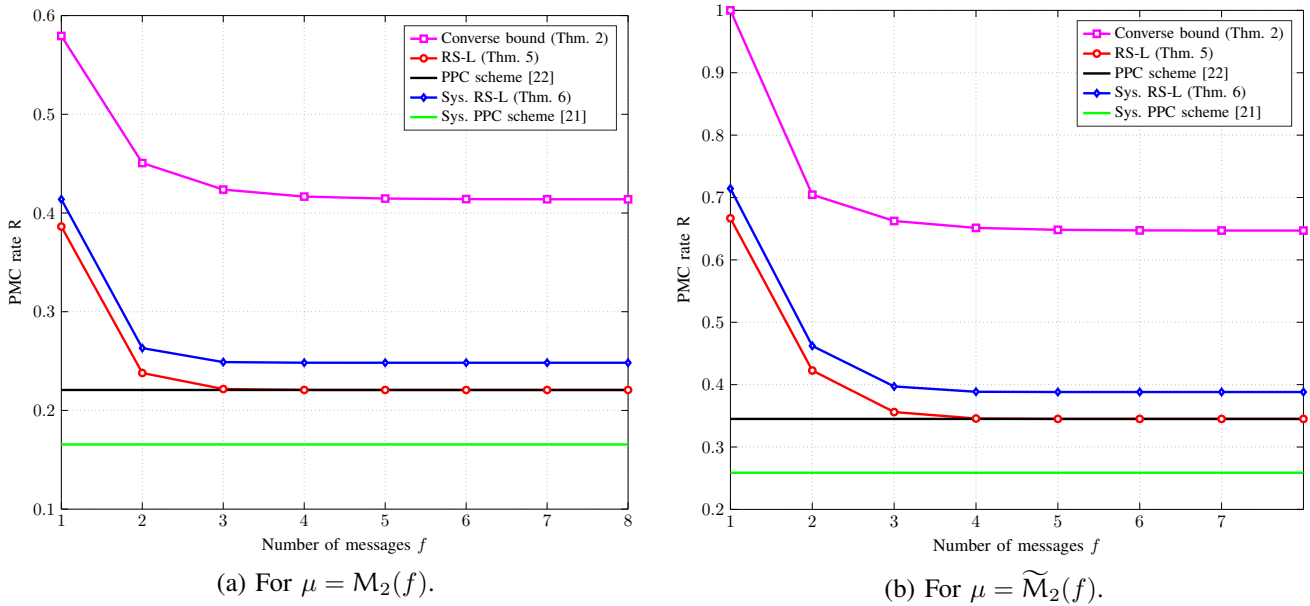

In Fig.~\ref{fig:PMC_all}, we compare the PMC rates of Theorems~\ref{thm:PMCrate_LagrangeCoded-DSS} and \ref{thm:RS_PPC_rate} to those of the schemes from \cite{Karpuk18_1,RavivKarpuk19_2} as well as the converse bound from Theorem~\ref{thm:converse_bound} (using the exact information-theoretic rate from \cref{def:def_info-PCrate}) for $n=7$, $k=2$, $g=2$, and computations over $\Field_3$.
 The scheme from Theorem~\ref{thm:PMCrate_LagrangeCoded-DSS} shows improved  performance for a low number of messages $f$, while the systematic scheme from Theorem~\ref{thm:RS_PPC_rate} shows improved performance for all values of $f$, even in the asymptotic case of $f \to \infty$. As the number of messages $f$ grows, the curve from Theorem~\ref{thm:PMCrate_LagrangeCoded-DSS} converges to the rate from \cite{RavivKarpuk19_2}, as can be seen from \cref{cor:PMCrate_LagrangeCoded-DSS}, while the asymptotic performance of the systematic scheme follows from  \cref{cor:RS_PPC_rate} (second case). The converse bound from Theorem~\ref{thm:converse_bound} shows a relatively large gap for all values of $f$.
For comparison, we also plot in Fig.~\ref{fig:PMC_unparl} the PMC rate when parallel monomials are excluded from the candidate function set. As for the  case when parallel monomials are included we observe improved performance for the systematic scheme from Theorem~\ref{thm:RS_PPC_rate} for all values of $f$. %

\begin{figure}[t]
	\centering
	\begin{subfigure}{.49\textwidth}
		\centering
		\resizebox{0.95\textwidth}{!} 
		{

\definecolor{mycolor1}{rgb}{0.00000,1.00000,1.00000}%
\definecolor{mycolor2}{rgb}{0.60000,0.00000,0.60000}%
\begin{tikzpicture}

\begin{axis}[%
width=4.967in,
height=4.329in,
at={(1.95in,0.779in)},
scale only axis,
every outer x axis line/.append style={black},
every x tick label/.append style={font=\color{black}},
every x tick/.append style={black},
xmin=0.05,
xmax=0.5,
xlabel={Storage code rate $\alpha=k/n$},
every outer y axis line/.append style={black},
every y tick label/.append style={font=\color{black}},
every y tick/.append style={black},
ymin=0.1,
ymax=0.65,
ylabel={PMC rate $\const{R}$},
axis background/.style={fill=white},
axis x line*=bottom,
axis y line*=left,
xmajorgrids,
ymajorgrids,
grid style={gray,opacity=0.75,dotted},
legend style={legend cell align=left, align=left, draw=white!15!black, fill=white}
]

\addplot [color=mycolor2, dashed, line width=1.5pt, mark=diamond*, mark options={solid, mycolor2}]
table[row sep=crcr]{%
	0.1	0.5\\
	0.105263157894737	0.5\\
	0.111111111111111	0.5\\
	0.117647058823529	0.5\\
	0.125	0.5\\
	0.133333333333333	0.5\\
	0.142857142857143	0.5\\
	0.153846153846154	0.5\\
	0.166666666666667	0.5\\
	0.181818181818182	0.5\\
	0.2	0.5\\
	0.222222222222222	0.5\\
	0.25	0.5\\
	0.285714285714286	0.5\\
	0.333333333333333	0.5\\
	0.4	0.5\\
	0.5	0.5\\
};
\addlegendentry{\small Trivial scheme: $\frac{1}{f} H_{\text{min}}$}

\addplot [color=red, line width=1.5pt, mark=o, mark options={red, fill=white}]
  table[row sep=crcr]{%
0.1	0.567442289345916\\
0.105263157894737	0.562341691586083\\
0.111111111111111	0.556701030927835\\
0.117647058823529	0.550432656504346\\
0.125	0.543429770226457\\
0.133333333333333	0.535561268209083\\
0.142857142857143	0.526665021523951\\
0.153846153846154	0.516539164089487\\
0.166666666666667	0.504930966469428\\
0.181818181818182	0.491523147681875\\
0.2	0.475918522748905\\
0.222222222222222	0.457627118644068\\
0.25	0.436069413392952\\
0.285714285714286	0.410637933983239\\
0.333333333333333	0.380952380952381\\
0.4	0.347801892042293\\
0.5	0.317224287484511\\
};
\addlegendentry{\small RS-L (Thm.~\ref{thm:PMCrate_LagrangeCoded-DSS})} 

\addplot [color=black, line width=1.5pt, mark=o, mark options={black,fill=white}]
  table[row sep=crcr]{%
0.1	0.566666666666667\\
0.105263157894737	0.56140350877193\\
0.111111111111111	0.555555555555556\\
0.117647058823529	0.549019607843137\\
0.125	0.541666666666667\\
0.133333333333333	0.533333333333333\\
0.142857142857143	0.523809523809524\\
0.153846153846154	0.512820512820513\\
0.166666666666667	0.5\\
0.181818181818182	0.484848484848485\\
0.2	0.466666666666667\\
0.222222222222222	0.444444444444444\\
0.25	0.416666666666667\\
0.285714285714286	0.380952380952381\\
0.333333333333333	0.333333333333333\\
0.4	0.266666666666667\\
0.5	0.166666666666667\\
};
\addlegendentry{\small PPC scheme \cite{RavivKarpuk19_2}}

\addplot [color=blue, line width=1.5pt, mark=diamond*, mark options={solid, blue}]
table[row sep=crcr]{%
	0.1	0.600679056468906\\
	0.105263157894737	0.589448150394178\\
	0.111111111111111	0.589448150394178\\
	0.117647058823529	0.589448150394178\\
	0.125	0.573815644509732\\
	0.133333333333333	0.573815644509732\\
	0.142857142857143	0.573815644509732\\
	0.153846153846154	0.550833781603012\\
	0.166666666666667	0.550833781603012\\
	0.181818181818182	0.550833781603012\\
	0.2	0.514830508474576\\
	0.222222222222222	0.514830508474576\\
	0.25	0.514830508474576\\
	0.285714285714286	0.457142857142857\\
	0.333333333333333	0.457142857142857\\
	0.4	0.457142857142857\\
	0.5	0.372699386503067\\
};

\addlegendentry{\small Sys.~RS-L (Thm.~\ref{thm:RS_PPC_rate})} 

\addplot [color=green, line width=1.5pt, mark=diamond*, mark options={green,fill=white}] 
table[row sep=crcr]{%
	0.1	0.1\\
	0.105263157894737	0.105263157894737\\
	0.111111111111111	0.111111111111111\\
	0.117647058823529	0.117647058823529\\
	0.125	0.125\\
	0.133333333333333	0.133333333333333\\
	0.142857142857143	0.142857142857143\\
	0.153846153846154	0.153846153846154\\
	0.166666666666667	0.166666666666667\\
	0.181818181818182	0.181818181818182\\
	0.2	0.2\\
	0.222222222222222	0.222222222222222\\
	0.25	0.25\\
	0.285714285714286	0.285714285714286\\
	0.333333333333333	0.333333333333333\\
	0.4	0.4\\
	0.5	0.25\\
};
\addlegendentry{\small Sys.~PPC scheme \cite{Karpuk18_1}}

\end{axis}
\end{tikzpicture}%
}
		\vspace{-0.5ex}
		\caption{For $f=2$ and $k=2$.}
		\label{fig:PMC_compare1}
	\end{subfigure}%
	\begin{subfigure}{.49\textwidth}
		\centering
		\resizebox{0.95\textwidth}{!} 
		{\input{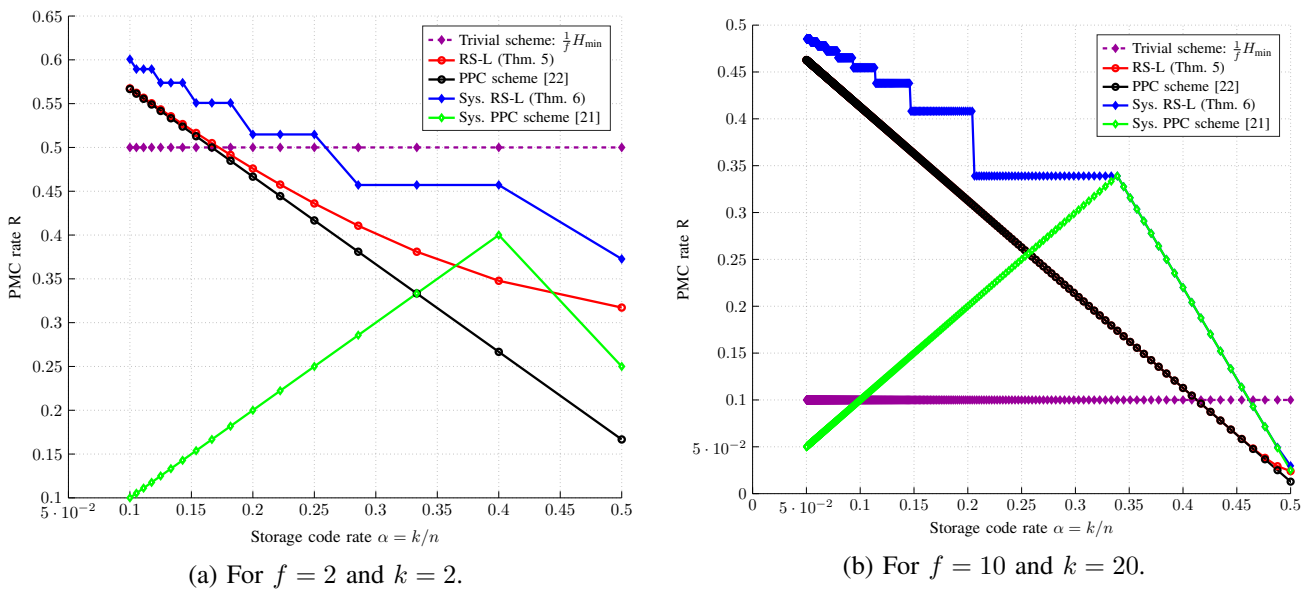}}
		\vspace{-0.5ex}
		\caption{For $f=10$ and $k=20$.}
		\label{fig:PMC_compare_asympt}
	\end{subfigure}%
	\caption{PMC rates as a function of the storage code rate $\alpha=k/n$ for fixed $f$, $k$, $g=2$, and $\mu = \const{M}_{2}(f)$. For the sake of simplicity, we assume $\HH_{\mathsf {min}}=1$.}
	\label{fig:PMC_compare}
\end{figure}

 In Fig.~\ref{fig:PMC_compare}, we compare the PMC rates of Theorems~\ref{thm:PMCrate_LagrangeCoded-DSS} and
	\ref{thm:RS_PPC_rate} and those of the schemes from \cite{Karpuk18_1,RavivKarpuk19_2} for various values of the storage code rate $\alpha=k/n$, fixed $k$, $g=2$, $\mu = \const{M}_2(f)$,   $f=2$ for \cref{fig:PMC_compare1}, and $f=10$ for \cref{fig:PMC_compare_asympt}. For a small number of files ($f=2$), the proposed schemes show improved performance for all code rates, while for a relatively large number of files ($f=10$), the systematic scheme from Theorem~\ref{thm:RS_PPC_rate} shows improved performance up to some code rate (see Remark~\ref{rem:1}). %
	Observe that when neglecting the computational cost at the user, the trivial scheme which downloads all the $f$ files and computes the desired function evaluation offline outperforms all considered PPC schemes when the code rate is above some threshold that depends on both $f$ and $g$. For $f=10$ the code rate needs to be close to $1/2$ for the trivial scheme to be the best. Note that the curve for the systematic scheme follows a staircase in which there are $\tilde{k}$ points on each horizontal line of the staircase. This follows directly from the term $\bigr \lfloor \frac{n}{\tilde{k}} \bigr \rfloor$ in the definition of $\hat{n}$ in  \eqref{eq:systematic_n}.
	%

\section{Conclusion}
\label{sec:conclusion}
We have provided the capacity of PLC from coded DSSs, where data is encoded and stored using an arbitrary linear code from the class of MDS-PIR capacity-achieving codes. Interestingly, the capacity of PLC is equal to the corresponding MDS-PIR capacity. Thus, privately retrieving arbitrary linear combinations of the stored messages does not incur any overhead in rate compared to retrieving a single message from the databases. For the PPC problem, we have presented two PPC schemes for RS-coded DSSs with Lagrange encoding  showing improved computation rates compared to the best known PPC schemes from the literature when the number of messages is small. %
Asymptotically, as the number of messages tends to infinity, the rate of our RS-coded nonsystematic PPC scheme approaches the rate of the best known nonsystematic PPC scheme. However, for systematically RS-coded DSSs, our scheme significantly outperforms all known PPC schemes up to some specific storage code rate that depends on the maximum degree of the candidate polynomials. Finally, a general converse bound on the PPC rate  was derived and compared to the achievable rates of the proposed schemes.


\appendices

\vspace{-0.5ex} 
\section{Proof of Lemma~\ref{lem:uniform_dist}}
\label{sec:proof_uniform-dist}

Since each linear function $\vmat{X}^{(v)}=\bigl(X^{(v)}_1,\ldots,X^{(v)}_\const{L}\bigr)$, $v\in[\mu]$, contains $\const{L}$ independent and identically distributed (i.i.d.) symbols, it is clear that $\forall\,l\in[\const{L}]$,
\begin{IEEEeqnarray*}{rCl}
\HP{\vmat{X}^{(1)},\ldots,\vmat{X}^{(\mu)}}& = & \const{L}\HP{X^{(1)}_l,\ldots,X^{(\mu)}_l},
\\
\HP{\vmat{W}^{(1)},\ldots,\vmat{W}^{(f)}}& = & \const{L}\HP{W^{(1)}_l,\ldots,W^{(f)}_l}.
\end{IEEEeqnarray*}
Let $\set{J}\eqdef\{j_1,\ldots,j_h\}$ for some $h\in [r]$. We have
\begin{IEEEeqnarray}{rCl}
\IEEEeqnarraymulticol{3}{l}{%
\Prv{X^{(i_1)}_l,\ldots,X^{(i_h)}_l}}\nonumber\\*\quad%
& = &\sum_{w^{\comp{\set{J}}}_l}\Prv{{W^{\comp{\set{J}}}_l}=w^{\comp{\set{J}}}_l}\cdot\Prvcond{X^{(i_1)}_l,\ldots,X^{(i_h)}_l}{W^{\comp{\set{J}}}_l=w^{\comp{\set{J}}}_l}
\nonumber\\
& = &\sum_{w^{\comp{\set{J}}}_l}\Prv{{W^{\comp{\set{J}}}_l}=w^{\comp{\set{J}}}_l}\cdot\Prvcond{W^{(j_1)}_l,\ldots,W^{(j_h)}_l}{W^{\comp{\set{J}}}_l=w^{\comp{\set{J}}}_l}
\label{eq:use_full-rank}
\\[1mm]
& = &
\sum_{w^{\comp{\set{H}}}}\Prv{{W^{\comp{\set{J}}}_l}=w^{\comp{\set{J}}}_l}
\Bigl(\frac{1}{q}\Bigr)^h   
=\Bigl(\frac{1}{q}\Bigr)^h,\label{eq:use_independent-uniform}
\end{IEEEeqnarray}
where \eqref{eq:use_full-rank} follows from the fact that there is a linear transformation between $X^{(i_1)}_l,\ldots,X^{(i_h)}_l$ and $W^{(j_1)}_l,\ldots,W^{(j_h)}_l$, and \eqref{eq:use_independent-uniform} holds since $W^{(j_1)}_l,\ldots,W^{(j_h)}_l$ are i.i.d.\ over $\Field_q$. Hence, $\bigHP{X^{(i_1)}_l,\ldots,X^{(i_h)}_l}=h$ (in $q$-ary units), which completes the proof.


\section{Sign Assignment Procedure}
	\label{sec:sign_assignment}
	For completeness, in the following we represent the sign assignment procedure formally introduced in \cite[Sec.~IV-B]{SunJafar19_2} and make the necessary adaptation to our linearly-coded storage setup. 
	To present the sign assignment procedure, we introduce the following notation. For a lexicographically ordered $\tau$-sum query $q$, i.e., $q \triangleq \sum_{\ell=1}^{\tau} U^{(v_{\ell})}$, $v_1<\dots<v_\tau$, where the segment indices and the database index 
	are suppressed to simplify the notation, let $\Delta_{\vmat{U}^{(v)}}(q)$ denote the position of the symbol associated with the desired function evaluation $\vmat{X}^{(v)}$ within $q$. Note that $\Delta_{\vmat{U}^{(v)}}(q) \leq v$ and takes values in $\{0\} \cup[\tau]$, where $\Delta_{\vmat{U}^{(v)}}(q)=0$ indicates that the query does not contain a symbol from the desired function evaluation. We can now proceed to the sign assignment procedure. 

\indent	{\bf %
    Preliminaries
  :} Before the sign assignment process we sort the queries for each database as follows. First the queries are sorted in an increasing order of rounds $\tau\in[\mu]$, where the queries of the $\tau$-th round contain only  $\tau$-sums. Then, the queries in each round are divided into subgroups indexed by $S=S(\Delta_{\vmat{U}^{(v)}}(q)) \in \{1,2,\ldots\}$ based on the position of the symbol associated with the desired function evaluation in a lexicographic order of the query. Note that since $\Delta_{\vmat{U}^{(v)}}(q) \leq v$, we have $S\in[v+1]$. Finally, the subgroups are ordered in a descending order. 
To illustrate the sorting process, we use the queries generated for  \cref{ex:PLCex_n4k2f4mu4} of \cref{sec:private-linear-computation_coded-DSSs}. In \cref{tab:answers-table_Ex1_sorted}, the queries are sorted based on the round $\tau$ and the subgroup index $S$.

\begin{table}[thbp!]
	\centering
	\caption{PLC query sets for $v=1$ of \cref{ex:PLCex_n4k2f4mu4} after ordering of the queries based on the round and the position of the symbols associated with the desired function evaluation.}
	\label{tab:answers-table_Ex1_sorted}
	\vskip -1mm
	\Resize[\columnwidth]{
		\begin{IEEEeqnarraybox}[
			\IEEEeqnarraystrutmode
			\IEEEeqnarraystrutsizeadd{3pt}{2pt}]{v/c/v/c/v/c/v/c/v/c/v}
			\IEEEeqnarrayrulerow\\
			& (\tau, S, \Delta_{\vmat{U}^{(1)}}(q)) && j=1 && j=2 && j=3 && j=4\\
			\hline\hline
			&  (1, \dots, \dots )
			&& x_{{ 1},1}, y_{1,1}, z_{1,1},w_{1,1} &&  x_{{ 2},2}, y_{2,2},  z_{2,2},  w_{2,2} && x_{{1},3}, y_{1,3},  z_{1,3},  w_{1,3}  && x_{{ 2},4}, y_{2,4},  z_{2,4} , w_{2,4} &
			\\*\cline{1-11}      
			& \multirow{3}{*}{$(2, 1, 1)$}
			&& x_{ 3,1}+y_{2,1}&&  x_{4,2}+y_{1,2} && x_{ 3,3}
			+y_{2,3} && x_{ 4,4}+y_{1,4} &
			\\ 
			& 
			&& x_{ 5,1}+z_{2,1} &&  x_{6,2}+z_{1,2} && x_{ 5,3}+z_{2,3} && x_{6,4}+z_{1,4} &
			\\ 
			& 
			&& x_{ 7,1}+w_{2,1}&&  x_{ 8,2}+w_{1,2} && x_{ 7,3}+w_{2,3} && x_{ 8,4}+w_{1,4} &
			\\*\cline{1-11}      
			& \multirow{3}{*}{$(2, 2, 0)$}  
			&& y_{5,1}+z_{3,1}&&  y_{6,2}+z_{4,2} && y_{5,3}+z_{3,3} && y_{6,4}+z_{4,4} &
			\\ 
			&  && y_{7,1}+w_{3,1} &&  y_{8,2}+w_{4,2} && y_{7,3}+w_{3,3} && y_{8,4}+w_{4,4} &
			\\ 
			&  && { z_{7,1}+w_{5,1}} &&  { z_{8,2}+w_{6,2}} && { z_{7,3}+w_{5,3}} && { z_{8,4}+w_{6,4}} &
			\\*\cline{1-11}      
			& \multirow{3}{*}{$(3, 1, 1)$}
			&& x_{{9},1}+y_{6,1}+z_{4,1}  && x_{{ 10},2}+y_{5,2}+z_{3,2}
			&& x_{{9},3}+y_{6,3}+z_{4,3} && x_{{ 10},4}+y_{5,4}+z_{3,4} &
			\\ 
			& && x_{{11},1}+y_{8,1}+w_{4,1} && x_{{ 12},2}
			+y_{7,2}+w_{3,2}
			&& x_{{11},3}+y_{8,3}+w_{4,3} && x_{{ 12},4}+y_{7,4}+w_{3,4} &
			\\ 
			& && x_{{13},1}+z_{8,1}+w_{6,1}  && x_{{ 14},2}+z_{7,2}+w_{5,2}
			&& x_{{13},3}+z_{8,3}+w_{6,3} && x_{{ 14},4}+z_{7,4}+w_{5,4} &
			\\*\cline{1-11}      
			& (3, 2, 0)
			&& y_{13,1}+z_{11,1}+w_{9,1}   && y_{14,2}+z_{12,2}+w_{10,2}
			&& y_{13,3}+z_{11,3}+w_{9,3} && y_{14,4}+z_{12,4}+w_{10,4} &   
			\\*\cline{1-11}      
			& (4, 1, 1)
			&& x_{{15},1}+y_{14,1}+z_{12,1}+w_{10,1} 
			&& x_{{16},2}+y_{13,2}+z_{11,2} +w_{9,2} && x_{{ 15},3}+y_{14,3}+z_{12,3}+w_{10,3}
			&& x_{{ 16},4}+y_{13,4}+z_{11,4} +w_{9,4}&   
			\\*\IEEEeqnarrayrulerow
	\end{IEEEeqnarraybox}}
\end{table}

Next, we describe the main steps of the actual sign assignment procedure.
\begin{enumerate}
  \item Consider the undesired queries, i.e., queries that do not include any symbols from the desired function evaluation. For such a query $q$, $\Delta_{\vmat{U}^{(v)}}(q)=0$, and each element of the query $\tau$-sum that occupies an even position is given a negative sign `$-$'. In \cref{ex:PLCex_n4k2f4mu4}, the queries $y_{5,1}+z_{3,1}$ and $y_{13,1}+z_{11,1}+w_{9,1}$ of database~$1$ in the subgroups indexed by $(2,2,0)$ and $(3,2,0)$ (see rows $4$ and $6$ of \cref{tab:answers-table_Ex1_sorted}), will be transformed into $y_{5,1}-z_{3,1}$ and $y_{13,1}-z_{11,1}+w_{9,1}$, respectively. 
 \item Now consider the desired queries, i.e., queries containing a symbol from the desired function evaluation. When the symbol of a candidate function evaluation is assigned a negative sign in step 1) it is assigned a negative sign everywhere it appears. Note that given the coded storage setup, undesired symbols given a negative sign in step 1) appear exactly once within the queries of $\nu-\kappa$ databases. 
 \item Consider the desired queries. For such a query $q$, $\Delta_{\vmat{U}^{(v)}}(q)>0$, and each element of the query $\tau$-sum is multiplied by $(-1)^{S+\mathds{1}(v\neq 1)}$,  where $S \in[v+1]$ is the subgroup index. 
 \item A symbol of the desired function evaluation in each query is assigned a negative sign `$-$' if it occupies an even position within the query and a positive sign `$+$' if it occupies an odd position within the query. 
 \end{enumerate}
This concludes the sign assignment procedure. Following these steps, the queries of  \cref{ex:PLCex_n4k2f4mu4}  presented in \cref{tab:answers-table_Ex1_sorted} are transformed to the queries in \cref{tab:answers-table_Ex1}.

\section{Proof of Lemma~\ref{lem:redundancy}}
\label{sec:proof_PLCredundancy}
Here we present the main components needed for the proof of Lemma~\ref{lem:redundancy}, however the detailed derivations are a direct application of the proof of \cite[Lem.~1, Sec.~V-B]{SunJafar19_2} and thus are omitted. The proof of  \cite[Lem.~1]{SunJafar19_2} is adapted to our setup with the following substitutions.

Let $\set{L}\triangleq\{\ell_1,\ldots,\ell_r\}\subseteq [\mu]$ be the set of candidate function indices that satisfies \eqref{eq:rank-function_PC}. For the PLC representation of \eqref{eq:linear-mappingV}, $\bigl\{\vect{v}_{\ell_1},\dots,\vect{v}_{\ell_r}\bigr\}$ is a row basis of the coefficient matrix $\mat{V}_{\mu \times f}$, and  $r\bigl(\vmat{X}^{[\mu]}\bigr)=\rank{\mat{V}}\leq\min\{\mu,f\}$. 
	Assume, without loss of generality, that the rows of the coefficient matrix are ordered such that the first $r$ rows constitute  the row basis, i.e., $(X^{(1)}_l,\dots,X^{(r)}_l)=(X_l^{(\ell_1)},\dots,X_l^{(\ell_r)})$, $l\in[\const{L}]$.
	Note that we can represent the candidate functions evaluations $\bigl(X^{(1)}_l,\ldots,X^{(\mu)}_l)$ in terms of the basis candidate functions evaluations $\bigl(X^{(\ell_1)}_l,\ldots,X^{(\ell_r)}_l)$ for $l\in[\const{L}]$ with a deterministic linear mapping $\mat{\hat{V}}_{\mu\times r}$ of size $\mu\times r$ as follows:
	\begin{IEEEeqnarray*}{rCl}
		\begin{pmatrix}
			X^{(1)}_l \\
			\vdots \\
			X^{(\mu)}_l
		\end{pmatrix}=\mat{\hat{V}}_{\mu\times r}
		\begin{pmatrix}
			X^{(\ell_1)}_l
			\\
			\vdots
			\\
			X^{(\ell_r)}_l
		\end{pmatrix}.\label{eq:linear-mappingVnew}
	\end{IEEEeqnarray*}
	As a result, we have $\trans{(\trans{\hat{\vect{v}}_{1}},\dots, \trans{\hat{\vect{v}}_{r}})}= \mat{I}_r$, where $\mat{I}_r$ is the $r\times r$ identity matrix and $\hat{\vect{v}}_i$ is the $i$-th row vector of the deterministic linear mapping matrix $\mat{\hat{V}}_{\mu\times r}$.   
	First, consider the case where the desired function evaluation index $v=1$. Consider the queries corresponding to undesired $\tau$-sums, i.e., $\tau$-sums that do not involve any symbols from the desired function evaluation $\vmat{U}^{(1)}$.  There are $\mu-1 \choose \tau$ different $\tau$-sum types corresponding to such queries which can be divided into two groups as follows.
	\begin{itemize}
		\item Group~1: ${\mu-1\choose \tau}- {\mu-r \choose \tau}$ $\tau$-sum types for which the corresponding $\tau$-sums involve at least one element from the set $\{ \vmat{U}^{(2)},\vmat{U}^{(3)}, \dots,\vmat{U}^{(r)} \}$.
		\item Group~2: ${\mu-r \choose \tau}$ $\tau$-sum types for which the corresponding $\tau$-sums do not involve any element from the set $\{ \vmat{U}^{(2)},\vmat{U}^{(3)},\dots,\vmat{U}^{(r)} \}$.
	\end{itemize} 
	Let $q(U^{(v_{[\tau]})})$ denote a $\tau$-sum as defined in \cref{def:tau-sum} after performing the sign assignment process, i.e.,  
		\begin{IEEEeqnarray*}{c}
                  q(U^{(v_{[\tau]})}) \triangleq \sum_{\ell=1}^{\tau} (-1)^{\ell-1}U^{(v_{\ell})},
		\end{IEEEeqnarray*}
		where $v_{[\tau]}=\{v_1,\dots, v_{\tau}\} \subseteq [\mu]$, $v_1<\dots<v_\tau$, are the indices of the functions evaluations, and where the segment indices and the database index are suppressed to simplify the notation. Let the \emph{type} of the $\tau$-sum be presented by the set of distinct indices of functions evaluations involved in the $\tau$-sum, i.e., the type of $q(U^{(v_{[\tau]})})$ is represented by $v_{[\tau]}=\{v_1, \dots, v_{\tau}\}$. 
		The key idea is to show that the symbols of the queries corresponding to Group 2 are deterministic linear functions of the queries corresponding to Group 1 when the symbols of the desired function evaluation $\vmat{U}^{(1)}$ are known. 
		Now, let 
                $q_0\eqdef q(U^{(v_{[\tau]})})$,
	where $r<v_1<\dots<v_\tau$, denote an arbitrary query corresponding to Group~2. 
	Specifically, we need to show that, when the symbols of $\vmat{U}^{(1)}$ queried by the given database are known, i.e., successfully decoded, the query $q_0$ can be written as a linear function of ${\tau+r-1\choose \tau}-1$ queries corresponding to Group~1. These ${\tau+r-1\choose \tau}-1$ queries contain elements of the row basis functions evaluations and elements included in the $\tau$-sum of $q_0$. These queries comprise all the $\tau$-sums of types in the set $\mathcal{I}\triangleq {[2:r]\cup v_{[\tau]}}$ except the type of $q_0$, i.e., $\{v_1, \dots, v_{\tau}\}$.  Now, let $\tilde{\mathcal{Q}} \triangleq \bigl\{ q(U^{(\hat{i}_{[\tau]})}) : \hat{i}_{[\tau]} \in 
        \mathcal{T} \bigr\}$ be a set of queries where there is exactly one query corresponding to  each of the ${\tau+r-1\choose \tau}-1$ $\tau$-sum types of Group~1, where 
	$\mathcal{T}\eqdef\bigl\{\hat{i}_{[\tau]}=\{\hat{i}_1,\hat{i}_2,\ldots,\hat{i}_\tau\}\in\set{I}\colon \hat{i}_{[\tau]}\neq v_{[\tau]}\bigr\}$
        . 
	Finally, assume, without loss of generality, that the subsets of distinct indices $\hat{i}_{[\tau]} \in \mathcal{T}$ are ordered in natural lexicographical order, i.e., $\hat{i}_{1}< \hat{i}_{2}<\dots<\hat{i}_{\tau}$. 
	Next, from the deterministic linear mapping between the candidate functions evaluations, $\hat{\mat{V}}_{\mu\times r}$, we have 
	\begin{IEEEeqnarray*}{c}
          U^{(v_\ell)}_{*}= \hat{v}_{v_\ell,1} U^{(1)}_{*}+\dots+ \hat{v}_{v_\ell,r} U^{(r)}_{*}, \quad  \ell\in[\tau],
        \end{IEEEeqnarray*}
where $(\hat{v}_{v_\ell,1} ,\dots, \hat{v}_{v_\ell,r})= \hat{\vect{v}}_{v_\ell} $. Now, we need to show that $q_0$ is a linear function of the queries of $\tilde{\mathcal{Q}}$ as follows:
	\begin{IEEEeqnarray}{c}
          \label{eq:redundacy_key}
	 q_0=\sum_{\hat{i}_{[\tau]} \in \mathcal{T}} h(U^{(\hat{i}_{[\tau]})}) q(U^{(\hat{i}_{[\tau]})}),
\end{IEEEeqnarray}
where $h(U^{(\hat{i}_{[\tau]})} )$ is a linear coefficient calculated as a function of the deterministic linear mapping coefficients represented by the matrix 
\begin{IEEEeqnarray*}{c}
	\label{eq:coef_matrix}
	\hat{\mat{V}}^{*}_{(r-1)\times\tau}= \begin{pmatrix}
		\hat{v}_{v_1,2} & \hat{v}_{v_2,2} & \cdots & \hat{v}_{v_\tau,2}\\
		\hat{v}_{v_1,3} & \hat{v}_{v_2,3} & \cdots & \hat{v}_{v_\tau,3}\\
		\vdots & \vdots & \cdots & \vdots \\
		\hat{v}_{v_1,r} & \hat{v}_{v_2,r} & \cdots & \hat{v}_{v_\tau,r}\\
	\end{pmatrix}
      \end{IEEEeqnarray*}
      as outlined in \cite[Sec.~V-B]{SunJafar19_2}.
 Given the above problem setup, notation, and definitions, one can verify that \eqref{eq:redundacy_key} holds for all queries corresponding to Group~2 (refer to \cite[Sec.~V-B]{SunJafar19_2} for the detailed derivation). Thus, a number of ${\mu-r \choose \tau}$ query types in Group~2 are redundant and can be removed from the query set.

	
\section{Proof of Lemma~\ref{lem:PPCredundancy}}
\label{sec:proof_PPCredundancy}
		The proof of \cref{lem:PPCredundancy} relies on two arguments as follows.
		\begin{enumerate}
			\item If the candidate polynomial functions set includes all monomials of degree less or equal to $g$, we can represent the PPC problem as an allied PLC problem over a set of dependent virtual messages, i.e., all monomials. Accordingly, we order the candidate polynomial functions according to their degree and find a deterministic linear mapping matrix $\mat{\hat{V}}_{\mu\times {\const M}_g(f)}$ of size $\mu\times {\const M}_g(f)$ that represents the candidate polynomial functions as linear combinations of their monomial basis. 
			Moreover, following the ordering of the candidate polynomial functions based on degree, we have $\trans{(\trans{\hat{\vect{v}}_{1}},\dots,\trans{\hat{\vect{v}}_{{\const M}_g(f)}})} = \mat{I}_{{\const M}_g(f)}$, where $\mat{I}_{{\const M}_g(f)}$ is the ${\const M}_g(f) \times {\const M}_g(f)$ identity matrix and $\hat{\vect{v}}_i$ is the $i$-th row vector of the allied polynomial coefficient matrix $\mat{\hat{V}}_{\mu\times {\const M}_g(f)}$. With this mapping one can show, by using a similar enumeration as in the proof of Lemma 3 and without loss of generality, for a desired polynomial indexed by $v=1$ that the types of $\tau$-sums corresponding to undesired queries, i.e., $\tau$-sums that do not involve any symbols from the desired function evaluation $\vmat{U}^{(1)}$   can be divided into two groups as follows.
			\begin{itemize}
				\item Group~1: ${\mu-1\choose \tau}- {\mu- {\const M}_g(f) \choose \tau}$ $\tau$-sum types for which the corresponding $\tau$-sums  involve at least one element from the set $\{ \vmat{U}^{(2)},\vmat{U}^{(3)}, \dots,\vmat{U}^{({\const M}_g(f))} \}$.
				\item Group~2: ${\mu-{\const M}_g(f) \choose \tau}$ $\tau$-sum types for which the corresponding $\tau$-sums do not involve any element from the set $\{ \vmat{U}^{(2)},\vmat{U}^{(3)},\dots,\vmat{U}^{({\const M}_g(f))} \}$
			\end{itemize}
			such that the symbols of the queries corresponding to Group~2 are functions of the symbols of the queries corresponding to Group~1 when the symbols of the desired function evaluation are known. 
			\item Following the above argument, we have for the first round ($\tau=1$)   $\mu-{\const M}_g(f)$ redundant $1$-sum types based on the linear dependencies between the polynomial functions and their monomial basis. However, given that in this round we query $1$-sums, we can further eliminate the nonlinear dependent symbols, i.e., ${\const M}_g(f)-f$ $1$-sum types regardless of the desired polynomial function evaluation. 
		\end{enumerate}
		Accordingly, with the mapping of the allied PLC problem, the  derivations of the first argument follow directly from \cref{lem:redundancy} and \cite[Lem.~1]{SunJafar19_2}, and thus are omitted for brevity.

\bibliographystyle{IEEEtran} 
\bibliography{references.bib,defshort1.bib, biblioHY.bib}

\end{document}